\documentclass[journal]{IEEEtran}
\usepackage{amsmath,amsthm,amssymb}
\usepackage{enumerate}
\usepackage{color}
\usepackage{xcolor}
\usepackage{url}
\usepackage{balance}
\usepackage{graphicx} 
\usepackage{mathrsfs}
\usepackage{epsfig}
\usepackage{verbatim}
\usepackage{setspace}
\usepackage{cite}

\newtheorem{theorem}{Theorem}
\newtheorem{lemma}[theorem]{Lemma}

\newtheorem{claim}[theorem]{Claim}
\newtheorem{remark}[theorem]{Remark}

\DeclareMathOperator*{\argmin}{arg\,min}
\DeclareMathOperator*{\argmax}{arg\,max}

\renewcommand{\vec}[1]{\mathbf{#1}}
\newcommand{\bX}{\mathbf{X}}
\newcommand{\bx}{\mathbf{x}}
\newcommand{\bY}{\mathbf{Y}}
\newcommand{\bZ}{\mathbf{Z}}
\newcommand{\by}{\mathbf{y}}

\newcommand{\bS}{\mathbf{S}}
\newcommand{\bs}{\mathbf{s}}
\newcommand{\bJ}{\mathbf{J}}
\newcommand{\bj}{\mathbf{j}}
\newcommand{\bU}{\mathbf{U}}
\newcommand{\bu}{\mathbf{u}}

\newcommand{\cU}{{\cal U}}
\newcommand{\cT}{{\cal T}}
\newcommand{\cX}{{\cal X}}
\newcommand{\cY}{{\cal Y}}
\newcommand{\cZ}{{\cal Z}}
\newcommand{\cS}{{\cal S}}

\newcommand{\cJ}{{\cal J}}
\newcommand{\cP}{{\cal P}}

\newcommand{\cM}{{\cal M}}

\newcommand{\td}[1]{\tilde{#1}}

\newcommand{\removed}[1]{}

\title{Communication in the Presence of a State-Aware Adversary}

\author{  Amitalok J. Budkuley,~\IEEEmembership{Member, ̃IEEE,
}  Bikash Kumar Dey,~\IEEEmembership{Member, ̃IEEE,\\
}  Vinod M. Prabhakaran,~\IEEEmembership{Member, ̃IEEE.  }
\\
\thanks{  A.~J.~Budkuley was with the Department of Electrical Engineering at the Indian Institute of Technology Bombay, Mumbai, India. He is now with the Department of Information Engineering, 	The Chinese University of Hong Kong, Sha Tin, Hong Kong (e-mail: amitalok@ie.cuhk.edu.hk).   }
\thanks{B.~K.~Dey is with the Department of Electrical Engineering at the Indian Institute of Technology Bombay, Mumbai, India (e-mail: bikash@ee.iitb.ac.in). }  
\thanks{V.~M.~Prabhakaran  is  with  the  Tata  Institute  of  Fundamental  Research,  Mumbai,  India (e-mail: vinodmp@tifr.res.in).} 

\thanks{This paper was presented in part at the IEEE Information Theory Workshop 2015 held at Jeju, South Korea.  }
\removed{
\thanks{   A.~J.~Budkuley and B.~K.~Dey were supported in part by Bharti Centre for Communication, IIT Bombay, and in part by Information Technology Research Academy (ITRA), Government of India under ITRA-Mobile grant ITRA/15(64)/Mobile/USEAADWN/01. In addition, B.~K.~Dey was supported in part by the Department of Science \& Technology, Government of India under a grant SB/S3/EECE/057/2013. V.~M.~Prabhakaran was supported in part by the Department of Science \& Technology, Government of India through the Ramanujan Fellowship and in part by Information Technology Research Academy (ITRA), Government of India under ITRA-Mobile grant ITRA/15(64)/Mobile/USEAADWN/01. } 
}
}

\begin{document}
\maketitle

\interdisplaylinepenalty=2500
\interfootnotelinepenalty=10
\removed{
\IEEEpubid{\begin{minipage}{\textwidth}\ \\[12pt] \centering\\
   \copyright~2017 IEEE. Personal use of this is permitted. \\
	However, permission to use this material for any other purposes must be obtained from the IEEE by sending a request to pubs-permissions@ieee.org.
\end{minipage}} 
}
\begin{abstract} We study communication systems over the state-dependent
channels in the presence of a malicious state-aware jamming adversary.
The channel has a memoryless state with an underlying distribution.
The adversary introduces a jamming signal into the channel. 
The message and the entire state sequence are known non-causally to both the encoder and the adversary.
This state-aware adversary may choose an arbitrary jamming vector depending on
the message and the state vector.
Taking an Arbitrarily Varying Channel (AVC) approach, we consider two
setups, namely, the discrete memoryless Gel'fand-Pinsker (GP) AVC and the
additive white 
Gaussian Dirty Paper (DP) AVC. We determine the randomized coding capacity
of both the AVCs under a maximum probability of error criterion.
Similar to other randomized coding setups, we show that the capacity
is the same even under the average probability of error criterion. Though
the adversary can choose an arbitrary vector jamming strategy, we prove that the
adversary cannot affect the rate any worse than when it employs a
memoryless strategy which depends only on the instantaneous state.  
Thus, the AVC capacity characterization is given in terms of the
capacity of the worst memoryless channels with state, induced by the
adversary employing such memoryless jamming strategies. For the DP-AVC, it
is further shown that among memoryless jamming strategies, none impact the
communication more than a memoryless
Gaussian jamming strategy which completely disregards the knowledge of the
state. Thus, the capacity of the DP-AVC equals that of a standard AWGN
channel with two independent sources of additive white Gaussian noise,
i.e., the channel noise and the jamming noise. 
\end{abstract}
%
\section{Introduction}\label{sec:introduction}

\begin{figure}[!ht]
  \begin{center}
    \includegraphics[trim=0cm 10cm 0cm 1cm, scale=0.3]{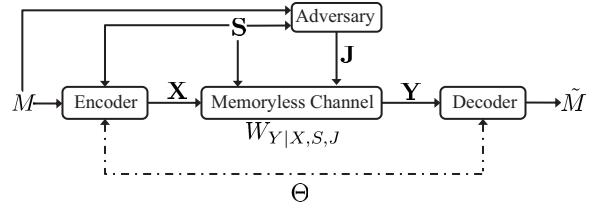}
    \caption{Our general  communication system setup with a state-aware jamming adversary.}
    \label{fig:general:setup}
  \end{center}
\end{figure}

We consider the problem of reliable communication over a state-dependent
channel in the presence of a jamming adversary. In our generic problem setup depicted
in Fig.~\ref{fig:general:setup}, a message $M$ is to be communicated reliably
over a channel with an independent and identically distributed (i.i.d.) state vector $\vec{S}$ and an adversarial 
jamming signal $\bJ$.  The state is known non-causally to the encoder.
The adversary too knows $M$ as well as state $\vec{S}$ non-causally. The
encoder and decoder share an unbounded amount of randomness, $\Theta$,
pre-shared and unknown to the adversary. We consider both the discrete memoryless 
channel version and the additive white Gaussian version of the setup as elaborated later.
Our aim is to determine the capacity of this
communication system. An allied interest is to understand the behaviour of
the adversary; specifically, its use of the knowledge of state $\vec{S}$ in the design
of its jamming strategy. 

State-dependent channels, where the state is known non-causally at the
transmitter, have been a subject of considerable interest since the seminal
work of Gel'fand and Pinsker~\cite{gelfand-pinsker}. In their work, the
capacity of the discrete memoryless channel version was established.
Henceforth, we refer to this channel as the `Gel'fand-Pinsker (GP) channel'.
Subsequently, using a coding scheme based on the technique
in~\cite{gelfand-pinsker}, called the \emph{dirty paper coding} scheme,
Costa~\cite{costa} determined the capacity of the Gaussian version of this
problem, i.e., the capacity of an AWGN channel with an additive white Gaussian state,
where the state is known non-causally to the encoder. Interestingly, Costa
showed that the effect of the additive state can be completely nullified.
Hence, the capacity of this {\it dirty paper}  channel was shown to be
equal to that of a standard AWGN channel without state. 
Thus, an intelligent use of the state knowledge, even when available only at the
encoder, enables the user to cancel its effect. In our setup,
we additionally assume  the presence of a \emph{state-aware} adversary,
i.e., an adversary with non-causal knowledge  of the state. An intelligent
adversary can use this knowledge to design a pernicious jamming strategy.
We study the impact of such an adversary on reliable communication. 

Our setup falls in the general framework of Arbitrarily Varying Channels
(AVC), and the interest lies in determining the randomized coding
capacity~\cite{lapidoth-narayan-it1998} of this setup. Note that many works on AVCs (for instance, see~\cite{csiszar-narayan-it1988,lapidoth-narayan-it1998}) refer to the adversary's
channel input $\vec{J}$ as \emph{state}. However, to avoid confusion, in this work we use the word
\emph{state} to refer exclusively  to the channel state $\vec{S}$ and
refer to $\vec{J}$ as the jamming signal. Thus, while state $\bS$ is probabilistic,
the jamming signal $\bJ$ is adversarial. \label{page:state} 
The adversary knows $M$ and $\bS$ prior to deciding the jamming vector
$\bJ$. So the process of coming up with $\bJ$ by the adversary can be
represented by a conditional distribution $Q_{\bJ|M,\bS}$, unknown
to the encoder and the decoder.
Such {\em stochastic} vector jamming strategies clearly include deterministic
jamming strategies which are functions of the message and the state vector. 
In addition, they capture possible randomization used by the adversary\footnote{
In fact, without loss of generality, we may restrict attention to deterministic jamming strategies; see footnote~\ref{footnote:deterministic_jammer_equivalence} on page~\pageref{footnote:deterministic_jammer_equivalence}. However, as in \cite{hughes-narayan-it1987}, in this paper we will consider stochastic jamming strategies. This is in the interest of simpler converse proofs.}.
\label{resp:jam:seq} 

For each message, we take the maximum value of probability of error over  all feasible
stochastic jamming strategies $Q_{\bJ|M,\bS}$. Further, our error criterion is the maximum (over messages) probability of error (see~\eqref{eq:maxpe:1} for the expression) and we determine the randomized capacity under
this error criterion. Note that the deterministic coding capacity problem in
this setting is a hard problem\footnote{ Even in the absence of the state $S$, the
deterministic coding capacity under maximum probability of error criterion
is related (cf.~\cite{ahlswede-ams1970,csiszar-korner-1981}) to Shannon's zero-error capacity~\cite{shannon-ire1958}, 
whose characterization is known to be a hard problem. }, and not
addressed in this work. Hence, unless stated otherwise, the term
\emph{capacity} will hereafter refer to the randomized 
capacity\footnote{In fact, this capacity remains unchanged for the average probability of error criterion. See Remark~\ref{rem:averagepe} on page~\pageref{rem:averagepe}. }. 
The adversary is aware of the state and the message, and furthermore, is assumed to also know the distribution of the randomized code. 
\label{resp:r:code}
 However, this randomized code is generated using randomness which is shared only between the encoder and decoder, and thus, its exact realization is unknown to the adversary. In particular, owing to the randomized encoding map the adversary does not know the transmitted codeword even though it knows the message. 
In this work, we consider two variants of the setup: the discrete memoryless Gel'fand-Pinsker AVC (GP-AVC) and the additive white Gaussian Dirty Paper AVC (DP-AVC), and determine their randomized capacity. As in many randomized coding
setups (for instance, see~\cite{csiszar-narayan-it1988,hughes-narayan-it1987}), we show that the capacity is the same even under an average (over messages) probability of error criterion.
 
Subsequent to~\cite{blackwell-ams1959}, where the AVC model was introduced,
several works analysed different AVC models. In
general, the capacity of an AVC communication system depends upon several
factors, viz., possibility of randomization (unknown to the adversary) at the
encoder/decoder, the probability of error criterion, assumptions on the adversary's
knowledge, etc.~\cite{lapidoth-narayan-it1998}. 
\label{avc:disc}
In the absence of state constraints, it is known that the deterministic coding capacity under
average error criterion of the AVC exhibits a dichotomy - it is 
zero if the AVC is symmetrizable 
 or is equal to the randomized coding capacity otherwise~\cite[Theorem~1]{csiszar-narayan-it1988-2}. 
An AVC $W_{Y|X,J}$, where $X\in\cX$, $J\in\cJ$ and $Y\in\cY$, is said
to be \emph{symmetrizable} if for some conditional distribution $V_{J|X}$
\\
\\
\begin{IEEEeqnarray*}{rCl}
&&\sum_{j} W_{Y|X,J}(y|x,j) V_{J|X}(j|x')\\
\>&&=\sum_{j} W_{Y|X,J}(y|x',j) V_{J|X}(j|x),\,\,\,\text{for every }x, x', y.
\end{IEEEeqnarray*}
However, the deterministic coding capacity under the maximum error criterion, 
of which Shannon's zero error capacity problem~\cite{shannon-ire1958} 
is a special case, is not known in 
general~\cite{ahlswede-ams1970,csiszar-korner-1981}. 
For a lucid  exposition on AVCs and a survey of many useful results, see~\cite{lapidoth-narayan-it1998}. 

To provide context to our work, we review certain
important results. In the standard point to point AVC setup under
randomized coding, models of adversary ranging from the oblivious adversary
(no knowledge of the codeword) to the codeword-aware adversary have been
considered~\cite{blackwell-ams1959,csiszar-narayan-it1988,langberg-it2008}. More generally, the myopic adversary which observes 
a noisy version of the codeword is analysed in~\cite{sarwate-itw2010,dey-isit2015}. 
The Gaussian versions of these problems~\cite{agarwal,hughes-narayan-it1987,sarwate-spcom2012} have
also been considered. An adversary with a causal view of the codeword~\cite{dey-causal-it2013} or a delayed view of the
codeword~\cite{dey-delayed-it2013}, has also been studied. The capacity
of an AVC version of the Gel'fand-Pinsker problem under deterministic coding is
determined by Ahlswede~\cite{ahlswede-1971}. Unlike our setup,
this model has only an adversarial state (known to the encoder), but
does not have an additional probabilistic state. 
The case where the decoder too is aware of the state is considered in~\cite{ahlswede-wolfowitz-elsevier1969}.
A model similar to our DP-AVC, but with a {\em state-oblivious}
adversary under  deterministic coding, is analysed
in~\cite{sarwate-isit2008}. The result under randomized coding also appears
there without proof. Our models have a stronger,
{\em state-aware,} adversary. Communication setups involving both jamming
and secrecy have been studied in~\cite{molavianjazi-etal-all2009,bjelakovic-etal-pit2013,dey-isit2015}. 
Achievability results for secret communication over the Gel'fand-Pinsker wiretap setup too have recently appeared~\cite{goldfeld-etal-arxiv2016}.

\begin{figure}[!ht]
  \begin{center}
    \includegraphics[trim=0cm 10cm 0cm 0cm, scale=0.3]{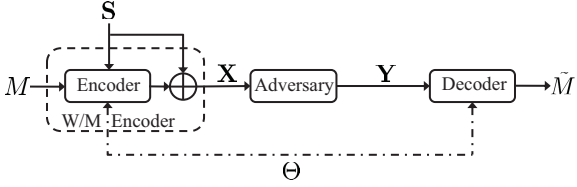}
    \caption{The watermarking game setup~\cite{cohen-lapidoth-it2002}.}
    \label{fig:info:hiding}
  \end{center}
\end{figure}

Closely related to our problem are also problems on information hiding.
Information hiding finds application in watermarking, fingerprinting,
steganography, etc. (cf.~\cite{swanson-ieee98,hernandez-ieee99}). An information-theoretic approach to the problem of
information hiding appears in~\cite{osullivan-moulin-ettinger-isit1998}, where
information hiding under distortion-attack adversaries is studied. Further
results on such \emph{watermarking games} can be found in subsequent works
like~\cite{cohen-lapidoth-it2002,moulin-osullivan-it2003,moulin-wang-it2007}
and some of the references therein. However, there are important differences
between these problems and our problem. In a generic watermarking game 
depicted in Fig.~\ref{fig:info:hiding} (also see, for instance,~\cite[Fig. 2]{cohen-lapidoth-it2002}),
the aim is to reliably communicate a message $M$ over a channel controlled by an adversary, by embedding it into a covertext
source (state $\vec{S}$), like an image. 
\label{watermarking}
The embedding process distorts $\vec{S}$, and the resulting data $\vec{X}$,
called stegotext, is directly observed by the adversary. The adversary, who
may or may not know $\vec{S}$, is capable of distortion attacks, and hence, can
further distort this text arbitrarily but within some overall distortion limit
(adversary's power constraint). In the watermarking game, unlike in our problem, the adversary knows
the distorted covertext, and thus, can correlate with and cancel it, partially
or fully, depending on its power. On the other hand, in our setup the adversary knows the covertext (i.e. state $\vec{S}$) but not $\bX$.
Equivalently, this means that the adversary is state-aware but not aware of the
transmitted vector. This difference has a major effect on the behaviour of the
adversary as well as the capacity of the system. In the Gaussian analogue of the watermarking problem,
considered for instance in~\cite{cohen-lapidoth-it2002}, it is seen that a
sufficiently strong adversary can force the capacity of this distortion attack
channel to zero. On the contrary, it will be shown that for our setup,
the capacity is always greater than zero for any finite value
of adversary's power. 
\subsection{Contribution and Organization of the Paper}
In Section~\ref{sec:problem}, we begin by describing the notation used in this
work, and then present our communication setups, viz., the discrete
Gel'fand-Pinsker AVC (GP-AVC) and the Gaussian Dirty Paper AVC (DP-AVC).
We state our main results in Section~\ref{sec:main:results}. Here is a summary of our contributions.  
\begin{itemize}
\item We present the capacity of the Gel'fand-Pinsker AVC (GP-AVC) in Theorem~\ref{thm:main:result:dmc}. 
Though the adversary is allowed to use an arbitrary vector jamming strategy, 
the AVC capacity is characterized through the capacity of the worst
Gel'fand-Pinsker channel that the adversary can induce using a memoryless
strategy. Towards proving our result, we also present a Refined Markov
Lemma (Lemma~\ref{lem:gen:markov:lemma}). This lemma
is a refined version of~\cite[Lemma~12.1]{elgamal-kim} and may also be useful in 
analysing other systems with adversaries (see Remark~\ref{rem:markov}(i) on page~\pageref{rem:markov}).
Our converse considers a memoryless (but not identically distributed)
jamming strategy, which depends on the encoder design, to upper bound the rate. 
\item We present the capacity of the Dirty Paper AVC (DP-AVC) in 
Theorem~\ref{thm:main:result:gaussian}. 
Interestingly, it is shown that the adversary, given its purpose, cannot do better than choosing an adversarial strategy that completely disregards the state knowledge and essentially performs i.i.d. Gaussian jamming, independent of the state.  
Here the user employs an appropriate dirty paper coding scheme. 
\label{resp:rev2}
As a consequence, the capacity of this channel 
is shown to be equal to that of a standard AWGN channel with no state which has two independent sources of zero mean additive white Gaussian noises, one with variance $\sigma^2$ and the other with $\Lambda$. 
Note that a result on the Gaussian version of the dirty
paper coding  setup with a {\it state-oblivious} adversary appears
in~\cite{sarwate-isit2008} without proof. However, we prove
the same capacity for a {\it state-aware}  adversary, and thus, our result subsumes the
result for the state-oblivious adversary. 
\item As known in other randomized coding setups, we observe
(see Remark~\ref{rem:averagepe} on page~\pageref{rem:averagepe}) that for both the GP-AVC and the DP-AVC, the capacity is
identical under both the  maximum and average probability of error criteria. 
\end{itemize}
The proofs of Theorems~\ref{thm:main:result:dmc} and~\ref{thm:main:result:gaussian} are given in Section~\ref{sec:proofs}. We discuss  some implications of our work and make  overall concluding remarks in Section~\ref{sec:conclusion}. The proofs of other auxiliary lemmas are given in the appendices.
\section{Notation and Problem Setup}\label{sec:problem}
\subsection{Notation} 
We denote random variables by upper case letters (e.g. $X$), the values they take by lower case letters (e.g. $x$) and their alphabets by calligraphic letters (e.g. $\mathcal{X}$). We assume all discrete random variables to have alphabets of finite size, unless stated otherwise. The continuous random variables take values in the set of real numbers $\mathbb{R}$. Let $\mathbb{R}^+$ denote the set of non-negative real numbers. We use boldface notation to denote random vectors (e.g. $\vec{X}$) and their values (e.g. $\vec{x}$). Here the vectors are of length $n$ (e.g. $\vec{X}=(X_1,X_2,\dots,X_n)$), where $n$ is the block length of operation. Let us also denote $\vec{X}^{i}=(X_1,X_2,\dots,X_i)$ and $\vec{x}^{i}=(x_1,x_2,\dots,x_i)$ as well as  $\bX_{i}^{k}=(X_i,X_{i+1},\dots,X_k)$ and $\bx_{i}^{k}=(x_i,x_{i+1},\dots,x_k)$. We use the $l_{\infty}$ norm for discrete vectors and the $l_2$ norm for continuous vectors. We denote the former by $\|.\|_{\infty}$ and the latter by $\|.\|$, where we drop the subscript. For a set $\mathcal{X}$, let $\cP(\mathcal{X})$ be the set of all probability distributions on $\mathcal{X}$. Similarly, let us write as $\cP(\mathcal{X}|\mathcal{Y})$, the set of all conditional distributions of a random variable with alphabet $\mathcal{X}$ conditioned on another random variable with alphabet $\mathcal{Y}$. Let $X$ and $Y$ be two  random variables. Then, we denote the distribution of $X$ by $P_{X}(\cdot)$, the joint distribution of $(X,Y)$ by $P_{XY}(\cdot,\cdot)$ and the conditional distribution of $X$ given $Y$ by $P_{X|Y}(\cdot|\cdot)$.
Distributions corresponding to strategies adopted by the adversary are denoted by $Q$ instead of $P$ for clarity. 
In cases where the subscripts are clear from the context, we sometimes omit them to keep the notation simple. 
For an event $E$, let $\mathbb{P}(E)$ denote the probability of $E$. Functions will be denoted in lowercase letters (e.g. $f$). A Gaussian distribution with mean $\mu$ and variance $\sigma^2$ is denoted by $\mathcal{N}(\mu,\sigma^2)$. 
All logarithms are with base $2$, and hence, all rates and capacities are expressed in bits. 
\subsection{The Gel'fand-Pinsker AVC (GP-AVC)}
In the communication setup depicted in Fig.~\ref{fig:general:setup},
there is an arbitrarily varying channel with input $X$, output $Y$, state $S$, and an input
$J$ of an adversary. These random variables take values in
the finite sets $\mathcal{X}$, $\mathcal{Y}$, $\mathcal{S}$, and $\mathcal{J}$
respectively. The states in different channel uses are i.i.d. with distribution $P_{S}$. We assume without loss of generality that $P_{S}(s)>0$, $\forall s\in\mathcal{S}$. The channel behaviour is given by the conditional
distribution $W_{Y|X,S,J}$. A standard block-coding framework is
considered where a message $M$ is communicated over $n$ channel
uses. Let $X_i$, $Y_i$, $S_i$ and $J_i$ denote the symbols of the respective
random variables associated with the $i$-th channel use. The encoder
as well as the adversary are assumed to know the state vector ${\bf S}$
non-causally before deciding their input vectors ${\bf X}$ and ${\bf J}$ respectively.
The encoder and the decoder share unlimited common randomness $\Theta$,
unknown to the adversary. Thus, the transmitted vector $\vec{X}$ is a function of 
$M$, ${\bf S}$ and $\Theta$. Hence, we consider {\em randomized coding.} 
Similarly, the \emph{state-aware} adversary chooses
its own channel input $\vec{J}$. Let the distribution used by the adversary be denoted by
$Q_{\vec{J}|M,\vec{S}}$. Note that the adversary does not have knowledge of
$\Theta$. For a given $\vec{x}$, $\vec{s}$ and $\vec{j}$, the channel
output $\vec{y}$ is observed over the channel $W_{Y|X,S,J}$ with
probability given by
\begin{IEEEeqnarray*}{rCl}
\mathbb{P}(\vec{Y}=\vec{y}|\bX=\vec{x},\bS=\vec{s},\bJ=\vec{j})=\prod_{i=1}^n W_{Y|X,S,J}(y_i|x_i,s_i,j_i).
\end{IEEEeqnarray*}
We call this channel the Gel'fand-Pinsker AVC (GP-AVC). 

An $(n,R)$ \emph{deterministic code} of block length $n$ and rate $R$ is a
pair $(\psi,\phi)$ of mappings with encoder $\psi:\{1,2,\dots,2^{nR}\}\times
{\mathcal{S}}^n\rightarrow {\mathcal{X}}^n$ and decoder $\phi:
{\mathcal{Y}}^n\rightarrow \{0,1,2,\dots,2^{nR}\}$, where an output of $0$ indicates that the decoder declares an error. Here we have assumed $2^{nR}$ to be an integer.
The vector transmitted on the channel is given by $\vec{X} = \psi(M,\vec{S})$. 

An $(n,R)$ \emph{randomized code} of block length $n$ and rate $R$ is a random
variable ($\Theta$ in this case) which takes values in the set of $(n,R)$
deterministic codes. Let the pair $\Theta=(\Psi,\Phi)$ denote the encoder-decoder for
the $(n,R)$ randomized code. In this case, the transmitted vector is given by
$\vec{X} = \Psi(M,\vec{S})$.

For this $(n,R)$ randomized code, the \emph{maximum probability of error}
is\footnote{\label{footnote:deterministic_jammer_equivalence}It is clear that, without loss of generality, we may assume that the adversary is deterministic, and not stochastic. Specifically, for a given $(m,\vec{s})$, the optimal jamming signal is given by
\[ \arg \max_{\vec{j}} \mathbb{P}(\Phi(\bY)\neq m| M=m,\vec{S}=\vec{s},\vec{J}=\vec{j}),\]
where the probability is over $\Theta$ and the channel $W_{Y|X,S,J}$ (if there are multiple maximizers, one among them may be chosen arbitrarily). Hence, capacity of GP-AVC defined here is the same under stochastic and deterministic jamming. This not withstanding, we will proceed to consider stochastic adversaries in this paper. This makes some of the converse proofs in the sequel slightly simpler.}  
\begin{equation}\label{eq:maxpe:1}
P_e^{(n)}=\max_{m} \max_{Q_{\vec{J}|M=m,\vec{S}}} \mathbb{P}(\Phi(\vec{Y})\neq m|M=m),
\end{equation}
where the probability is  over the state $\vec{S}$, the adversary's 
action $\vec{J}$, the channel behavior and $\Theta=(\Psi, \Phi)$.
The rate $R$ is \emph{achievable} if for any $\epsilon>0$, there exists an $(n,R)$ randomized code for some $n$ such that the corresponding  $P_e^{(n)}$ is less than $\epsilon$. We define the \emph{capacity} of the GP-AVC as the supremum of all achievable rates.
\subsection{The Dirty Paper AVC (DP-AVC) }
\begin{figure}[!ht]
  \begin{center}
    \includegraphics[trim=0cm 10cm 0cm 1cm, scale=.3]{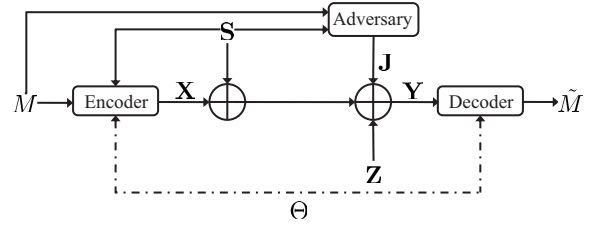}
    \caption{The Dirty Paper AVC (DP-AVC)  communication setup.}
    \label{fig:gaussian:main:setup}
  \end{center}
\end{figure}
The communication channel depicted in Fig.~\ref{fig:gaussian:main:setup} is a Gaussian arbitrarily varying channel with an additive white Gaussian state and an additive jamming interference. The encoder and decoder share an unbounded amount of common randomness $\Theta$, unknown to the adversary. Let us denote by $\vec{Y}=(Y_1,Y_2,\ldots,Y_n)$, the signal received at the decoder. Then,
\begin{equation*}\label{eq:Y}
\vec{Y}=\vec{X}+\vec{S}+\vec{J}+\vec{Z},
\end{equation*}
where $\vec{X}$, $\vec{S}$, $\vec{J}$ and $\vec{Z}$ are the encoder's input to
the channel, the additive white Gaussian state, adversary's channel input and the
channel noise respectively. The components of $\vec{S}$ are i.i.d. with
$S_i\sim\mathcal{N}(0,\sigma_S^2)$ for $i=1,2,\dots,n$. The components of
$\vec{Z}$ are i.i.d. with $Z_i\sim \mathcal{N}(0,\sigma^2)$, $\forall i$.
Similar to the GP-AVC, the state vector $\vec{S}$ is known non-causally to both the encoder and the
adversary, but it is not known to the decoder. Hence, the encoder's output $\vec{X}$ is a function of $M$, $\vec{S}$ and $\Theta$. We call this channel the Dirty Paper AVC (DP-AVC). The encoder has a power constraint $P$, i.e.
$\|\vec{X}\|^2 \leq n P$. Similarly, the adversary's power constraint is $\Lambda$,
such that $\|\vec{J}\|^2 \leq n \Lambda$. Let $\mathcal{J}(\Lambda)=\left\{
\vec{j}: \|\vec{j}\|^2 \leq n \Lambda\right\}$ denote the set of feasible jamming signals.

An $(n,R,P) $ \emph{deterministic code} of block length $n$, rate $R$ and
average power $P$ is a pair $(\psi,\phi)$ of encoder map
$\psi:\{1,2,\dots,2^{nR}\}\times \mathcal{\mathbb{R}}^n\rightarrow
\mathcal{\mathbb{R}}^n$, such that $\|\psi(m,\vec{s})\|^2\leq n P$, $\forall
m,\vec{s}$, and decoder map $\phi: \mathcal{\mathbb{R}}^n\rightarrow \{0,1,2,\dots,2^{nR}\}$, where an output of $0$ indicates that the decoder declares an error. Here we have assumed $2^{nR}$ to be an integer. 
The transmitted vector is given by $\vec{X} = \psi(M,\vec{S})$.

An $(n,R,P)$ \emph{randomized code} is a random variable $(\Psi,\Phi)$ which forms the shared randomness $\Theta$ and takes values in the set of $(n,R,P)$ deterministic codes. Here the transmitted vector is given by $\vec{X} = \Psi(M,\vec{S})$.
For an $(n,R,P)$ randomized code with encoder-decoder pair $(\Psi,\Phi)$, 
the \emph{maximum probability of error} is
\begin{equation}
\label{eq:maxpe:2}
P^{(n)}_e=\max_{m} \max_{Q_{\vec{J}|M=m,\vec{S}}:\vec{J}\vec\in \mathcal{J}(\Lambda) } \mathbb{P}\left( \Phi(\vec{Y})\neq m|M=m\right) ,
\end{equation}
where the probability is  over the state $\vec{S}$, the adversary's action $\vec{J}$, the channel behavior and $\Theta=(\Psi, \Phi)$.
The rate $R$ is \emph{achievable} if for every $\epsilon>0$, there exists an
$(n,R,P)$ randomized code for some $n$ such that $P^{(n)}_e$ is less than $\epsilon$ . We define
the \emph{capacity} of the DP-AVC as the supremum of all
achievable rates.
\section{The Main Results}\label{sec:main:results}
In this section, we present our main results.
Theorem~\ref{thm:main:result:dmc} characterizes the  capacity of the GP-AVC while 
Theorem~\ref{thm:main:result:gaussian} determines the capacity of the DP-AVC. 

Given a state distribution $P_S$ and for fixed distributions $P_{U,X|S}$ and $Q_{J|S}$, let $I(U;Y)$ and $I(U;S)$ denote respectively,
the mutual information quantities evaluated with respect to the corresponding marginals $P_{U,Y}$ and $P_{U,S}$.
In the following theorem, let $\cU$ denote the alphabet of $U$. 
\begin{theorem}\label{thm:main:result:dmc}[GP-AVC Capacity]
The capacity of the Gel'fand Pinsker AVC is\footnote{The max-min exists as mutual information  $I(U;Y)-I(U;S)$ is a continuous function of these variables which take values over a compact set.}
\begin{equation}\label{eq:capacity:dmc:1}
C=\max_{  P_{U|S},\ x(\cdot,\cdot)  } \min_{Q_{J|S}} \left(I(U;Y)-I(U;S)\right),
\end{equation}
where $P_{U|S}\in\cP(\mathcal{U}|\mathcal{S})$, $x:\cU \times \cS\rightarrow \cX$, $Q_{J|S}\in \cP(\mathcal{J}|\mathcal{S})$, and $|\cU|\leq |\cX|^{|\cS|}$.
\end{theorem}
The proof of this result is presented in Section~\ref{sec:proofs}.
\begin{remark}\label{rem:averagepe}
Though the capacity is stated for the maximum probability
of error criterion, the converse is proved for the average probability
of error as defined in~\eqref{eq:avgpe}. On the other hand, the achievability 
under maximum probability of error also implies the achievability under 
average probability of error. Thus, the GP-AVC capacity under the average probability of error criterion
is the same as in~\eqref{eq:capacity:dmc:1}.

\label{resp:footnote}
This fact can also be seen directly from the definition itself. Clearly, capacity under average probability of error criterion cannot be smaller than that under maximum probability of error criterion. To see that the capacities must be the same, given a code with a certain average probability of error $\bar{P_e}$, we can obtain a code whose probability of error under each message is $\bar{P_e}$ and hence whose maximal probability is $\bar{P_e}$. This can be done by simply using a part of the shared randomness $\Theta$ to uniformly permute the messages. Specifically, a uniformly random permutation $\Pi:\{1,\ldots,2^{nR}\}\rightarrow\{1,\ldots,2^{nR}\}$ is chosen using a part of $\Theta$, and to send message $m$, the permuted message $\Pi_m$ is sent using the encoder which guarantees average probability of error $\bar{P_e}$. At the receiver the inverse map $\Pi^{-1}$ is applied to the output of the decoder. \label{rem:averagepe:comments}

The above argument also shows how the adversary's knowledge of $M$ can be rendered essentially useless. Indeed, for an encoder-decoder pair which uses a random permutation as above, the optimal $Q_{\bJ|M,\bS}$ must be such that it does not depend on $M$, in other words, it must be of the form $Q_{\bJ|\bS}$. Since the above random permutation can always be used without resulting in an increase in the maximal (and average) probability of error, it is clear that the capacity under a state-aware adversary who also knows the message $M$ is the same as that under a state-aware adversary who does not know the message.
\end{remark}
\begin{remark}\label{rem:average:ch}
Every memoryless jamming strategy $Q_{J|S}$ induces some GP channel $V_{Y|X,S}$. Thus,~\eqref{eq:capacity:dmc:1} can be expressed through the capacity of the worst memoryless channel that the adversary can induce through a memoryless strategy, i.e, 
\begin{equation*}\label{eq:capacity:dmc:2}
C=\max_{  P_{U|S},\ x(\cdot,\cdot)    } \min_{V_{Y|X,S}} \left(I(U;Y)-I(U;S)\right).
\end{equation*}
Here 
\begin{IEEEeqnarray*}{rCl}
V_{Y|X,S}(y|x,s)=\sum_{j} W_{Y|X,S,J}(y|x,s,j) Q_{J|S}(j|s),
\end{IEEEeqnarray*}
where $Q_{J|S}\in\cP(\cJ|\cS)$. 
\end{remark}
\begin{remark}\label{rem:shannon:str}
Recall that the standard GP channel capacity~\cite{gelfand-pinsker} is given by
\begin{IEEEeqnarray*}{rCl}
C&=&\max_{P_{U,X|S}} I(U;Y)-I(U;S)\\
&=& \max_{P_{U|S},\ x(\cdot,\cdot)} I(U;Y)-I(U;S).\yesnumber\label{eq:capacity:gp}
\end{IEEEeqnarray*}
The standard argument for the reduction to~\eqref{eq:capacity:gp} uses the fact that $\left(I(U;Y)-I(U;S)\right)$ is a convex function of 
$P_{X|U,S}$ for a fixed distribution $P_{U|S}$~\cite{elgamal-kim}. For the GP-AVC, though, such an approach fails 
as $\min_{Q_{J|S}} \left(I(U;Y)-I(U;S)\right)$ is not necessarily a convex function of
$P_{X|U,S}$ for a fixed $P_{U|S}$. However, in the proof of the converse of Theorem~\ref{thm:main:result:dmc}, we use a different approach to  show  that such a  simplification is still possible for the GP-AVC.  
\end{remark}
\begin{remark}
Our bound on $|\cU|$ in Theorem~\ref{thm:main:result:dmc} follows from the
set of Shannon strategies~\cite[Remark~7.6]{elgamal-kim} at the encoder as
there exist up to $|\cX|^{|\cS|}$ functions from $\cS$ to $\cX$.  
The details can be seen in the proof of the converse. In the
standard GP channel, where the GP channel is fixed, 
a stronger bound of $|\cU|\leq |\cX|\cdot |\cS|$ is known to hold using
Support lemma~\cite[Lemma~15.4]{csiszar-korner-book2011} (which uses Carath\'eodory's
theorem). However, we cannot use the Support lemma for the GP-AVC because the statistics
of $U$ depend upon the statistics of the output $Y$ and the adversary can
induce any of the infinitely many GP channels. 
\end{remark}
Our next result gives the capacity of the Dirty Paper AVC. 
\label{main:results}
\begin{theorem}\label{thm:main:result:gaussian}[DP-AVC Capacity]
The capacity\footnote{For the same reason as explained in Remark~\ref{rem:averagepe}, the capacity
is the same under both maximum error probability and average error
probability criteria. } 
of the Dirty Paper AVC is
\begin{equation}\label{eq:capacity:gaussian}
C=\frac{1}{2}\log \left( 1+\frac{P}{\Lambda+\sigma^2}\right).
\end{equation}
\end{theorem}
The proof is given in Section~\ref{sec:proofs}. This result again implies that essentially a memoryless strategy is optimal for the adversary. Unlike in the case of the GP-AVC, here the adversary completely disregards the knowledge of the
state. The adversary essentially inputs i.i.d. Gaussian jamming noise independent of the state. The effect of the additive random state ($S$) is completely eliminated as in the standard dirty paper channel, and the capacity of the DP-AVC equals that of the
dirty paper channel considered by Costa in~\cite{costa} where the noise variance is $(\Lambda+\sigma^2)$.
\section{Proofs}\label{sec:proofs}
\subsection{Proof of Theorem~\ref{thm:main:result:dmc}: The Gel'fand-Pinsker AVC Capacity}
In this section, we first discuss the converse for the Gel'fand-Pinsker AVC capacity theorem and then give a proof of achievability.
\subsubsection{Converse}
\label{gpavc:converse}
In the following, we prove the converse for an average probability of error criterion instead of the maximum probability of error criterion. 
For this stronger version of the converse, let the average probability of error be
\begin{equation}\label{eq:avgpe}
P^{(n)}_e=\frac{1}{2^{nR}} \sum_{m=1}^{2^{nR}} P^{(n)}_{e,m},
\end{equation}
where
\begin{equation*}
P^{(n)}_{e,m}=\max_{Q_{\vec{J}|M=m,\vec{S}}} \mathbb{P}\left( \Phi(\vec{Y})\neq m|M=m\right).
\end{equation*}
To prove our converse, we will consider a specific memoryless (but not i.i.d.)
jamming strategy, which depends on the randomized code (although as discussed in Section~\ref{sec:introduction}, the actual realization of the encoding map is unknown to the adversary), and upper bound the rate
of reliable communication possible under this strategy of the adversary. 

Our proof starts along the lines of the standard Gel'fand-Pinsker converse~\cite{gelfand-pinsker}.
Let us consider any sequence of codes with rate $R$ and $P^{(n)}_e\rightarrow
0$ as $n\rightarrow \infty$. We know from Fano's inequality that for such a
sequence of codes, we have $H(M|\vec{Y},\Theta)\leq n\epsilon_n$, where
$\epsilon_n \rightarrow 0$ as $n\rightarrow \infty$. Then,
\begin{IEEEeqnarray*}{rCl}
nR&\stackrel{}{=}& H(M)\\
&\stackrel{}{=}& I(M;\vec{Y},\Theta)+H(M|\vec{Y},\Theta)\\
&\stackrel{}{\leq }& I(M;\vec{Y},\Theta)+n\epsilon_n\\
&\stackrel{}{= }& I(M;\Theta)+\sum_{i=1}^n I(M;Y_{i}|\vec{Y}^{i-1},\Theta)+n\epsilon_n\\
&\stackrel{(a)}{= }& \sum_{i=1}^n I(M;Y_{i}|\vec{Y}^{i-1},\Theta)+n\epsilon_n\\
&\stackrel{}{\leq }& \sum_{i=1}^n I(M,\vec{Y}^{i-1},\Theta;Y_{i})+n\epsilon_n\\
&\stackrel{}{= }&  \sum_{i=1}^n \Big(I(M,\bS_{i+1}^n,\vec{Y}^{i-1},\Theta;Y_{i})  \\
&&\qquad - I(\bS_{i+1}^n;Y_{i}|M,\vec{Y}^{i-1},\Theta) \Big) +n\epsilon_n\\
&\stackrel{(b)}{= }& \sum_{i=1}^n (I(M,\bS_{i+1}^n,\vec{Y}^{i-1},\Theta;Y_{i})    \\
&&\qquad- I(\vec{Y}^{i-1};S_i|M,\bS_{i+1}^n,\Theta) )+n\epsilon_n\\
&\stackrel{(c)}{= }& \sum_{i=1}^n (I(M,\bS_{i+1}^n,\vec{Y}^{i-1},\Theta;Y_{i})\\
&&\qquad- I(M, \bS_{i+1}^n,\vec{Y}^{i-1},\Theta;S_i) )+n\epsilon_n\\
&\stackrel{(d)}{= }& \sum_{i=1}^n (I(U_{i};Y_{i})- I(U_{i};S_i))+n\epsilon_n\\
&\stackrel{}{= }& n \left(\sum_{i=1}^n \frac{1}{n}\left(I(U_{i};Y_{i})- I(U_{i};S_i)\right)\right)+n\epsilon_n\yesnumber\label{eq:zz}
\end{IEEEeqnarray*}
Here we get $(a)$ as $M$ and $\Theta$ are independent, and $(b)$ follows from
Csisz\'ar's sum identity~\cite[pg. 25]{elgamal-kim}. The independence of $(M, \bS_{i+1}^n,\Theta)$ and $S_i$ gives $(c)$, and $(d)$ follows by denoting $U_{i}=(M,\bS_{i+1}^n,\vec{Y}^{i-1},\Theta)$. 

Given the randomized encoding map, we analyze the performance under a memoryless jamming
strategy of the form
\label{jam:mem}
\begin{align*}
Q_{\bJ|M,\bS}(\vec{j}|m,\vec{s}) := \prod_{i=1}^n Q_{J_i|S_i}(j_i|s_i),
\end{align*}
where $Q_{J_i|S_i}$ are described sequentially for $i=1,2,\dots,n$
below. Note that under such a memoryless jamming strategy, 
$U_{i}\rightarrow (X_{i},S_i)\rightarrow Y_{i}$ is a Markov chain for all $i$.
Before specifying $Q_{J_i|S_i}$, we first note that, given $Q_{J_k|S_k}$ for $k=1,2,	
\dots, i-1$,
\begin{align}
&P_{U_i,X_i|S_i}(U_i=u_i,X_i=x_i|S_i=s_i)\notag\\
&\stackrel{(a)}{=} P_{(M,\Theta,\bS_{i+1}^n,\vec{Y}^{i-1}),X_i|S_i}((m,\theta,\bs_{i+1}^n,\vec{y}^{i-1}),x_i|s_i)\label{eq:P:usx}
\end{align}
\newcounter{storeeqcounter}
\newcounter{tempeqcounter}
Here we have substituted $U_i=(M,\Theta,\bS_{i+1}^n,\vec{Y}^{i-1})$ and 
$u_i=(m,\theta,\bs_{i+1}^n,\vec{y}^{i-1})$ in $(a)$. Simplifying~\eqref{eq:P:usx} further we get~\eqref{eq:usr:strategy}, given on top of the next page,
%
\addtocounter{equation}{1}%
\setcounter{storeeqcounter}%
{\value{equation}}%
%
\begin{figure*}[!t]
\normalsize
\setcounter{tempeqcounter}{\value{equation}} 
\begin{IEEEeqnarray*}{rCl}
\setcounter{equation}{\value{storeeqcounter}}
P_{U_i,X_i|S_i}(U_i=u_i,X_i=x_i|S_i=s_i)
&=& \sum_{\vec{s}^{i-1}, \vec{x}^{i-1}}  P_{M,\Theta,\vec{S}^{i-1},\bS_{i+1}^n,\vec{Y}^{i-1},\vec{X}^{i-1},X_i|S_i}(m,\theta,\vec{s}^{i-1},\bs_{i+1}^n,\vec{y}^{i-1},\vec{x}^{i-1},x_i|s_i) \\
&=& \sum_{\vec{s}^{i-1}, \vec{x}^{i-1}}  P_M(m) P_{\Theta}(\theta) P_{\vec{S}^{i-1}}(\vec{s}^{i-1}) P_{\bS_{i+1}^n}(\bs_{i+1}^n) P_{\vec{X}^{i}|M,\Theta,\vec{S}} ((\vec{x}^{i-1},x_i)|m,\theta,\vec{s})\\
&&\hspace{20mm}\cdot~\left[\prod_{m=1}^{i-1}\left[\sum_{j_m} P_{Y|X,S,J}(y_m|x_m,s_m,j_m) Q_{J_m|S_m}(j_m|s_m)\right]\right].\yesnumber \label{eq:usr:strategy}
%
\end{IEEEeqnarray*}
\setcounter{equation}{\value{tempeqcounter}} 
\hrulefill
\vspace*{4pt}
\end{figure*}
from which it clearly follows that $P_{U_i,X_i|S_i}$ depends on the randomized encoding map 
(in particular, on $P_{\vec{X}^{i}|M,\Theta,\vec{S}}$) as well as  
$Q_{J_k|S_k}:\,k=1,2,\cdots,i-1$, but it does not depend on $Q_{J_i|S_i}$. 
We now define $Q_{J_i|S_i}$ inductively as follows. Given $Q_{J_k|S_k}:\,k=1,2,\cdots,i-1$ and $P_{U_i,X_i|S_i}$, let $Q_{J_i|S_i}$ be the minimizer of $(I(U_i;Y_i)-I(U_i;S_i))$.
Hence, from~\eqref{eq:zz} we have,
\begin{IEEEeqnarray*}{rCl}
nR&\stackrel{}{\leq}& n \left(\sum_{i=1}^n \frac{1}{n}\min_{Q_{J_i|S_i}}\left(I(U_{i};Y_{i})- I(U_{i};S_i)\right)\right)+n\epsilon_n,
\end{IEEEeqnarray*}
for $P_{U_i,X_i|S_i}$, $i=1,2,\dots, n$.
\label{resp:min}
Further, note that for $i=1,2, \ldots,n$,
\begin{IEEEeqnarray*}{rCl}
&&\min_{Q_{J_i|S_i}}\left(I(U_{i};Y_{i})- I(U_{i};S_i)\right)\nonumber\\
&&\>\leq \max_{P_{U_i,X_i|S_i}}\min_{Q_{J_i|S_i}}\left(I(U_i;Y_i)- I(U_i;S_i)\right). \yesnumber\label{eq:U:countable}
\end{IEEEeqnarray*}
Here the maximization is over all conditional distributions $P_{U_i,X_i|S_i}$ 
with finite alphabet $\cU$ of $U_i$.
This inequality holds because the fixed $P_{U_i,X_i|S_i}$ (induced by
the code) on the LHS is such a distribution.
Since the channel is memoryless, the RHS in~\eqref{eq:U:countable} does not depend on $i$, and thus we have 
\begin{IEEEeqnarray*}{rCl}
R&\stackrel{}{\leq}& \max_{P_{U,X|S}}\min_{Q_{J|S}}\left(I(U;Y)- I(U;S)\right)+\epsilon_n.
\end{IEEEeqnarray*}
Since this holds for all $n$, and $\epsilon_n\rightarrow 0$ as $n\rightarrow \infty$, we have
\begin{IEEEeqnarray}{rCl}\label{eq:u:no}	
R&\stackrel{}{\leq}& \max_{P_{U,X|S}}\min_{Q_{J|S}}\left(I(U;Y)- I(U;S)\right).
\end{IEEEeqnarray}

We now show that it is sufficient to perform the maximization in~\eqref{eq:u:no} over distributions $P_{U|S}$ and  functions $x:\cU\times \cS\rightarrow \cX$, i.e., 
\begin{IEEEeqnarray*}{rCl}
&&\max_{P_{U,X|S}}\min_{Q_{J|S}}\left(I(U;Y)- I(U;S)\right)\nonumber\\
\>&&=\max_{P_{U|S},\ x(\cdot,\cdot)}\min_{Q_{J|S}}\left(I(U;Y)- I(U;S)\right).\yesnumber\label{eq:max}
\end{IEEEeqnarray*}
Let us fix the conditional distribution $P_{U|S}$. We know from the functional representation lemma~\cite[pg.~626]{elgamal-kim} that there exists a random variable $W$ which is independent of $(U,S)$ such that $X$ is a function of $(W,U,S)$.  Let us define $U'=(U,W)$ and denote its alphabet by $\mathcal{U}'$, then we have $P_{U'|S}((u,w)|s)=P_{U|S}(u|s)P_W(w)$. Let the function be denoted by $x:\mathcal{U}'\times\mathcal{S}\rightarrow\mathcal{X}$. Note that $U'\rightarrow (X,S)\rightarrow Y$ is a Markov chain. Then, 
\begin{IEEEeqnarray*}{rCl}
I(U';S)&=&I(U,W;S)\\
&=& I(U;S)+I(W;S|U)\\
&=& I(U;S),\yesnumber\label{eq:us}
\end{IEEEeqnarray*}
where the last equality follows from $W\perp\!\!\!\perp (U,S)$. Further, for any $Q_{J|S}\in\cP(\cJ|\cS)$,
\begin{IEEEeqnarray*}{rCl}
I(U';Y)&=&I(U,W;Y)\\
&=& I(U;Y)+I(W;Y|U)\\
&\geq & I(U;Y),
\end{IEEEeqnarray*}
and hence, 
\begin{IEEEeqnarray*}{rCl}
\min_{Q_{J|S}} I(U';Y)\geq  \min_{Q_{J|S}}I(U;Y).\yesnumber\label{eq:uy}
\end{IEEEeqnarray*}
From~\eqref{eq:us} and~\eqref{eq:uy}, it then follows that 
\begin{IEEEeqnarray*}{rCl}
\min_{Q_{J|S}} I(U;Y)-I(U;S)&\leq& \min_{Q_{J|S}} I(U';Y)-I(U';S).\label{eq:u:u'}
\end{IEEEeqnarray*}
Here the LHS is evaluated under a  conditional distribution $P_{X|U,S}$ and the RHS under the corresponding $P_{U'|S}$ and $x:\mathcal{U}'\times\mathcal{S}\rightarrow\mathcal{X}$.
Since the inequality holds for any $P_{X|U,S}$, we have~\eqref{eq:max}, and thus
\begin{IEEEeqnarray}{rCl}\label{eq:R:1}
R&\stackrel{}{\leq}& \max_{P_{U|S},\ x(u,s)}\min_{Q_{J|S}}\left(I(U;Y)- I(U;S)\right).
\end{IEEEeqnarray}
For the bound on the cardinality of $\cU$, we use the \emph{Shannon strategy} approach
in a similar manner, for example, as in the context of 
channels with state with causal knowledge of the state at the
encoder~\cite[Remark~7.6]{elgamal-kim}. In particular, the maximization over
functions $x(u,s)$ in~\eqref{eq:R:1} can be equivalently viewed as a
maximization over functions $x_u:\cS\rightarrow \cX; u\in \cU$.
Since there are exactly $|\cX|^{|\cS|}$
such distinct functions, without loss of generality, we can restrict $\cU$ to be of cardinality
at most $|\cX|^{|\cS|}$.
This completes the proof of the converse.
\subsubsection{Achievability}\label{sec:achievability}
To begin, let us introduce some useful notation. Given $\vec{x}$, $\vec{y}$, the type of $\vec{x}$ will be denoted by $T_\vec{x}$, the joint type of $(\vec{x},\vec{y})$ by $T_{\vec{x},\vec{y}}$ and the conditional type of $\vec{x}$ given $\vec{y}$ by $T_{\vec{x}|\vec{y}}$. Here $\forall (x,y)$ such that $T_{\vec{y}}(y)>0$,
\begin{IEEEeqnarray*}{rCl}
T_{\vec{x}|\vec{y}}(x,y)=\frac{T_{\vec{x},\vec{y}}(x,y)}{T_{\vec{y}}(y)}. 
\end{IEEEeqnarray*}
For any $\epsilon\in(0,1)$, the set of $\epsilon$-typical sequences $\bx$ for a distribution $P_X$ is
\begin{equation}\label{eq:typicality}
\mathcal{T}^{n}_{\epsilon}(P_X)=\{\vec{x}:\|T_\vec{x}-P_X\|_{\infty}\leq \epsilon\},
\end{equation}
where $\|.\|_{\infty}$ is the $l_{\infty}$ norm. 
For a joint distribution $P_{X,Y}$ and $\vec{x}\in\cX^n$, the set of conditionally $\epsilon$-typical sequences $\by$, conditioned on $\vec{x}$, is defined as 
\begin{equation*}
\mathcal{T}^{n}_{\epsilon}(P_{X,Y}|\vec{x})=\{\vec{y}:\|T_{\vec{x},\vec{y}}-P_{X,Y}\|_{\infty}\leq \epsilon\}.
\end{equation*}
We use randomized Gel'fand-Pinsker coding scheme~\cite{gelfand-pinsker}, which
involves an auxiliary random variable denoted by $U$. We choose a rate $R<C$,
where $C$ is as given in~\eqref{eq:capacity:dmc:1}. Consider a conditional
distribution $P_{U|S}$ and a function  $x:\cU\times\cS\rightarrow \cX$ with $X=x(U,S)$ such
that 
\begin{IEEEeqnarray*}{rCl}
R<\min_{Q_{J|S}} I(U;Y)-I(U;S).
\end{IEEEeqnarray*}
Note that here $Q_{J|S}$ takes values 
from all conditional distributions in $\cP(\cJ|\cS)$, and the encoder
and the decoder clearly know this set.  

\noindent\emph{Code construction:}
\begin{itemize}
\item We generate a binned codebook $\mathcal{C}$ comprising $2^{n
R_U}=2^{n(R+\tilde{R})}$ vectors  $\vec{U}_{j,k}$, where $j=1,2,\dots,2^{nR}$
and $k=1,2,\dots,2^{n\tilde{R}}$. $\td{R}\geq 0$ will be defined later. Here $j$ 
indicates the bin index while $k$ indicates the position within the bin.  There are $2^{nR}$ bins with each bin
containing $2^{n\tilde{R}}$ codewords. Every codeword $\vec{U}_{j,k}$ is chosen independently and 
uniformly at random from $\mathcal{T}^n_{\delta}(P_U)$ (the choice of $\delta>0$ will be discussed later), where
\begin{IEEEeqnarray*}{rCl}
P_U(u)=\sum_{s} P_{U|S}(u,s)P_S(s), \forall u.
\end{IEEEeqnarray*}
The codebook is shared between the encoder and  decoder as
the shared randomness $\Theta$. 
\end{itemize}
\label{code:desc} 
\emph{Encoding:}
\begin{itemize}
\item Given a message $m$ and having observed the state $\vec{S}$, the encoder looks within the bin $m$ for some $\vec{U}_{m,k}$ such that 
\begin{equation}\label{eq:encoder:condition:dmc}
\|T_{\vec{U}_{m,k},\vec{S}}-P_{U,S}\|_{\infty}\leq \delta_1(\delta),
\end{equation}
for some $\delta_1(\delta)>0$ (the choice of $\delta_1(\delta)$ will be discussed later). Here $P_{U,S}=P_{U|S}P_S$. 
The condition~\eqref{eq:encoder:condition:dmc} implies that $\vec{U}_{m,k}$ and $\vec{S}$ are jointly typical according to $P_{U,S}$. 
If no such $\vec{U}_{m,k}$ is found, then the encoder selects $\vec{U}_{1,1}$. If more than one $\vec{U}_{m,k}$ satisfying~\eqref{eq:encoder:condition:dmc} exists, then the encoder chooses one uniformly at random from amongst them. Let $\vec{U}$ denote the chosen  codeword. 
\item The encoder then generates $\vec{X}$, where $X_i=x(U_i,S_i)$, $i=1,2,\dots,n$ are independent, and transmits it over the channel. 
\end{itemize}
\emph{Decoding:}
\begin{itemize} 
\item When $\vec{y}$ is
received at the decoder and given some fixed parameter $\gamma (\delta) >0$ (the choice of $\gamma(\delta)$ will be discussed later), 
the decoder determines the set 
\begin{IEEEeqnarray*}{rCl}
L(\vec{y},\gamma(\delta))= \Big\{ \vec{u}\in \mathcal{C}:&&\exists Q_{J|S}\in\cP(\mathcal{J}|\mathcal{S}) \\
\>&&\text{   s.t. } \|T_{\vec{u},\vec{y}}-P^{(Q)}_{U,Y}\|_{\infty}\leq\gamma(\delta) \Big\}, 
\end{IEEEeqnarray*}
where for  $Q_{J|S}\in\mathcal{P}(\mathcal{J}|\mathcal{S})$, 
\begin{IEEEeqnarray*}{rCl}
P^{(Q)}_{U,Y}(u,y)&=&\sum_{x,s,j}P_{S}(s) P_{U|S}(u|s)  \vec{1}_{\{X=x(U,S)\}} \\
&&\> \cdot~W_{Y|X,S,J}(y|x,s,j)  Q_{J|S}(j|s) , \forall (u,y).
\end{IEEEeqnarray*}
Here the  decoder lists all codewords $\vec{u}\in \mathcal{C}$ which are jointly typical with
$\vec{y}$ according to $P_{U,Y}^{(Q)}$, for some
$Q_{J|S}\in\mathcal{P}(\mathcal{J}|\mathcal{S})$. \label{decoder}  
\item If $L(\vec{y},\gamma(\delta))$ is not empty and all the bin indices of
the codewords in it are identical, then the decoder outputs the common bin
index $\tilde{m}$. Otherwise, it declares an error by setting $\td{m}=0$.
\end{itemize}
\noindent \emph{Probability of error analysis:}\\
A decoding error occurs if either the chosen  codeword $\bU_{m,k}$ is not jointly typical
with $\vec{Y}$ or  some other codeword $\vec{U}_{m',k'}$, for some $m'\neq m$ and 
$k'\in \{1,2,\dots, 2^{nR}\}$, is jointly typical with $\vec{Y}$. Here the
typicality is according to $P_{U,Y}^{(Q)}$,
for some $Q_{J|S}\in \cP(\mathcal{J}|\mathcal{S})$.
We show that the probability of this decoding error event is vanishing as $n\rightarrow \infty$.
Let $\epsilon>0$ be such that 
\begin{IEEEeqnarray*}{rCl}
R=\min_{Q_{J|S}} (I(U;Y)-I(U;S))-\epsilon,
\end{IEEEeqnarray*}
and 
\begin{IEEEeqnarray*}{rCl}
\tilde{R}=I(U;S)+\epsilon/2.
\end{IEEEeqnarray*}
Recall from earlier that $R_U=R+\tilde{R}$, and hence, 
\begin{IEEEeqnarray*}{rCl}
R_U=\min_{Q_{J|S}} I(U;Y)-\epsilon/2.
\end{IEEEeqnarray*}
Let $\mathcal{E}=\{\td{M}\neq M\}$ denote the decoding error event. Let the message sent be $M=m$ and 
let $\vec{U}=\vec{U}_{m,k}$ denote the chosen codeword. 
Then, we have
\begin{IEEEeqnarray*}{rCl}
&\mathbb{P}(&\mathcal{E}|M=m)\\
&=& \mathbb{P}(\vec{U}\not \in L(\vec{Y},\gamma(\delta))|M=m)\\
&& +  \mathbb{P}(\exists m', k' : m'\neq m, \vec{U}_{m',k'}\in L(\vec{Y},\gamma(\delta))|M=m  ). 
\end{IEEEeqnarray*}
From \eqref{eq:maxpe:1}, we have 
\begin{IEEEeqnarray*}{rCl}
P_e^{(n)}=\max_m \max_{Q_{\bJ|m,\bS}}\mathbb{P}(\mathcal{E}|M=m),
\end{IEEEeqnarray*}
and thus,
\begin{IEEEeqnarray}{rCl}
P_e^{(n)}&\leq& \max_m \max_{Q_{\bJ|m,\bS}}\mathbb{P}(\vec{U}\not \in L(\vec{Y},\gamma(\delta))|M=m)\notag\\
&& + \,\, \max_m \max_{Q_{\bJ|m,\bS}}\mathbb{P}(\exists m', k': m'\neq m, \nonumber\\
&&\hspace{16mm}   \, \vec{U}_{m',k'}\in L(\vec{Y},\gamma(\delta))|M=m  ). \label{eq:P:Em:6:dmc}
\end{IEEEeqnarray}
We will show that for any $\epsilon>0$, we can find a $\delta>0$ such that 
both the terms go to zero as $n\rightarrow \infty$.

We now state some useful results  which are required to bound the terms in the RHS of~\eqref{eq:P:Em:6:dmc}. Recall from~\eqref{eq:encoder:condition:dmc} that $\delta_1(\delta)$ is the parameter which appears in the definition of the encoder. The following claim specifies this $\delta_1(\delta)$ parameter.
\begin{claim}\label{lem:binning:rate:DMC}
If $\tilde{R}>I(U;S)$, then there exists $\delta_1:\mathbb{R}^+\rightarrow \mathbb{R}^+$, where $\delta_1(\delta)\rightarrow 0$ as $\delta\rightarrow 0$, such that the probability that the encoder finds at least one $\vec{U}_{m,k}$ such that $(\vec{U}_{m,k},\vec{S})\in \cT^n_{\delta_1}(P_{U,S})$ approaches 1 as $n\rightarrow\infty$.
\end{claim}
This result follows from the use of the covering lemma, the proof of which is along the lines of the proof of~\cite[Lemma 3.3]{elgamal-kim}. 
To bound the first term in \eqref{eq:P:Em:6:dmc}, we will consider the conditional type
$T_{\bJ|\bS}$ of $\bJ$ given $\bS$~\footnote{ In fact, it will be be clear through the proof  that even though the adversary can employ arbitrary vector jamming strategies of the form $Q_{\bJ|M,\bS}$, its impact is completely captured through the conditional type $T_{\bJ|\bS}\in\cP(\cJ|\cS)$. See the proof of Lemma~\ref{lem:u:in:L} for details.   }\label{pg:defn1}.
As $\bS$ is i.i.d. with $S_i\sim P_S$, $\forall i$, it follows that the pair $(\bS, \bJ)$ will be jointly typical according to $P_S T_{\bJ|\bS}$ with high probability.
We next present a lemma which is a refined version of the Markov Lemma
in \cite[Lemma~12.1]{elgamal-kim}. This lemma will be used later 
(with $X\rightarrow J$, $Y \rightarrow S$, and $Z\rightarrow U$) to
conclude that $(\bS,\bJ,\bU)$ are jointly typical
according to $P_ST_{\bJ|\bS}P_{U|S}$ with high probability. 
\begin{lemma}[Refined Markov Lemma]\label{lem:gen:markov:lemma}
Suppose $X\rightarrow Y\rightarrow Z$ is a Markov chain, i.e., $P_{X,Y,Z}=P_{Y}P_{X|Y} P_{Z|Y} $. Let $(\vec{x},\vec{y})\in \mathcal{T}^n_{\delta_0}\left(P_{X,Y}\right)$ and $\vec{Z} \sim P_{\vec{Z}}$ be such that
\begin{enumerate}[(a)]
\item for some $\epsilon>0$,
\begin{IEEEeqnarray*}{rCl}
\mathbb{P}\left((\vec{y},\vec{Z})\not\in
\mathcal{T}^n_{\delta_0}\left(P_{Y,Z}\right)\right)\leq \epsilon,
\end{IEEEeqnarray*}
\item for every $\vec{z}\in \mathcal{T}^n_{\delta_0}\left(P_{Y,Z}|\vec{y}\right)$,
\begin{equation*}\label{eq:condition:markov}
2^{-n(H(Z|Y)+g(\delta_0))}\leq P_{\vec{Z}}(\vec{z})\leq 2^{-n(H(Z|Y)-g(\delta_0))},
\end{equation*}
for some $g:\mathbb{R}^{+}\rightarrow\mathbb{R}^{+}$, where $g(\delta_0)\rightarrow 0$ as $\delta_0\rightarrow 0$.
\end{enumerate}
Then, there exists $\delta:\mathbb{R}^{+}\rightarrow\mathbb{R}^{+}$, where $\delta (\delta_0) 
\rightarrow 0$ as $\delta_0 \rightarrow 0$, such that
\begin{equation*}
\mathbb{P}\left((\vec{x},\vec{y},\vec{Z})\not \in \mathcal{T}^n_{\delta(\delta_0)}\left(P_{X,Y,Z}\right)   \right)\leq 2|\cX||\cY||\cZ| e^{-n K} +\epsilon.
\end{equation*}
Here $K>0$ and $K$ does not depend on $n$, $P_{X,Y}$, $P_{\vec{Z}}$ or $(\vec{x},\vec{y})$ but does depend on $\delta_0$, $g$ and $P_{Z|Y}$. Further, the $\delta$ function does not depend on $(\bx,\by)$, $P_{X,Y}$ or $P_{\bZ}$. 
\end{lemma}
The proof of the lemma is presented in Appendix~\ref{app:lem:gen:markov:lemma}.
\begin{remark}\label{rem:markov}
(i) The Refined Markov lemma is
a refinement of the Markov lemma~\cite{elgamal-kim}. 
Markov lemma gives the bound (see the proof in~\cite[Appendix 12A]{elgamal-kim})
\begin{IEEEeqnarray*}{rCl}
&\mathbb{P}(&(\vec{x},\vec{y},\vec{Z})\not \in
\mathcal{T}^n_{\delta}\left(P_{X,Y,Z}\right)   )\\
&\leq& 2(n+1)2^{2ng(\delta_0)}e^{-n(\delta - g(\delta_0))^2 P^{\min}_{X,Y,Z} /(3(1+g(\delta_0)))},
\end{IEEEeqnarray*}
where 
\begin{IEEEeqnarray*}{rCl}
P^{\min}_{X,Y,Z}:=\min_{(x,y,z):P_{X,Y,Z}(x,y,z)>0}P(x,y,z),
\end{IEEEeqnarray*}
and $\delta>0$
is a constant. On the other hand, Lemma~\ref{lem:gen:markov:lemma} gives a bound which does not depend on $P_{X,Y}$. 

This refinement is crucial in our proof of achievability. Here 
the lemma will be used (in the proof of Claim~\ref{lem:j:typ:u:s}) replacing $X\rightarrow J$, $Y\rightarrow S$ and
$Z\rightarrow U$. Thus, we have the Markov chain $J\rightarrow S\rightarrow U$ with $P_{J,S,U}=P_{J,S} P_{U|S}$. For a given $\bj$ and $\bs$, we will take their joint
type $T_{\bj,\bs}$ as the distribution $P_{J,S}$. Since $\bj$ is
decided by the adversary based on their non-causal knowledge of $\bs$, the joint
type $T_{\bj,\bs}$ can have non-zero components as small as $1/n$. This can
be easily caused by the adversary by enforcing a pair of values $(j,s)$
only once in the length-$n$ pair of vectors. In such cases, the original
Markov lemma does not guarantee any useful bound on the probability
$\mathbb{P}\left((\vec{j},\vec{s},\vec{U})\not \in \mathcal{T}^n_{\delta}\left(P_{J,S,U}\right)\right)$.
We believe that for similar reasons, our version of the Markov lemma
may also be useful in achievability proofs in other systems with adversaries.

(ii) Another minor difference from the Markov lemma is
that we use a slightly different notion of typicality~\eqref{eq:typicality}
than the one used in~\cite{elgamal-kim}. This makes the analysis easier in the
second part of the proof of Lemma~\ref{lem:u:in:L}. However, the Refined 
Markov lemma can also be proved under the typicality notion used
in~\cite{elgamal-kim} along the lines of our proof.
\end{remark}
The following lemma bounds the two components of the probability of
error in~\eqref{eq:P:Em:6:dmc}. 
\begin{lemma}\label{lem:u:in:L}
Let the message be $M=m$.
There exist $\gamma,\tilde{\gamma}:\mathbb{R}^+\rightarrow \mathbb{R}^+$,
where $\gamma(\delta),\tilde{\gamma}(\delta)\rightarrow 0$ as $\delta\rightarrow 0$, 
such that for any jamming strategy $Q_{\bJ|M,\bS}$,
\begin{enumerate}[(i)]
\item for $\epsilon_n$ independent of $m$, where $\epsilon_n\rightarrow 0$
as $n\rightarrow \infty$
\begin{IEEEeqnarray*}{rCl}
\mathbb{P}(\vec{U}\not \in L(\vec{Y},\gamma(\delta))|M=m)\leq \epsilon_n,
\end{IEEEeqnarray*}
%
%
\item if $\vec{U}'\sim \text{Unif} \left(\mathcal{T}^n_{\delta}(P_U)\right)$, independent of $(\vec{U},\vec{S},\vec{X},\vec{J},\vec{Y})$, then
\begin{IEEEeqnarray*}{rCl}\label{eq:U:in:L:def}
\mathbb{P}\big(\vec{U}'\in  && ~L(\vec{Y},\gamma(\delta))|M=m) \nonumber\\
&&\leq 2^{ -n\left( \min_{Q_{J|S}} I(U;Y)-\tilde{\gamma}(\delta)\right) }.
\end{IEEEeqnarray*} 
\end{enumerate}
\end{lemma}
Before we prove this lemma, we complete the proof of achievability. The proof of Lemma~\ref{lem:u:in:L} follows immediately after  and concludes this section. Claim~\ref{lem:binning:rate:DMC}  and Lemma~\ref{lem:gen:markov:lemma} (Refined Markov lemma)  are used in the proof of Lemma~\ref{lem:u:in:L}.
Coming back to the probability of error analysis, note that the first part of Lemma~\ref{lem:u:in:L} implies that as $n\rightarrow \infty$, the first term in the RHS in~\eqref{eq:P:Em:6:dmc} goes to zero. For the second RHS term, we have 
for any $Q_{\bJ|M,\bS}$,
\begin{IEEEeqnarray*}{rCl}
&\mathbb{P}(\exists& m', k': m'\neq m,  \vec{U}_{m',k'}\in L(\vec{Y},\gamma(\delta))|M=m)\\
&\stackrel{(a)}{\leq}& \sum_{m'\neq m,\,  k'}\mathbb{P}\left(\vec{U}_{m',k'}\in L(\vec{Y},\gamma(\delta))|M=m\right)\\
&\stackrel{}{\leq}& 2^{nR_U} \mathbb{P}\left(\vec{U}_{m',k'}\in L(\vec{Y},\gamma(\delta))|M=m\right)\\
&\stackrel{(b)}{\leq}& 2^{nR_U}2^{ -n\left( \min_{Q_{J|S}} I(U;Y)-\tilde{\gamma}(\delta)\right)}.
\end{IEEEeqnarray*}
Here we get $(a)$ using the union bound while $(b)$ follows from the second part of  Lemma~\ref{lem:u:in:L}. 
Thus, by choosing a small enough $\delta$  such that
\begin{equation*}
R_U<\min_{Q_{J|S}} I(U;Y)-\tilde{\gamma}(\delta),
\end{equation*}
it follows that the second term in the RHS of~\eqref{eq:P:Em:6:dmc} can be made
to go to $0$ as $n\rightarrow \infty$. This implies that
$P_e^{(n)}\rightarrow 0$ as $n\rightarrow 0$, and hence,
concludes the proof of achievability. 

Now, it only remains to prove Lemma~\ref{lem:u:in:L}. For the proof of the first part of this lemma,
we begin by stating a few useful claims. Recall our assumption that $P_{S}(s)>0$, $\forall s$. In the following, when we write $\vec{s}\in\cT^n_{\delta_0}(P_S)$ we assume that $\delta_0$ is small enough and $n$ large enough such that $T_{\vec{s}}(s)>0$, $\forall s$. Hence, we may write 
\begin{IEEEeqnarray*}{rCl}
T_{\vec{j}|\vec{s}}(j|s)=\frac{T_{\vec{s},\vec{j}}(s,j)}{T_{\vec{s}}(s)}, \forall (s,j).
\end{IEEEeqnarray*}
It will be seen through the following claims that the effect of the jamming input given the underlying adversarial strategy is completely captured through this conditional type $T_{\bj|\bs}$.
\begin{claim}\label{lem:j:typ:s}
Let $(\vec{s},\vec{j})$ be a pair of vectors where $\vec{s}\in\mathcal{T}_{\delta_0}^n(P_S)$. 
Then, $(\vec{s},\vec{j})\in \mathcal{T}_{\delta_0}^n(P_{S} T_{\vec{j}|\vec{s}})$.
\end{claim}
\label{resp:straight}
The proof is straightforward, and hence, omitted. Now, we note that
under the event that the encoder succeeds in finding a typical $\bU$ codeword,
$\vec{U}\sim \text{Unif}\left(\mathcal{T}^n_{\delta_1}(P_{U,S}|\vec{s})\right)$
conditioned on $M=m$.
\begin{claim}\label{lem:j:typ:u:s}
Let $(\vec{s},\vec{j})\in \mathcal{T}^n_{\delta_1}(P_{S} T_{\vec{j}|\vec{s}})$. Then, there exists some $\delta_2(\delta_1)>0$, where $\delta_2(\delta_1)\rightarrow 0$ as $\delta_1\rightarrow 0$, such that if $\vec{U}\sim \text{Unif}\left(\mathcal{T}^n_{\delta_1}(P_{U,S}|\vec{s})\right)$, where $P_{U,S}= P_{S} P_{U|S} $, then 
\begin{equation*}
\mathbb{P}\left( \vec{U}\not \in \mathcal{T}^n_{\delta_2} (P_{S}P_{U|S} T_{\bj|\bs}|\vec{s},\vec{j})|M=m  \right)\leq \epsilon_n,
\end{equation*}
where $\delta_2$ and $\epsilon_n$ do not depend\footnote{The fact that these do not depend on
$(\bs,\bj)$ is crucial, and it follows from our Refined Markov
Lemma. They also do not depend on $m$, as the distribution of $\bU$ does
not depend on $m$.} on $(\vec{s},\vec{j},m)$ and $\epsilon_n\rightarrow 0$ as $n\rightarrow \infty$.  
\end{claim}
\begin{IEEEproof}
We use Lemma~\ref{lem:gen:markov:lemma} with $X\rightarrow J$, $Y\rightarrow S$ and $Z\rightarrow U$. Further, replace $\delta_0\rightarrow \delta_1$ and $\delta(\delta_0)\rightarrow \delta_2(\delta_1)$ here. 
Next, the distribution $P_{S,J}=P_S T_{\vec{j}|\vec{s}}$ and $P_{\bf U} = \text{Unif}
(\mathcal{T}^n_{\delta_1}(P_{U,S}|\vec{s}))$.
As $\vec{U}\sim \text{Unif}\left(\mathcal{T}^n_{\delta_1}(P_{U,S}|\vec{s})\right)$, it follows that both the conditions of Lemma~\ref{lem:gen:markov:lemma} are satisfied. In particular, the first condition is met with $\epsilon=0$ as $\vec{U}\in \cT^n_{\delta_1}(P_{U,S}|\vec{s})$, while the second condition is met as there exists some $g(\delta_1)>0$, where $g(\delta_1)\rightarrow 0$ as $\delta_1\rightarrow 0$, such that 
\begin{IEEEeqnarray*}{rCl}
2^{n(H(U|S)-g(\delta_1))}\leq |\mathcal{T}^n_{\delta_1}(P_{U,S}|\vec{s})|\leq 2^{n(H(U|S)+g(\delta_1))}.
\end{IEEEeqnarray*}
The claim now follows.
\end{IEEEproof}
The following two claims follow from the conditional typicality lemma, where the proof of the latter is along the lines of the one which appears in~\cite[pg.~27]{elgamal-kim}.
\begin{claim}\label{lem:x:typ:u:s:j}
Let $(\vec{u},\vec{s},\vec{j}) \in \mathcal{T}^n_{\delta_2} 
(P_{U,S,J})$, and let $\vec{X}$ be generated from $(\vec{u},\vec{s})$ 
through the memoryless distribution $\vec{1}_{\{X=x(U,S)\}}$. 
Then there exists $\delta_3(\delta_2)>0$, where $\delta_3(\delta_2)\rightarrow 0$ as $\delta_2\rightarrow 0$, such that
\begin{align*}
\mathbb{P}\left((\vec{u},\vec{s},\vec{j},\vec{X}) \not \in {\mathcal{T}}^n_{\delta_3}
(P_{U,S,J}  \vec{1}_{\{X=x(U,S)\}})|M=m    \right) \leq \epsilon_n,
\end{align*}
where $\delta_3$ and $\epsilon_n$ do not depend on $(\vec{u},\vec{s},\vec{j},m)$, and
$\epsilon_n\rightarrow 0$ as $n\rightarrow \infty$. 
\end{claim}
\begin{claim}\label{lem:y:typ:u:s:x:j}
Let $(\vec{u},\vec{s},\vec{x},\vec{j})  \in \mathcal{T}^n_{\delta_3} 
(P_{U,S,X,J})$, and let $\vec{Y}$ be generated from $(\vec{x},\vec{s},\vec{j})$ through the channel $W_{Y|X,S,J}$. Then there exists $\delta_4(\delta_3)>0$, where $\delta_4(\delta_3)\rightarrow 0$ as $\delta_3\rightarrow 0$, such that
\begin{align*}
\mathbb{P}\left((\vec{u},\vec{s},\vec{x},\vec{j}, \vec{Y}) \not \in {\mathcal{T}}^n_{\delta_4}
(P_{U,S,X,J}W_{Y|X,S,J})|M=m\right) \leq \epsilon_n,
\end{align*}
where $\delta_4$ and $\epsilon_n$ do not depend on $(\vec{u},\vec{s},\vec{j},\vec{x},m)$,
and $\epsilon_n\rightarrow 0$ as $n\rightarrow \infty$. 
\end{claim}

To proceed with the proof of Lemma~\ref{lem:u:in:L}, let us define the following error event.
\begin{IEEEeqnarray*}{rCl}
E&=&\{ \vec{U} \not \in L(\vec{Y},\gamma(\delta))\}
\end{IEEEeqnarray*}
From the definition of the decoder, it follows that this event $E$ occurs if there does not exist any $Q_{J|S}\in\cP(\cJ|\cS)$, and  correspondingly any resulting distribution $P_{UY}$, such that the chosen $\bU$ codeword and the received output $\bY$ are jointly typical. 
 In the following, we show that for the specific choice of 
$Q_{J|S}=T_{\bJ|\bS}$, the correct codeword $\bU$ will satisfy the
decoding criterion w.h.p.. 
Toward this, we define some events:
\label{pg:E:event}
\begin{IEEEeqnarray*}{rCl}
E_1&=&\{ \vec{S} \not \in \cT^{(n)}_{\delta_0}(P_S)\},\\
E_2&=&\{ (\vec{S},\vec{J}) \not \in \cT^{(n)}_{\delta_0}(P_S T_{\vec{J}|\vec{S}})\},\\
E_3&=&\{ (\vec{S},\vec{U}) \not \in \cT^{(n)}_{\delta_1}(P_S P_{U|S} )\},\\
E_4&=&\{ (\vec{S},\vec{J},\vec{U}) \not \in \cT^{(n)}_{\delta_2}(P_S T_{\vec{J}|\vec{S}} P_{U|S} )\},\\
E_5&=& \{ (\vec{S},\vec{J},\vec{U},\vec{X}) \not \in \cT^{(n)}_{\delta_3}(P_S T_{\vec{J}|\vec{S}} P_{U|S}  \vec{1}_{\{X=x(U,S)\}}  )\},   \\
E_6&=&       \{ (\vec{S},\vec{J},\vec{U},\vec{X},\vec{Y})   \\
&&  \hspace{10mm} \not \in \cT^{(n)}_{\gamma(\delta)}(P_S T_{\vec{J}|\vec{S}} P_{U|S}  \vec{1}_{\{X=x(U,S)\}}   W_{Y|X,S,J} )\}.  
\end{IEEEeqnarray*}
Here $\delta_0(\delta)=\frac{\delta}{2}$ and $\delta_i$, $i=1,2,3$ and $\gamma$ will be chosen such that as functions of $\delta$, they approach $0$ as $\delta\rightarrow 0$. Using the union bound, we have
\begin{IEEEeqnarray*}{rCl}
&\mathbb{P}(E&|M=m)\\
&\leq& \mathbb{P}(E_1|M=m)+\mathbb{P}(E_2|E_1^c,M=m)\\
&&+\mathbb{P}(E_3|E_2^c,E_1^c,M=m)+\mathbb{P}(E_4|E_3^c,E_2^c,E_1^c,M=m)\\
&&+\mathbb{P}(E_5|E_4^c,E_3^c,E_2^c,E_1^c,M=m)\\
&&+\mathbb{P}(E_6|E_5^c,E_4^c,E_3^c,E_2^c,E_1^c,M=m)\\
&\stackrel{(a)}{=}&\mathbb{P}(E_1|M=m)+\mathbb{P}(E_3|E_1^c,M=m)\\
&&+\mathbb{P}(E_4|E_3^c,E_2^c,M=m)+\mathbb{P}(E_5|E_4^c,E_3^c,E_2^c,M=m) \\
&&+\mathbb{P}(E_6|E_5^c,E_4^c,E_3^c,E_2^c,M=m).\yesnumber\label{eq:P:E:ext}
\end{IEEEeqnarray*}
Here $(a)$ follows from Claim~\ref{lem:j:typ:s}. This is because given $\vec{s}\in\cT^{(n)}_{\delta_0}(P_S)$ and any $\vec{j}$,  we have $(\vec{s},\vec{j})\in \mathcal{T}^n_{\delta_0}(P_ST_{\vec{j}|\vec{s}})$, which implies $E_1=E_2$, and thus, $\mathbb{P}(E_2|E_1^c,M=m)=0$. We now analyse each of the terms in the RHS of~\eqref{eq:P:E:ext}.

As $\vec{S}$ is the output of an i.i.d. source with distribution $P_S$ irrespective of $m$,  it follows that 
\begin{IEEEeqnarray*}{rCl}
\mathbb{P}(\vec{S}\in \mathcal{T}^n_{\delta_0}(P_S)|M=m)\rightarrow 1
\end{IEEEeqnarray*}
as $n\rightarrow \infty$. Hence, $\mathbb{P}(E_1|M=m)\rightarrow 0$ as $n\rightarrow \infty$.

For the second term, 
Claim~\ref{lem:binning:rate:DMC} guarantees
that there exists $\delta_1(\delta_0)>0$, $\delta_1(\delta_0)\rightarrow 0$ as
$\delta_0\rightarrow 0$, such that
\begin{IEEEeqnarray*}{rCl}
\mathbb{P}\left((\vec{S},\vec{U})\not\in\cT^{(n)}_{\delta_1}(P_{S} P_{U|S})|\vec{S}\in\cT^{(n)}_{\delta_0}(P_S),M=m\right)\rightarrow 0
\end{IEEEeqnarray*}
as $n\rightarrow \infty$. We choose $\delta_1>\delta_0$. Thus,
\begin{IEEEeqnarray*}{rCl}
\mathbb{P}(E_3|E_1^c,M=m)=\mathbb{P}(E_3|E_2^c,M=m)\rightarrow 0
\end{IEEEeqnarray*}
as $n\rightarrow \infty$.

For the third term, let $(\vec{s},\vec{j})\in \cT^{(n)}_{\delta_0}(P_{S}T_{\vec{j}|\vec{s}})$. Then conditioned on $(\bS,\bJ,M)=(\bs,\bj,m)$  as well as conditioned on $E_3^c$, the distribution of $\bU$
is $\text{Unif}\left(\mathcal{T}^n_{\delta_1}(P_{U,S}|\vec{s})\right)$. Note that $(\bs,\bj)\in \cT^{(n)}_{\delta_0}(P_{S}T_{\vec{j}|\vec{s}})\Rightarrow (\bs,\bj)\in \cT^{(n)}_{\delta_1}(P_{S}T_{\vec{j}|\vec{s}})$ since $\delta_1>\delta_0$. We now use Claim~\ref{lem:j:typ:u:s} which guarantees that there exists $\delta_2(\delta_1)>0$, where $\delta_2(\delta_1)\rightarrow 0$ as $\delta_1\rightarrow 0$, such that 
\begin{IEEEeqnarray*}{rCl}
\mathbb{P}(E_4|E_3^c,(\bS,\bJ,M)=(\bs,\bj,m)) \leq \epsilon_n,
\end{IEEEeqnarray*}
where $\epsilon_n\rightarrow 0$ as $n\rightarrow \infty$ (here $\delta_2$ as well as $\epsilon_n$ do not depend on $(\bs,\bj,m)$). Then,
\begin{align*}
&\mathbb{P}(E_4|E_3^c,E_2^c,M=m)\\
&=\sum_{(\bs,\bj)\in \cT^{(n)}_{\delta_0}(P_{S}T_{\vec{j}|\vec{s}})} \mathbb{P}(E_4|E_3^c,(\bS,\bJ,M)=(\bs,\bj,m)) \\
&\qquad\hspace{25mm}\cdot~\mathbb{P}((\bS,\bJ)=(\bs,\bj)|M=m)\\
&\leq\sum_{(\bs,\bj)\in \cT^{(n)}_{\delta_0}(P_{S}T_{\vec{j}|\vec{s}})} \epsilon_n~\mathbb{P}((\bS,\bJ)=(\bs,\bj)|M=m)\\
&\leq\epsilon_n.
\end{align*}
Hence, we can conclude that $\mathbb{P}(E_4|E_3^c,E_2^c,M=m) \rightarrow 0$ as $n\rightarrow \infty$. 

For the fourth term, let 
$(\vec{s},\vec{j},\vec{u})\in
\cT^{(n)}_{\delta_2}(P_{S}P_{U|S}T_{\vec{j}|\vec{s}})$. Now conditioned on $(\vec{S},\vec{J},\vec{U},M)=(\vec{s},\vec{j},\vec{u},m)$, let $\vec{X}$ be generated using the memoryless distribution $\vec{1}_{\{X=x(U,S)\}}$.
 Then, Claim~\ref{lem:x:typ:u:s:j} guarantees that there exists $\delta_3(\delta_2)>0$, where
$\delta_3(\delta_2)\rightarrow 0$ as $\delta_2\rightarrow 0$, such that
\begin{IEEEeqnarray*}{rCl}
\mathbb{P}(E_5|(\bS,\bJ,\bU,M)=(\bs,\bj,\bu,m))\leq \epsilon_n,
\end{IEEEeqnarray*}
where $\epsilon_n\rightarrow 0$ as $n\rightarrow \infty$ (here $\delta_3$ and $\epsilon_n$ do not depend on $(\bs,\bj,\bu,m)$). 
Let us now define the set 
\begin{IEEEeqnarray*}{rCl}
\cM=\{(\bs,\bj,\bu):&&(\bs,\bj,\bu)\in
\cT^{(n)}_{\delta_2}(P_{S}P_{U|S}T_{\vec{j}|\vec{s}}),\\
&& (\vec{s},\vec{u})  \in \cT^{(n)}_{\delta_1}(P_S P_{U|S} ),\\
&& (\vec{s},\vec{j})  \in \cT^{(n)}_{\delta_0}(P_S T_{\bj|\bs} )\}.
\end{IEEEeqnarray*}
Then,
\begin{align*}
&\mathbb{P}(E_5|E_4^c,E_3^c,E_2^c,M=m)\\
&=\sum_{(\bs,\bj,\bu)\in \cM} \hspace*{-1.5mm}\mathbb{P}(E_5|(\bS,\bJ,\bU,M)=(\bs,\bj,\bu,m))\\
&\quad\hspace{15mm} \cdot~ \mathbb{P}((\bS,\bJ,\bU)=(\bs,\bj,\bu)|M=m)\\
&\leq\sum_{(\bs,\bj,\bu)\in \cM} \epsilon_n~\mathbb{P}((\bS,\bJ,\bU)=(\bs,\bj,\bu)|M=m)\\
&\leq\epsilon_n.
\end{align*}
Hence, it follows that $\mathbb{P}(E_5|E_4^c,E_3^c,E_2^c,M=m)\rightarrow 0$ as $n\rightarrow \infty$. 

Similarly, for the final term,  let 
\begin{IEEEeqnarray*}{rCl}
(\vec{s},\vec{j},\vec{u},\vec{x})\in \cT^{(n)}_{\delta_3}(P_{S}P_{U|S}T_{\vec{j}|\vec{s}}  \vec{1}_{\{X=x(U,S)\}}).
\end{IEEEeqnarray*}
Then, conditioned on $(\vec{S},\vec{J},\vec{U},\vec{X},M)=(\vec{s},\vec{j},\vec{u},\vec{x},m)$, let $\bY$ be generated using the memoryless distribution $W_{Y|X,S,J}$. From Claim~\ref{lem:y:typ:u:s:x:j}, we know that there exists $\delta_4(\delta_3)>0$, where $\delta_4(\delta_3)\rightarrow 0$ as $\delta_3\rightarrow 0$, such that
\begin{IEEEeqnarray*}{rCl}
&&\mathbb{P}((\bS,\bJ,\bU,\bX,\bY)\not \in \cT^n_{\delta_4}(P_{S}P_{U|S}T_{\vec{j}|\vec{s}} \vec{1}_{\{X=x(U,S)\}}    \\
&&\hspace{10mm}\cdot~W_{Y|X,S,J})|(\bS,\bJ,\bU,\bX)=(\bs,\bj,\bu,\bx))\leq \epsilon_n,
\end{IEEEeqnarray*}
where $\epsilon_n\rightarrow 0$ as $n\rightarrow \infty$ (here $\delta_4$ and $\epsilon_n$ do not depend on $(\bs,\bj,\bu,\bx,m)$). We now assume $\gamma(\delta)=\delta_4(\delta_3(\delta_2(\delta_1(\delta_0(\delta)))))$ in the definition
of $E_6$. Then, by an argument similar to that of the fourth
term, it follows that 
\begin{IEEEeqnarray*}{rCl}
\mathbb{P}(E_6|E_5^c,E_4^c,E_3^c,E_2^c,M=m)\rightarrow 0
\end{IEEEeqnarray*}
as $n\rightarrow \infty$.

As each term in the RHS of~\eqref{eq:P:E:ext} is vanishing as $n\rightarrow
\infty$, we can conclude that $\mathbb{P}(E|M=m)\rightarrow 0$ as
$n\rightarrow \infty$.  
Thus, we have shown that, conditioned on $M=m$,
$\vec{U}\in L(\vec{Y},\gamma(\delta))$ with probability approaching $1$ as
$n\rightarrow \infty$.  
\label{pg:resp:c:typ}
In particular, we have shown that the correct codeword satisfies the
decoding condition w.r.t. $Q_{J|S}=T_{\bJ|\bS}$.
This completes the proof of part (i) of the lemma.

We prove the second part using some well-known properties of types~\cite{csiszar-korner-book2011,csiszar-it1998,cover-thomas}. We begin by introducing some notation and useful quantities. Let $H_{P_{U,Y}}(U|Y)$ denote the conditional entropy of $U$ given $Y$ under the joint distribution $P_{U,Y}$. As discussed at the beginning of Section~\ref{sec:problem}, to keep the notation simple, we drop the subscript in $P_{U,Y}$ and denote this conditional entropy by $H_P(U|Y)$ henceforth. Similarly, the mutual information between $U$ and $Y$ is denoted as $I_P(U;Y)$. Let $\mathscr{T}$ denote the set of all types of length-$n$ sequences $(\vec{u},\vec{y})$. For any type $P_{U,Y}\in\mathscr{T}$, we define
\begin{equation*}\label{eq:ball}
B_{\delta}(P_{U,Y})=\{\tau\in\mathscr{T}: \|\tau-P_{U,Y} \|_{\infty}\leq \delta\}.
\end{equation*}
By definition, if $T_{\vec{u},\vec{y}}\in B_{\delta}(P_{U,Y})$, then $(\vec{u},\vec{y})\in \mathcal{T}^n_{\delta}(P_{U,Y})$. We know that if $(\vec{u},\vec{y})\in \mathcal{T}^n_{\delta}(P_{U,Y})$, then
\begin{enumerate}[(I)]
\item[$(\alpha)$] $\vec{u}\in \mathcal{T}^n_{\delta}(P_{U,Y}|\vec{y})$.
\item[$(\beta)$] there exists $g(\delta)>0$, where $g(\delta)\rightarrow 0$ as $\delta\rightarrow 0$ and $g(\delta)$ does not depend on $P_{U,Y}$, such that 
\begin{IEEEeqnarray*}{rCl}
|\mathcal{T}^n_{\delta}(P_{U,Y}|\vec{y})|\leq 2^{n(H_{P_{U,Y}}(U|Y)+g(\delta))}.
\end{IEEEeqnarray*}
\end{enumerate}
Thus, given $(\vec{u},\vec{y})\in \mathcal{T}^n_{\delta}(P_{U,Y})$ and for any $\tau\in B_{\delta}(P_{U,Y})$,
\begin{IEEEeqnarray*}{rCl}
\big|  \{ \vec{\td{u}}: T_{\vec{\td{u}},\vec{y}}=\tau  \} \big| &\leq & \big|  \{ \vec{\td{u}}: T_{\vec{\td{u}},\vec{y}}\in B_{\delta}(P_{U,Y})  \} \big|\\
&=& \big|  \{ \vec{\td{u}}: (\vec{\td{u}},\vec{y})\in \mathcal{T}^n_{\delta}(P_{U,Y})  \} \big|\\
&\stackrel{(a)}{=}& \big|  \{ \vec{\td{u}}: \vec{\td{u}}\in \mathcal{T}^n_{\delta}(P_{U,Y}|\vec{y})  \} \big|\\
&\stackrel{(b)}{\leq}& 2^{n(H_{P}(U|Y)+g(\delta))},\yesnumber\label{eq:size:u}
\end{IEEEeqnarray*}
where $(a)$ follows from $(\alpha)$ above while $(b)$ follows from $(\beta)$. 
Let 
\begin{IEEEeqnarray*}{rCl}\label{eq:P:UY:notation}
&&P^{(Q)}_{U,Y}(u,y)=\sum_{x,s,j} P_S (s)P_{U|S}(u|s) \vec{1}_{\{X=x(u,s)\}}  \\
&&\hspace{30mm}\cdot~ W_{Y|X,S,J} Q_{J|S}(j|s) \,\,\,\,\, \forall (u,y),
\end{IEEEeqnarray*}
be the joint distribution for $(U,Y)$ under the memoryless strategy $Q_{J|S}\in\cP(\mathcal{J}|\mathcal{S})$ of the adversary.
Finally, let us denote  
\begin{equation}\label{eq:Q:*}
Q_{J|S}^*=\argmin_{Q_{J|S}\in\cP(\mathcal{J}|\mathcal{S})} I_{P^{(Q)}}(U;Y).
\end{equation}
Note that the above minimum  is achieved, and hence, at least one exists. 
If there are more than one minimizers, pick one arbitrarily from amongst them. 

We now get a bound on the size of $L(\vec{y},\gamma(\delta))$.
\begin{IEEEeqnarray}{rCl}
|L(\vec{y},\gamma(\delta))|&\stackrel{}{=}&  \Big| \Big\{  \vec{u}: \|T_{\vec{u},\vec{y}}-P^{(Q)}_{U,Y}\|_{\infty}\leq \gamma(\delta),\nonumber\\
&&\hspace{20mm} \text{ for some } Q_{J|S}\in\cP(\mathcal{J}|\mathcal{S}) \Big\}  \Big|\nonumber\\
&=& \left| \left\{  \vec{u}: T_{\vec{u},\vec{y}}\in \bigcup_{Q_{J|S}\in\cP(\mathcal{J}|\mathcal{S})} B_{\gamma(\delta)}\left(P^{(Q)}_{U,Y}\right) \right\}  \right|\nonumber\\
&=& \left| \bigcup_{\tau\in\bigcup_{Q_{J|S}\in\cP(\mathcal{J}|\mathcal{S})} B_{\gamma(\delta)}\left(P^{(Q)}_{U,Y}\right) } \{ \vec{u}: T_{\vec{u},\vec{y}}=\tau  \} \right|\nonumber\\
&\stackrel{(a)}{\leq}& (n+1)^{|\mathcal{U}||\mathcal{Y}|}2^{n(H_{P^{(Q^*)}}(U|Y)+g(\delta))}\nonumber\\
&\leq& 2^{n(H_{P^{(Q^*)}}(U|Y)+\tilde{g}(\delta))},\label{eq:L:size}
\end{IEEEeqnarray}
where $\tilde{g}(\delta)>0$ and $\tilde{g}(\delta)\rightarrow 0$ as $\delta\rightarrow 0$. Here $(a)$ follows from noting that there exist at most $(n+1)^{|\mathcal{U}||\mathcal{Y}|}$ types of $(\vec{u},\vec{y})$ as well as  using~\eqref{eq:size:u} and~\eqref{eq:Q:*}. Note that $|L(\vec{y},\gamma(\delta))|$ does not depend on $\by$.
Hence, we have
\begin{align*}
&\mathbb{P}\big(\vec{U}'\in L(\vec{Y},\gamma(\delta))|M=m\big)\\
&=\frac{\left|L(\vec{Y},\gamma(\delta))\right|}{\left|\cT^n_{\delta}(P_{U})\right|}\\
&\stackrel{(a)}{\leq} 2^{-n(H(U)-f(\delta))} 2^{n(H_{P^{(Q^*)}}(U|Y)+\tilde{g}(\delta))}\\
&= 2^{-n(I_{P^{(Q^*)}}(U;Y)-\tilde{\gamma}(\delta))},
\end{align*}
where $\tilde{\gamma}(\delta)=f(\delta)+\td{g}(\delta)>0$ and $\tilde{\gamma}(\delta)\rightarrow 0$ as $\delta\rightarrow 0$. We get $(a)$ from noting that 
\begin{IEEEeqnarray*}{rCl}
|\cT^n_{\delta}(P_U)|\geq 2^{n(H(U)-f(\delta)}
\end{IEEEeqnarray*}
for some $f(\delta)>0$, where $f(\delta)\rightarrow 0$ as $\delta\rightarrow 0$, and from~\eqref{eq:L:size}. This completes the proof of the second part, and concludes the proof of Lemma~\ref{lem:u:in:L}. 
\subsection{Proof of Theorem~\ref{thm:main:result:gaussian}: The Dirty Paper AVC Capacity}
We first analyse an achievable scheme followed by the converse. Before we proceed, let us introduce some useful notation. For any $\vec{x}\in \mathbb{R}^n$, $\|\vec{x}\|\neq 0$, let $\hat{\vec{x}}=\vec{x}/\|\vec{x}\|$ denote the unit vector in the direction of $\vec{x}$. Next, given two vectors $\vec{x},\vec{y} \in \mathbb{R}^n$, $\left<\vec{x},\vec{y}\right>\in \mathbb{R}$ denotes their dot (inner) product.  
\subsubsection{Achievability}
Our code uses the dirty paper coding scheme, which involves an auxiliary random variable denoted as $U$ and a fixed parameter $\alpha$. We choose a rate $R<C$, where $C$ is as defined in~\eqref{eq:capacity:gaussian}. \\
\emph{Code construction:}
\begin{itemize}
\item The encoder generates a binned codebook comprising $2^{n R_U}=2^{n(R+\tilde{R})}$ vectors  $\{\vec{U}_{j,k}\}$, $j=1,2,\dots,2^{nR}$ and $k=1,2,\dots,2^{n\tilde{R}}$. Here there are $2^{nR}$ bins which are indexed by $j$, where each bin contains $2^{n\tilde{R}}$ codewords with $k$ indexing these codewords. $\td{R}>0$ will be specified later. For $\epsilon_1>0$, define $P'=P-\epsilon_1$. Every codeword $\vec{U}_{j,k}$ is chosen independently and uniformly at random over the surface of the $n$-sphere of radius $\sqrt{n P_U}$, where 
\begin{IEEEeqnarray*}{rCl}
P_U&=&P'+\alpha^2 \sigma^2_S,\\
\alpha &=&P'/(P'+\Lambda+\sigma^2).
\end{IEEEeqnarray*}
The codebook is shared between the encoder and  decoder as the shared randomness $\Theta$. 
\end{itemize}
\emph{Encoding:}
\begin{itemize} 
\item Given a message $m$ and having observed the state $\vec{S}$, the encoder looks within the bin $m$ for some $\vec{U}_{m,k}$, $k\in 1,2,\dots,2^{n\tilde{R}}$, such that 
\begin{equation}\label{eq:encoder:condition:gaussian}
|\left<\vec{U}_{m,k}-\alpha \vec{S},\vec{S}\right>| \leq n\delta_1,
\end{equation}
for some $\delta_1>0$ (the choice of $\delta_1$ will be discussed later in Lemma~\ref{lem:Y:U:unit:expr}). If no such $\vec{U}_{m,k}$ is found, then the encoder chooses $\vec{U}_{1,1}$. If more than one $\vec{U}_{m,k}$ satisfying~\eqref{eq:encoder:condition:gaussian} exists, the encoder chooses one uniformly at random from amongst them. Let $\vec{U}$ denote the chosen codeword. 
\item If $\|\vec{U}-\alpha \vec{S}\|\leq \sqrt{n P}$, then the encoder transmits $\vec{X}=\vec{U}-\alpha \vec{S}$ over the channel. Otherwise, it transmits the zero vector. 
\end{itemize}
\emph{Decoding:}
\begin{itemize}
\item We employ the minimum angle decoder. When $\vec{y}$ is received at the decoder, its message estimate $\tilde{m}$ is the solution of the following optimization problem.
\begin{equation*}
\tilde{m}=\argmax_{1\leq j\leq 2^{nR}}  \left(\max_{1\leq k\leq 2^{n\tilde{R}} } \left < \vec{\hat{y}}, \vec{\hat{u}}_{j,k}\right>\right).
\end{equation*}
Here the decoder finds the codeword $\vec{u}\in\mathcal{C}$ closest in angle to $\vec{y}$.
\item If no unique solution exists, the decoder declares an error by setting $\td{m}=0$.
\end{itemize}
\noindent \emph{Probability of error analysis:}\\
Fix some $\epsilon_1$, $\epsilon>0$, and let 
\begin{IEEEeqnarray*}{rCl}
R=\frac{1}{2}\log\left(1+P'/(\Lambda+\sigma^2) \right)-\epsilon.
\end{IEEEeqnarray*}
Note that $R<C$ and $R$ approaches $C$ as $\epsilon_1$, $\epsilon\rightarrow 0$. Next, let 
\begin{IEEEeqnarray*}{rCl}
\tilde{R}=\frac{1}{2}\log(P_U/P')+\epsilon/2.
\end{IEEEeqnarray*}
Recall that  $R_U=R+\tilde{R}$, and hence, we have
\begin{equation}\label{eq:R:U}
R_U=\frac{1}{2}\log\left(\frac{(P'+\Lambda+\sigma^2)P_U}{(\Lambda+\sigma^2)P'}      \right) -\epsilon/2.
\end{equation}
Before we proceed, here is a brief outline of the analysis. Given any $\delta>0$, we establish in Lemma~\ref{lem:Y:U:unit:expr} that irrespective of the adversary's strategy, the inner product $\left<\vec{\hat{Y}},\vec{\hat{U}}\right>$ is at least $(\theta-\delta)$ (here $\theta$ is given in~\eqref{eq:theta}) w.h.p. for sufficiently large $n$. Now regardless of the strategy the adversary employs, a decoding error occurs only if either $\left<\vec{\hat{Y}},\vec{\hat{U}}\right>< (\theta-\delta)$ or some other codeword $\vec{U}_{m',k'}$, for $m'\neq m$ and $k'\in \{1,2,\dots, 2^{nR}\}$, satisfies $\left<\vec{\hat{Y}},\vec{\hat{U}}_{m',k'}\right>\geq (\theta-\delta)$. Our aim will be to show that the probability of this decoding error event goes to zero as $n\rightarrow\infty$.

Let us denote the decoding error event by $\mathcal{E}$. Then, we have $\mathcal{E}=\{\td{M}\neq M\}$. Let $M=m$ be the message sent.
Given $\theta$ and for any $\delta>0$, we  then have
\begin{IEEEeqnarray*}{rCl}
\mathbb{P}(\mathcal{E}|M=m)&\leq& \mathbb{P}\left(\left<\vec{\hat{Y}},\vec{\hat{U}}\right><\theta-\delta\Big|M=m\right)\\
&&+\mathbb{P}\Big(\exists m', k': m'\neq m, \\
&&\hspace{10mm} \left<\vec{\hat{Y}},\vec{\hat{U}}_{m',k'}\right> \geq \theta-\delta\Big| M=m  \Big).
\end{IEEEeqnarray*}
Using~\eqref{eq:maxpe:2}, it follows that 
\begin{equation*}
P_e^{(n)}=\max_m \max_{Q_{\bJ|m,\bS}:\bJ\in \cJ(\Lambda)}\mathbb{P}(\mathcal{E}|M=m).
\end{equation*}
Hence,
\begin{IEEEeqnarray}{rCl}
P_e^{(n)}&\leq& \max_m \max_{Q_{\bJ|m,\bS}:\bJ\in \cJ(\Lambda)}\mathbb{P}\left(\left<\vec{\hat{Y}},\vec{\hat{U}}\right><\theta-\delta\Big|M=m\right)\notag\\
&& + \,\, \max_m \max_{Q_{\bJ|m,\bS}:\bJ\in \cJ(\Lambda)}\mathbb{P}\Big(\exists m', k': m'\neq m, \nonumber\\
&&\hspace{20mm} \left<\vec{\hat{Y}},\vec{\hat{U}}_{m',k'}\right> \geq \theta-\delta\Big| M=m\Big). \label{eq:P:Em:RHS:2}
\end{IEEEeqnarray}
We will show that given any $\epsilon_1$, $\epsilon>0$, we can find a $\delta>0$ such that both the RHS terms above converge to $0$ as $n\rightarrow \infty$.

We now state some important lemmas which are needed to proceed with the probability of error analysis. We first state a lemma which directly follows from ~\cite[Lemma 2]{csiszar-narayan-it1991}. 
\begin{lemma}\label{lem:csiszar:narayan}
Consider any $\vec{\hat{r}}$ on the unit $n$-sphere and suppose an independent random vector $\vec{\hat{R}}$ is uniformly distributed on this sphere. Then for any $\gamma$ satisfying $1/\sqrt{2\pi n}<\gamma<1$, we have
\begin{equation*}\label{eq:CN}
\mathbb{P}\{ \left<\vec{\hat{r}},\vec{\hat{R}}\right>\geq \gamma\}\leq  2^{(n-1)\frac{1}{2}\log\left(1-\gamma^2\right)}. 
\end{equation*}
\end{lemma}
The above lemma is used in the proof of the next lemma, which guarantees encoding success with high probability.
\begin{lemma}\label{lem:binning:rate}
For any $\delta_1>0$ and message $M=m$, the probability that the encoder finds at least one $\vec{U}_{m,k}$ satisfying~\eqref{eq:encoder:condition:gaussian} approaches 1 as $n\rightarrow \infty$.
\end{lemma}
The proof of this lemma appears in Appendix~\ref{app:lems:2}. 
In the following lemma, we show that $\vec{U}-\alpha \bS$ satisfies the encoder power constraint, and hence, $\vec{X}=\vec{U}-\alpha \bS$ with high probability.
\begin{lemma}\label{lem:enc}
For any $\delta_2$ satisfying $0<\delta_2<\epsilon_1$ and message $M=m$,
\begin{equation*}
\mathbb{P}\left(\left|\|\vec{U}-\alpha \vec{S}\|^2-n P' \right| > n\delta_2 |M=m\right)\rightarrow 0,
\end{equation*}
as $n\rightarrow\infty$.
\end{lemma}
Refer Appendix~\ref{app:lems:2} for the proof of this lemma.
The following lemma captures the correlation that an adversary can induce with the chosen codeword through the choice of its jamming signal. We use Lemma~\ref{lem:csiszar:narayan} in the proof of this lemma as well.
\begin{lemma}\label{lem:J:U}
For any $\delta_3>0$ and message $M$=m, under any jamming strategy $Q_{\vec{J}|M,\vec{S}}:\vec{J}\in\mathcal{J}(\Lambda)$, 
\begin{equation*}
\mathbb{P}\left(\left| \left<\vec{J},\vec{U} \right>-\left<\vec{J},\vec{\hat{S}} \right>\left<\vec{\hat{S}},\vec{U} \right>\right| > n\delta_3\Big|M=m \right)\rightarrow 0,
\end{equation*}
as $n\rightarrow\infty$.
\end{lemma}
The proof can be found in Appendix~\ref{app:lems:2}. 
The following is the main lemma. We use Lemmas~\ref{lem:binning:rate},~\ref{lem:enc} and~\ref{lem:J:U} towards proving it.
This lemma shows that given any $\delta>0$, the inner product $\left<\vec{\hat{Y}},\vec{\hat{U}}\right>$ is at least $(\theta-\delta)$ with high probability irrespective of the adversary's strategy $Q_{\vec{J}|M,\vec{S}}:\vec{J}\in \mathcal{J}(\Lambda)$. Recall that $\delta_1$ is the parameter which appears in the definition of the encoder (see~\eqref{eq:encoder:condition:gaussian}).
\begin{lemma}\label{lem:Y:U:unit:expr}
There is a function $\delta_1:\mathbb{R}^+\rightarrow \mathbb{R}^+$, where $\delta_1(\delta)\rightarrow 0$ as $\delta\rightarrow 0$, such that for every message $M=m$, under any jamming strategy $Q_{\vec{J}|M,\vec{S}}:\vec{J}\in\mathcal{J}(\Lambda)$ and for any $\delta>0$, if the parameter $\delta_1$ in the definition of the encoder is chosen as $\delta_1(\delta)$, then
\begin{equation*}\label{eq:tilde:Y:U:theta}
\mathbb{P}\left(\left<\vec{\hat{Y}},\vec{\hat{U}}\right><\left(\theta-\delta\right) \Big|M=m\right)\rightarrow 0,
\end{equation*}
as $n\rightarrow\infty$, where
\begin{equation}\label{eq:theta}
\theta=\sqrt{\frac{\alpha(P'+\alpha \sigma_S^2)}{P_U  }}. 
\end{equation}
\end{lemma}
The proof of this lemma is in Appendix~\ref{app:lems:2}. Note that $\theta$ also depends on $\epsilon_1$. Coming back to the error analysis, note that Lemma~\ref{lem:Y:U:unit:expr} implies that the first RHS term in~\eqref{eq:P:Em:RHS:2} can be made arbitrarily small by choosing a sufficiently large $n$, provided the encoder parameter
$\delta_1$ is chosen suitably depending on $\delta$. Now, the second RHS term in~\eqref{eq:P:Em:RHS:2} can be bounded using the union bound, and hence, for any $Q_{\bJ|M,\bS}:\bJ\in\cJ(\Lambda)$ we have, 
\begin{align}
&\mathbb{P}\left(\exists m',\,  k': m'\neq m, \left<\vec{\hat{Y}},\vec{\hat{U}}_{m',k'}\right> \geq \theta-\delta \Big| M=m \right)\nonumber\\
&\stackrel{}{\leq} \sum_{m'\neq m,\,  k'}\mathbb{P}\left(\left<\vec{\hat{Y}},\vec{\hat{U}}_{m',k'}\right> \geq \theta-\delta \Big|M=m \right).\yesnumber\label{eq:P:m':k':all}
\end{align}
For any $m'\neq m$ and $k'$, we have
\begin{align}
&\mathbb{P}\left(\left<\vec{\hat{Y}},\vec{\hat{U}}_{m',k'}\right> \geq \theta-\delta \Big| M=m \right)\notag\\
&\leq 2^{(n-1)\frac{1}{2}\log\left(1-(\theta-\delta)^2\right)},\label{eq:P:m':k'}
\end{align}
by Lemma~\ref{lem:csiszar:narayan}, where we replace $(\vec{\hat{r}},\vec{\hat{R}})$ by $(\vec{\hat{Y}},\vec{\hat{U}}_{m',k'})$ and $\gamma$ by $(\theta-\delta)$. 
Using~\eqref{eq:P:m':k'} in~\eqref{eq:P:m':k':all} and noting that the total number of codewords is $2^{n R_U}$, we can conclude that for any $Q_{\bJ|M,\bS}:\bJ\in\cJ(\Lambda)$
%
\begin{align}
&\mathbb{P}\left(\exists m',\,  k': m'\neq m, \left<\vec{\hat{Y}},\vec{\hat{U}}_{m',k'}\right> \geq \theta-\delta \Big| M=m \right)\nonumber\\
&\stackrel{}{\leq} 2^{nR_U} 2^{(n-1)\frac{1}{2}\log\left(1-(\theta-\delta)^2\right)}.\label{eq:main:Pe:upbd}
\end{align}
%
We now give an alternate expression for $R_U$ in terms of $\theta$. Toward this, consider the following.
\begin{IEEEeqnarray*}{rCl}
1-\theta^2&\stackrel{(a)}{=}&1-\frac{\alpha (P'+\alpha \sigma_S^2)}{P_U}\\
&\stackrel{}{=}& \frac{P_U-\alpha^2 \sigma_S^2-\alpha P'}{P_U}\\
&\stackrel{(b)}{=}& \frac{P'-\alpha P'}{P_U}\\
&\stackrel{}{=}& \frac{(1-\alpha)P'}{P_U}\\
&\stackrel{(c)}{=}& \frac{(\Lambda+\sigma^2)P'}{(P'+\Lambda+\sigma^2)P_U}, \yesnumber\label{eq:1:minus:theta}
\end{IEEEeqnarray*}
where~\eqref{eq:theta} gives $(a)$, while $(b)$ follows from noting that $P_U=P'+\alpha^2 \sigma_S^2$. We get $(c)$ as $\alpha=P'/(P'+\Lambda+\sigma^2)$.
Recall from earlier in~\eqref{eq:R:U} our choice of $R_U$. Using~\eqref{eq:1:minus:theta}, we observe that $R_U$ can be also expressed as
\begin{IEEEeqnarray*}{rCl}
R_U= -\frac{1}{2}\log\left(1-\theta^2\right)-\epsilon/2.\nonumber\label{eq:R:u:alt}
\end{IEEEeqnarray*}
Now choosing a small enough $\delta>0$ in~\eqref{eq:main:Pe:upbd} such that\footnote{Note that there exists $\delta>0$ such that~\eqref{eq:R_U:delta} is satisfied. To see this, define $f(\delta)=-1/2 \log(1-(\theta-\delta)^2)$. It can be easily verified that $f$ is a continuous and  monotonically decreasing function of $\delta$.}
\begin{equation}\label{eq:R_U:delta}
R_U<-\frac{1}{2}\log\left(1-(\theta-\delta)^2\right),
\end{equation}
the RHS in~\eqref{eq:main:Pe:upbd}, and hence, the second term in the RHS of~\eqref{eq:P:Em:RHS:2}, approaches $0$ as $n\rightarrow\infty$. Thus, $P^{(n)}_e$ goes to $0$ as $n\rightarrow \infty$, and this completes the proof of achievability. 
\subsubsection{Converse} 
\label{dpavc:converse}
We prove the converse for an average probability of error criterion instead of
the maximum probability of error criterion. For this stronger version of the
converse, we define the average probability of error
(similarly as in~\eqref{eq:avgpe}) by
\begin{equation}\label{eq:avgdp}
P^{(n)}_e=\frac{1}{2^{nR}} \sum_{m=1}^{2^{nR}} P^{(n)}_{e,m},
\end{equation}
where
\begin{equation}\label{eq:avgdp:1}
P^{(n)}_{e,m}=\max_{Q_{\vec{J}|M=m,\vec{S}}:\vec{J}\vec\in \mathcal{J}(\Lambda) } \mathbb{P}\left( \Phi(\vec{Y})\neq m|M=m\right).
\end{equation}

Now let us consider any sequence of codes with rate $R$ and $P^{(n)}_e\rightarrow 0$ as $n\rightarrow \infty$. 
Even though the adversary can choose an arbitrary feasible 
vector jamming strategy
$Q_{\vec{J}|M,\vec{S}}:\vec{J}\vec\in \mathcal{J}(\Lambda)$, we analyze the performance  of the encoder-decoder
pair under an 
i.i.d. Gaussian jamming strategy. For an arbitrarily
small $\delta >0$, let $\Lambda'=\Lambda-\delta$. We define
$\bJ'$ to be a vector of length $n$ generated 
i.i.d. with $J'_i\sim\mathcal{N}(0,\Lambda')$, $\forall i$. We emphasize that $\bJ'$ is not
a feasible jamming strategy as $\|\bJ'\|$ can be greater than $\sqrt{n\Lambda}$.
We also define a feasible jamming strategy $\bJ$ whose distribution
is the same as the conditional distribution of $\bJ'$, conditioned
on $\vec{J}'\in\mathcal{J}(\Lambda)$. Let $\epsilon>0$ here.
Under the jamming strategy $\bJ'$, let ${P}'^{(n)}_{e}$ be the average probability of error achieved
by the given sequence of randomized codes. 
Then,
\begin{IEEEeqnarray*}{rCl}
&{P}'^{(n)}_{e}&\\
&\stackrel{}{=}&\frac{1}{2^{nR}} \sum_{i=1}^{2^{nR}} \mathbb{P} \left( \Phi(\Psi(i,\bS)+\vec{S}+\vec{J}'+\vec{Z}  )\neq i \right)\\
&\stackrel{}{\leq}&\frac{1}{2^{nR}} \sum_{i=1}^{2^{nR}} \mathbb{P} \left( \Phi(\Psi(i,\bS)+\vec{S}+\vec{J}'+\vec{Z}  )\neq i\Big| \vec{J}'\in\mathcal{J}(\Lambda) \right) \\
&&\hspace{15mm} \cdot~\mathbb{P}\left( \vec{J}'\in\mathcal{J}(\Lambda)\right)+\mathbb{P} \left( \|\vec{J}'\|^2> n\Lambda \right)\\
&\stackrel{(a)}{\leq}&\frac{1}{2^{nR}} \sum_{i=1}^{2^{nR}}  \mathbb{P} \left( \Phi(\Psi(i,\bS)+\vec{S}+\vec{J}'+\vec{Z}  )\neq i \Big| \vec{J}' \in\mathcal{J}(\Lambda)\right)\\
&&\hspace{15mm}+  \epsilon \hspace{30mm}\mbox{(for large enough }n)\\
&= &\frac{1}{2^{nR}} \sum_{i=1}^{2^{nR}}  \mathbb{P} \left( \Phi(\Psi(i,\bS)+\vec{S}+\vec{J}+\vec{Z}  )\neq i \right)+  \epsilon\\
&\stackrel{(b)}{\leq}& \frac{1}{2^{nR}} \sum_{i=1}^{2^{nR}} P_{e,i}^{(n)}+  \epsilon \\
&\stackrel{(c)}{=}&P_{e}^{(n)}+  \epsilon  \\
&\stackrel{(d)}{<}& 2\epsilon.
\end{IEEEeqnarray*}
Here the probability is over the shared randomness, the channel, the state and
adversary's (i.i.d. Gaussian) action. As $\vec{J}'$ is i.i.d Gaussian with
$J'_i\sim\mathcal{N}(0,\Lambda')$, $\forall i$, we have
$\mathbb{P}(\|\vec{J}'\|^2> n\Lambda)\rightarrow 0$ as $n\rightarrow \infty$. We choose $n$ large enough such that $\mathbb{P}(\|\vec{J}'\|^2> n\Lambda)\leq \epsilon$, which gives $(a)$.
Then, $(b)$ follows from~\eqref{eq:avgdp:1} since $\bJ$ is a feasible jamming strategy, while
 $(c)$ follows from~\eqref{eq:avgdp}. We now choose $n$  large enough
such that the probability $P_e^{(n)}$ is less than $\epsilon$, where $\epsilon>0$. This gives us $(d)$. Thus, we have
shown that for any $\epsilon >0$, under i.i.d. Gaussian (variance $\Lambda'$) jamming, 
the given sequence of randomized encoder-decoder pairs achieve
${P}'^{(n)}_{e}<2\epsilon$ for large enough $n$.

Under the jamming strategy $\bJ'$, the resulting channel is a dirty paper channel with noise variance $\Lambda'+\sigma^2$. Hence, the rate $R$ must be smaller than the capacity of this channel, i.e.,
\begin{equation*}
C\leq \frac{1}{2} \log \left(1+\frac{P}{\Lambda'+\sigma^2}\right).
\end{equation*}
Since this holds for any $\Lambda' < \Lambda$, we have
\begin{equation*}
C\leq \frac{1}{2} \log \left(1+\frac{P}{\Lambda+\sigma^2}\right).
\end{equation*}
This completes the proof of the converse.
\section{Discussion and Conclusion}\label{sec:conclusion}
In this work, we analysed the performance of a communication system over a
state-dependent channel in the presence of an adversary. Here  both the
encoder and the adversary were state-aware, i.e., they possessed non-causal
knowledge of the state. The adversary induced an AVC through its jamming 
interference into the channel, where the interference could be designed
using the non-causal knowledge of the state. 
We studied two versions, the discrete memoryless GP-AVC and the additive white Gaussian DP-AVC, 
and determined
their randomized coding capacity under a maximum probability of error
criterion. As in other randomized coding setups, we showed that the capacity for both our AVC setups was the same under the average probability of error criterion as well. Owing to the presence of shared randomness, it was seen that
even with the non-causal knowledge of the state vector and the ability to
use vector jamming strategies, the adversary could impact
the communication rate no worse than by choosing
memoryless strategies. Thus,  the capacity of both the AVCs
was characterized as that of the worst memoryless channel with state
that the adversary could induce through some
memoryless strategy. Furthermore, in the DP-AVC 
it was shown that the adversary, given its purpose, could do no better than to 
disregard the state knowledge entirely and introduce state-independent
white Gaussian noise. Both deterministic coding capacity and the effect of limited
shared randomness are natural next steps to this work. It would be interesting
to know if, like for standard AVCs~\cite{ahlswede-1978,csiszar-korner-book2011,hughes-thomas-it1996}, $O(\log n)$ bits of randomness (in a block length of $n$)
are sufficient to achieve randomized capacity. Finally, the results presented in this work could be similarly extended to
state-dependent channels, where, in addition to the encoder and adversary,
the decoder too is state-aware. 
\appendices
\section{Proof of Lemma~\ref{lem:gen:markov:lemma}}\label{app:lem:gen:markov:lemma}
The given distribution $P_\bZ$ is `close' to the uniform distribution over 
$\mathcal{T}^n_{\delta_0}(P_{Y,Z}|\vec{y})$ due to the properties
$(a)$ and $(b)$. Hence, in a two part proof, we 
first bound the probability $\mathbb{P}\left( \vec{Z}\not \in
\mathcal{T}^n_{\delta} (P_{X,Y,Z}|\vec{x},\vec{y})  \right)$
for $\vec{Z}\sim \allowbreak\text{Unif}(\mathcal{T}^n_{\delta_0}(P_{Y,Z}|\vec{y}))$.
Then, in the second part, we appropriately modify this bound to obtain a bound on
$\mathbb{P}\left( \vec{Z}\not \in
\mathcal{T}^n_{\delta} (P_{X,Y,Z}|\vec{x},\vec{y})  \right)$ under
the given distribution $P_\bZ$. 

\newcounter{storeeqcounter2}
\newcounter{tempeqcounter2}
To prove the first part, we begin by assuming that
$\vec{Z}\sim \allowbreak\text{Unif}(\mathcal{T}^n_{\delta_0}(P_{Y,Z}|\vec{y}))$. Then, as given on the next page, we can simplify $\mathbb{P}( \bZ  \not \in \mathcal{T}^n_{\delta} (P_{X,Y,Z}|\vec{x},\vec{y})  )$ to~\eqref{eq:1}, where~\eqref{eq:1:a} follows from the union bound, and~\eqref{eq:1:b} follows by relaxing the strict inequality. 
\addtocounter{equation}{1}%
\setcounter{storeeqcounter2}%
{\value{equation}}%
%
\begin{figure*}[!t]
\normalsize
\begin{IEEEeqnarray*}{rCl}
\setcounter{equation}{\value{storeeqcounter2}}
\mathbb{P}( \bZ  \not \in \mathcal{T}^n_{\delta} (P_{X,Y,Z}|\vec{x},\vec{y})  )&=&\mathbb{P}\Bigg(  \bigcup_{(x,y,z)} \Bigg\{\Bigg|  \frac{N(x,y,z |\vec{x},\vec{y},\vec{Z})}{n} -P_{X,Y}(x,y) P_{Z|Y}(z|y) \Bigg| > \delta\Bigg\} \Bigg)\\
&\stackrel{}{\leq}&  \sum_{x,y,z} \mathbb{P}\Bigg(  \Bigg|  \frac{N(x,y,z|\vec{x},\vec{y},\vec{Z})}{n}-P_{X,Y}(x,y) P_{Z|Y}(z|y)\Bigg| > \delta \Bigg)\yesnumber\label{eq:1:a}\\
&\stackrel{}{\leq}&    \sum_{x,y,z} \mathbb{P}\Bigg(  \Bigg|  \frac{N(x,y,z|\vec{x},\vec{y},\vec{Z})}{n}-P_{X,Y}(x,y) P_{Z|Y}(z|y)\Bigg| \geq \delta \Bigg)\yesnumber   \label{eq:1:b}\\
&\stackrel{}{=}&  \sum_{x,y,z} \mathbb{P}\left(  \frac{N(x,y,z|\vec{x},\vec{y},\vec{Z})}{n} \geq P_{X,Y}(x,y) P_{Z|Y}(z|y)+\delta    \right)\\
&&+ \sum_{x,y,z} \mathbb{P}\bigg(  \frac{N(x,y,z|\vec{x},\vec{y},\vec{Z})}{n} \leq P_{X,Y}(x,y) P_{Z|Y}(z|y)-\delta    \bigg)\yesnumber\label{eq:1}
\setcounter{tempeqcounter2}{\value{equation}}
\end{IEEEeqnarray*}
\setcounter{equation}{\value{tempeqcounter2}} 
\hrulefill
\vspace*{4pt}
\end{figure*}
Since $\vec{Z}\in \mathcal{T}^n_{\delta_0}(P_{Y,Z}|\vec{y})$ (with probability one), we have
$\forall (y,z)\in \mathcal{Y}\times \mathcal{Z}$ 
\begin{IEEEeqnarray}{rCl}
\left|\frac{N(y,z|\vec{y},\vec{Z})}{n}-P_{Y}(y)P_{Z|Y}(z|y)\right| < \delta_0.\label{eq:y:z:typ} 
\end{IEEEeqnarray}
For every $(y,z)$ such that $P_{Z|Y}(z|y)=0$, 
$N(y,z|\vec{y},\vec{Z})\leq n\delta_0$ using \eqref{eq:y:z:typ}. This
further implies that $N(x,y,z|\vec{x},\vec{y},\vec{Z})\leq n\delta_0$.
By choosing $\delta$ large enough such that $\delta>\delta_0$, we can
guarantee that $N(x,y,z|\vec{x},\vec{y},\vec{Z})< n\delta$, and hence, it
follows that the probability of both the terms in the summation in~\eqref{eq:1} is zero.

For other values of $(y,z)$, for which $P_{Z|Y}(z|y) >0$, we first note that $P_{Z|Y}(z|y) \geq P^{\min}_{Z|Y}$, where
\begin{IEEEeqnarray*}{rCl}
P^{\min}_{Z|Y}:=\min_{(y,z) : P_{Z|Y}(z|y)>0}  P_{Z|Y}(z|y).
\end{IEEEeqnarray*}
We 
define $\delta_0'' = \delta_0 + \sqrt{\delta_0} < 2\sqrt{\delta_0}$, and we assume
that $\delta > 3\sqrt{\delta_0}$.
If $P_Y(y) < \delta_0''$, then $P_{Y,Z}(y,z) < \delta_0''$. This implies that
\begin{IEEEeqnarray*}{rCl}
N(y,z|\vec{y},\vec{Z})&\leq& n(\delta_0'' + \delta_0) \\
&<& n(3\sqrt{\delta_0})\\
&<& n\delta.
\end{IEEEeqnarray*}
This again implies that
\begin{IEEEeqnarray*}{rCl}
N(x,y,z|\vec{x},\vec{y},\vec{Z})\leq n \delta,
\end{IEEEeqnarray*}
and thus, the probability of the first term in~\eqref{eq:1}  
 is zero. Further, if $P_Y(y) < \delta_0''$, then 
\begin{IEEEeqnarray*}{rCl}
P_{Y}(y)P_{X|Y}(x|y)P_{Z|Y}(z|y)\leq \delta_0''.
\end{IEEEeqnarray*}
This implies that 
\begin{IEEEeqnarray*}{rCl}
-\delta+P_{Y}(y)P_{X|Y}(x|y)P_{Z|Y}(z|y)<0,
\end{IEEEeqnarray*}
and hence, the probability of the second term in~\eqref{eq:1} is zero. We have, thus, shown that the probability terms in both the summations in the RHS of~\eqref{eq:1} are equal to zero. Based on the above observations, we now consider 
those $(y,z)\in\mathcal{Y}\times\mathcal{Z}$ such that $P_{Z|Y}(z|y) \geq P^{\min}_{Z|Y}$
and $P_Y(y) \geq \delta_0''$.

We know that $(\vec{x},\vec{y})\in \mathcal{T}^n_{\delta_0}(P_{X,Y})$. Hence,  
\begin{IEEEeqnarray}{rCl}\label{eq:s:j:typ}
\left| \frac{N(x,y|\vec{x},\vec{y})}{n} -P_{X,Y}(x,y)\right| \leq \delta_0 \,\,\,\,\, \forall (x,y).
\end{IEEEeqnarray}
We now make the following claim. 
\begin{claim}
If $\vec{z}\in \mathcal{T}^n_{\delta_0}(P_{Y,Z}|\vec{y})$, then
\begin{IEEEeqnarray}{rCl}\label{eq:yz:y:typ}
\left| \frac{N(y,z|\vec{y},\vec{z})}{N(y|\vec{y})} -P_{Z|Y}(z|y)\right| \leq \delta'_0 \,\,\,\,\,\, \forall (y,z),
\end{IEEEeqnarray}
where $\delta'_0(\delta_0)=2\sqrt{\delta_0}$.
\end{claim}
\begin{IEEEproof}
Since $(\vec{y},\vec{z})\in \mathcal{T}^{n}_{\delta_0}(P_{Y,Z})$, we have $\forall (y,z)$,
\begin{IEEEeqnarray*}{rCl}
\frac{N(y,z|\vec{y},\vec{z})}{n} -P_{Y,Z}(y,z) \leq \delta_0 .
\end{IEEEeqnarray*}
As $P_{Y}(y)\geq \delta_0''$ and from~\eqref{eq:s:j:typ}, it follows that $N(y|\vec{y})/n>0$.
Thus, 
\begin{IEEEeqnarray*}{rCl}
\frac{N(y,z|\vec{y},\vec{z})}{N(y|\vec{y})}\leq \frac{P_{Y,Z}(y,z)+ \delta_0}{\frac{N(y|\vec{y})}{n}}.\label{eq:yz:11}
\end{IEEEeqnarray*}
But, we know that
\begin{IEEEeqnarray*}{rCl}\label{eq:y:typ}
\left|\frac{N(y|\vec{y})}{n} -P_{Y}(y)\right| \leq \delta_0 \hspace{5mm}\forall y.
\end{IEEEeqnarray*}
Hence, it follows that
\begin{IEEEeqnarray*}{rCl}
\frac{N(y,z|\vec{y},\vec{z})}{N(y|\vec{y})}-P_{Z|Y}(z|y)
&\leq & \frac{P_{Y,Z}(y,z)+ \delta_0}{P_{Y}(y)-\delta_0} -P_{Z|Y}(z|y)\\
&= & \frac{\delta_0(1+P_{Z|Y}(z|y))}{P_{Y}(y)-\delta_0}\\
&\leq &\frac{2\delta_0}{P_{Y}(y)-\delta_0}\\
&\stackrel{(a)}{\leq}& \frac{2\delta_0}{\delta_0''-\delta_0}\\
&\stackrel{(b)}{=}& 2\sqrt{\delta_0}.
\end{IEEEeqnarray*}
Here $(a)$ follows from $P_{Y}(y)\geq \delta_0''$, and $(b)$ is true as $\delta_0''=\delta_0+\sqrt{\delta_0}$.
Similarly, it can be shown that
\begin{IEEEeqnarray*}{rCl}\label{eq:yz:1}
\frac{N(y,z|\vec{y},\vec{z})}{N(y|\vec{y})}-P_{Z|Y}(z|y)
\geq-2\sqrt{\delta_0}  .
\end{IEEEeqnarray*}
This completes the proof of the claim.
\end{IEEEproof}
Continuing the analysis further, we consider a term inside the first sum in~\eqref{eq:1}. We first recall that if $N(x,y|\vec{x},\vec{y}) \allowbreak <n\delta$, then $N(x,y,z|\vec{x},\vec{y},\vec{Z}) <n\delta$, and thus, the probability under consideration is
zero. Hence, in the following, we assume w.l.o.g. that $N(x,y|\vec{x},\vec{y}) \geq n\delta$. 
We now get~\eqref{eq:prob:1}, as given on top of the next page,
\newcounter{storeeqcounter3}
\newcounter{tempeqcounter3}
%
\addtocounter{equation}{1}%
\setcounter{storeeqcounter3}%
{\value{equation}}%
%
\begin{figure*}[!t]
\normalsize
\begin{IEEEeqnarray}{rCl}
\setcounter{equation}{\value{storeeqcounter3}}
\mathbb{P}\bigg( \frac{N(x,y,z|\vec{x},\vec{y},\vec{Z})}{n} \geq& \delta&+ P_{X,Y}(x,y) P_{Z|Y}(z|y)    \bigg)\nonumber\\
&\stackrel{}{=}&\mathbb{P}\bigg(  \frac{N(x,y,z|\vec{x},\vec{y},\vec{Z})}{N(x,y|\vec{x},\vec{y}) }\frac{  N(x,y|\vec{x},\vec{y})}{n} \geq \delta+ P_{X,Y}(x,y) P_{Z|Y}(z|y)    \bigg)\nonumber\\
 &\stackrel{}{\leq}&  \mathbb{P}\left(  \frac{N(x,y,z|\vec{x},\vec{y},\vec{Z})}{N(x,y|\vec{x},\vec{y}) } \geq \frac{(\delta+ P_{X,Y}(x,y) P_{Z|Y}(z|y))}{(\delta_0+P_{X,Y}(x,y) )}     \right)\label{eq:prob:1:a}\\
&\stackrel{}{=}&  \mathbb{P}\bigg(  \frac{N(x,y,z|\vec{x},\vec{y},\vec{Z})}{N(x,y|\vec{x},\vec{y}) } \geq \frac{P_{Z|Y}(z|y)}{P_{Z|Y}(z|y)}\frac{(\delta+ P_{X,Y}(x,y) P_{Z|Y}(z|y))}{(\delta_0+P_{X,Y}(x,y) )}    \bigg)\nonumber\\
  &\stackrel{}{\leq}&  \mathbb{P}\bigg(  \frac{N(x,y,z|\vec{x},\vec{y},\vec{Z})}{N(x,y|\vec{x},\vec{y}) }\geq P_{Z|Y}(z|y)\frac{(\delta+ P_{X,Y}(x,y) P_{Z|Y}(z|y))}{(\delta_0+P_{X,Y}(x,y) P_{Z|Y}(z|y))}     \bigg)\nonumber\\	
	&\leq&  \mathbb{P}\left(  \frac{N(x,y,z|\vec{x},\vec{y},\vec{Z})}{N(x,y|\vec{x},\vec{y}) } \geq P_{Z|Y}(z|y)\left(\frac{\delta+1}{\delta_0+1}\right)       \right)\nonumber\\
	&=&  \mathbb{P}\left(  \frac{N(x,y,z|\vec{x},\vec{y},\vec{Z})}{N(x,y|\vec{x},\vec{y}) } \geq P_{Z|Y}(z|y)\left(1+\frac{\delta-\delta_0}{1+\delta_0}\right)       \right)\nonumber\\
 &\stackrel{}{=}&  \mathbb{P}\bigg(  \frac{N(x,y,z|\vec{x},\vec{y},\vec{Z})}{N(x,y|\vec{x},\vec{y}) }-(P_{Z|Y}(z|y)+\delta'_0) \geq P_{Z|Y}(z|y)\left(\frac{\delta-\delta_0}{1+\delta_0}\right)-\delta'_0      \bigg)\nonumber\\
&\stackrel{}{\leq}&  \mathbb{P}\bigg(  \frac{N(x,y,z|\vec{x},\vec{y},\vec{Z})}{N(x,y|\vec{x},\vec{y}) }-(P_{Z|Y}(z|y)+\delta'_0) \geq P^{\min}_{Z|Y}(z|y)\left(\frac{\delta-\delta_0}{1+\delta_0}\right)-\delta'_0      \bigg)\label{eq:prob:1:b}\\
&\stackrel{}{=}&  \mathbb{P}\left(  \frac{N(x,y,z|\vec{x},\vec{y},\vec{Z})}{N(x,y|\vec{x},\vec{y}) }-(P_{Z|Y}(z|y)+\delta'_0)\geq  t_1      \right),\label{eq:prob:1}
\setcounter{tempeqcounter3}{\value{equation}}
\end{IEEEeqnarray}
\setcounter{equation}{\value{tempeqcounter3}} 
\hrulefill
\vspace*{4pt}
\end{figure*}
where  
\begin{IEEEeqnarray*}{rCl}
t_1=P^{\min}_{Z|Y}\left(\left(\frac{\delta-\delta_0}{1+\delta_0}\right)-\frac{\delta'_0}{P^{\min}_{Z|Y}}\right),
\end{IEEEeqnarray*}
and does not depend on $n$. We choose $\delta$ such that $t_1>0$. Recall 
that we have earlier required $\delta >3\sqrt{\delta_0}$ already. Note that~\eqref{eq:prob:1:a} (given on the next page) 
follows from the upper bound for $N(x,y|\vec{x},\vec{y})/n$ in~\eqref{eq:s:j:typ}, while~\eqref{eq:prob:1:b} (given on the next page)  follows as $\forall (y,z)$ under consideration, $P_{Z|Y}(z|y)\geq P_{Z|Y}^{\min}$.   
The following claim now gives an exponentially decaying bound on the term appearing in~\eqref{eq:prob:1}.  %
\begin{claim}\label{claim:hoeffding}
If $N(x,y|\vec{x},\vec{y})\geq n\delta$ and $t_1>0$,
\begin{equation*}
\mathbb{P}\left(  \frac{N(x,y,z|\vec{x},\vec{y},\vec{Z})}{N(x,y|\vec{x},\vec{y}) }- (P_{Z|Y}(z|y)+\delta'_0) \geq t_1  \right)\leq e^{-2n\delta t_1^2}.
\end{equation*}
\end{claim}
\begin{IEEEproof}
Let $S{(x,y|\vec{x},\vec{y})}$ denote the indices of $(\vec{x},\vec{y})$
with the value $(x,y)$ and $S{(y|\vec{y})}$ denote the indices of 
$\vec{y}$ with the value $y$. We now consider a different but equivalent random
experiment for generating $\vec{Z}$. First $\vec{\tilde{Z}}$ is chosen uniformly
at random from $\mathcal{T}^n_{\delta_0}(P_{Y,Z}|\vec{y})$, where $P_{Y,Z}=P_{Y} P_{Z|Y}$, and then, for each $y$, its components at $S{(y|\vec{y})}$ are subjected to a permutation chosen
uniformly at random from the set of all permutations of 
$S{(y|\vec{y})}$. Since the set of sequences in 
$\mathcal{T}^n_{\delta_0}(P_{Y,Z}|\vec{y})$ are invariant under such permutations,
this two-step process results in the same final distribution of
$\vec{\tilde{Z}}$, i.e., uniform over $\mathcal{T}^n_{\delta_0}(P_{Y,Z}|\vec{y})$.
From~\eqref{eq:yz:y:typ}, $N(y,z|\vec{y},\vec{\tilde{Z}})$ is bounded by
\begin{align}\label{eq:N:u:s:ub}
&N(y,z|\vec{y},\vec{\tilde{Z}}) \leq N{(y|\vec{y})}(P_{Z|Y}(z|y)+\delta'_0).
\end{align}
For a given $S{(y|\vec{y})}$ and conditioned on $N(y,z|\vec{y},\vec{\tilde{Z}})=k$, 
the number $N(x,y,z|\vec{x},\vec{y},\vec{\tilde{Z}})$ can be considered as the
number of positions in $S{(x,y|\vec{x},\vec{y})}$ at which the letter $z$ is assigned by 
the random permutation in the components in $S{(y|\vec{y})}$. 
Thus, $N(x,y,z|\vec{x},\vec{y},\vec{\tilde{Z}})$ is the number of times $z$ is obtained
when a total of $|S{(x,y|\vec{x},\vec{y})}|=N(x,y|\vec{x},\vec{y})$ samples are drawn without replacement
from a collection of $|S{(y|\vec{y})}|$ components, of which 
$k$ components have value $z$. Now using Hoeffding's inequality for
sampling without replacement~\cite{hoeffding-1963}, 
\begin{IEEEeqnarray*}{rCl}
&& \mathbb{P}\bigg( \frac{N(x,y,z|\vec{x},\vec{y},\vec{\tilde{Z}})}{|S{(x,y|\vec{x},\vec{y})}|}-\frac{k}{|S{(y|\vec{y})}|}>t_1 \Big| N(y,z|\vec{y},\vec{\tilde{Z}})=k\bigg)\\
&&\hspace{25mm}\leq e^{-2|S{(x,y|\vec{x},\vec{y})}|t_1^2}\nonumber\\
\Rightarrow &&\mathbb{P}\bigg( \frac{N(x,y,z|\vec{x},\vec{y},\vec{\tilde{Z}})}{N{(x,y|\vec{x},\vec{y})}}-\frac{N{(y,z|\vec{y},\vec{\tilde{Z}})}}{N{(y|\vec{y})}}>t_1)\bigg)\\
&&\hspace{25mm} \leq e^{-2N{(x,y|\vec{x},\vec{y})}t_1^2}\nonumber\\
\Rightarrow
&&\mathbb{P}\bigg(  \frac{N(x,y,z|\vec{x},\vec{y},\vec{\tilde{Z}})}{N(x,y|\vec{x},\vec{y}) }- (P_{Z|Y}(z|y)+\delta'_0) \geq t_1      \bigg)\\
&&\hspace{25mm} \stackrel{}{\leq} e^{-2n\delta t_1^2},  
\end{IEEEeqnarray*}
where the last step follows from~\eqref{eq:N:u:s:ub} and  $N(x,y|\vec{x},\vec{y}) \geq n\delta$.
This completes the proof of Claim~\ref{claim:hoeffding}.
\end{IEEEproof}

We will now get a similar bound for each term inside the second sum 
in~\eqref{eq:1}. Recall that $(y,z)\in \mathcal{Y}\times\mathcal{Z}$ such that 
\begin{IEEEeqnarray*}{rCl}
P_{Y}(y)&>&\delta_0''\\
&=&\delta_0+\sqrt{\delta_0}
\end{IEEEeqnarray*}
and $P_{Z|Y}(z|y)\geq P^{\min}_{Z|Y}$. Note that if $N(x,y|\vec{x},\vec{y}) \leq (1/4) n\delta$, then from~\eqref{eq:s:j:typ}, 
\begin{IEEEeqnarray*}{rCl}
P_{X,Y}(x,y)&\leq& N(x,y|\vec{x},\vec{y})/n+\delta_0\\
&\leq& \delta/4+\delta_0.
\end{IEEEeqnarray*}
Hence, 
\begin{IEEEeqnarray*}{rCl}
P_{X,Y}(x,y) P_{Z|Y}(z|y)\leq\frac{\delta}{4}+\delta_0.
\end{IEEEeqnarray*}
Then, the probability under consideration is zero if $\delta_0< 3\delta/4$. Hence, for the rest of the analysis we assume that 
\begin{IEEEeqnarray*}{rCl}
N(x,y|\vec{x},\vec{y})\geq (1/4) n\delta.
\end{IEEEeqnarray*}
Note that this implies 
\begin{IEEEeqnarray*}{rCl}
P_{X,Y}(x,y)-\delta_0&\stackrel{(a)}{\geq}& \left(\frac{N(x,y|\vec{x},\vec{y})}{n}-\delta_0\right)-\delta_0\nonumber\\
&\stackrel{}{\geq}& \frac{\delta}{4}-2\delta_0\\
&\stackrel{(b)}{>}& 0.\yesnumber\label{eq:P:xy:geq:0}
\end{IEEEeqnarray*}
Here, $(a)$ follows from~\eqref{eq:s:j:typ}, and $(b)$ follows by choosing $\delta> 8\delta_0$. 
\newcounter{storeeqcounter4}
\newcounter{tempeqcounter4}
We now get~\eqref{eq:prob:2}, given on top of the next page,
%
\addtocounter{equation}{1}%
\setcounter{storeeqcounter4}%
{\value{equation}}%
%
\begin{figure*}[!t]
\normalsize
\begin{IEEEeqnarray}{rCl}
\setcounter{equation}{\value{storeeqcounter4}}
\mathbb{P}\bigg(  \frac{N(x,y,z|\vec{x},\vec{y},\vec{Z})}{n} \leq &-\delta&+ P_{X,Y}(x,y) P_{Z|Y}(z|y)   \bigg)\nonumber\\
&\stackrel{}{=}&\mathbb{P}\bigg(  \frac{N(x,y,z|\vec{x},\vec{y},\vec{Z})}{N(x,y|\vec{x},\vec{y}) }\frac{  N(x,y|\vec{x},\vec{y})}{n}\leq -\delta+ P_{X,Y}(x,y) P_{Z|Y}(z|y)     \bigg)\nonumber\\
 &\stackrel{(a)}{\leq}&  \mathbb{P}\left(  \frac{N(x,y,z|\vec{x},\vec{y},\vec{Z})}{N(x,y|\vec{x},\vec{y}) } \leq \frac{(-\delta+ P_{X,Y}(x,y) P_{Z|Y}(z|y))}{(-\delta_0+P_{X,Y}(x,y) )}     \right)\label{eq:prob:2:a}\\
&\stackrel{}{=}&  \mathbb{P}\bigg(  \frac{N(x,y,z|\vec{x},\vec{y},\vec{Z})}{N(x,y|\vec{x},\vec{y}) } \leq \frac{P_{Z|Y}(z|y)}{P_{Z|Y}(z|y)}\frac{(-\delta+ P_{X,Y}(x,y) P_{Z|Y}(z|y))}{(-\delta_0+P_{X,Y}(x,y) )}     \bigg)\nonumber\\
  &\stackrel{}{\leq}&  \mathbb{P}\bigg(  \frac{N(x,y,z|\vec{x},\vec{y},\vec{Z})}{N(x,y|\vec{x},\vec{y}) } \leq P_{Z|Y}(z|y)\frac{(-\delta+ P_{X,Y}(x,y) P_{Z|Y}(z|y))}{(-\delta_0+P_{X,Y}(x,y) P_{Z|Y}(z|y))}     \bigg)\nonumber\\
 &\stackrel{}{\leq}&  \mathbb{P}\left(  \frac{N(x,y,z|\vec{x},\vec{y},\vec{Z})}{N(x,y|\vec{x},\vec{y}) } \leq P_{Z|Y}(z|y)\frac{(-\delta+1)}{(-\delta_0+1)}     \right)\nonumber\\
		&=&  \mathbb{P}\left(  \frac{N(x,y,z|\vec{x},\vec{y},\vec{Z})}{N(x,y|\vec{x},\vec{y}) } \leq P_{Z|Y}(z|y)\left(1-\frac{\delta-\delta_0}{1-\delta_0}\right)       \right)\nonumber\\
 &\stackrel{}{=}&  \mathbb{P}\bigg(  \frac{N(x,y,z|\vec{x},\vec{y},\vec{Z})}{N(x,y|\vec{x},\vec{y}) }-P_{Z|Y}(z|y)+\delta'_0  \leq -P_{Z|Y}(z|y)\left(\frac{\delta-\delta_0}{1-\delta_0}\right)+\delta'_0      \bigg)\nonumber\\
&\stackrel{(b)}{\leq}&  \mathbb{P}\Bigg(  \frac{N(x,y,z|\vec{x},\vec{y},\vec{Z})}{N(x,y|\vec{x},\vec{y}) }-(P_{Z|Y}(z|y)-\delta'_0) \leq -P^{\min}_{Z|Y}\left(\left(\frac{\delta-\delta_0}{1-\delta_0}\right)-\frac{\delta'_0}{P^{\min}_{Z|Y}}\right)      \Bigg)\label{eq:prob:2:b}\\
&\stackrel{}{=}&  \mathbb{P}\left(  \frac{N(x,y,z|\vec{x},\vec{y},\vec{Z})}{N(x,y|\vec{x},\vec{y}) }-(P_{Z|Y}(z|y)-\delta'_0)\leq  -t_2      \right),\label{eq:prob:2}%
\setcounter{tempeqcounter4}{\value{equation}} 
\end{IEEEeqnarray}
\setcounter{equation}{\value{tempeqcounter4}} 
\hrulefill
\vspace*{4pt}
\end{figure*}
where \begin{IEEEeqnarray*}{rCl}
t_2=P^{\min}_{Z|Y}\left(\left(\frac{\delta-\delta_0}{1-\delta_0}\right)-\frac{\delta'_0}{P^{\min}_{Z|Y}}\right)
\end{IEEEeqnarray*}
and does not depend on $n$. Once again, we choose $\delta$ so as to ensure that $t_2>0$. 
Observe that ~\eqref{eq:prob:2:a} (given on the next page) follows from the lower bound for $N(x,y|\vec{x},\vec{y})/n$
in~\eqref{eq:s:j:typ} as well as by choosing $\delta>8\delta_0$ so that~\eqref{eq:P:xy:geq:0} is true. We get~\eqref{eq:prob:2:b} (given on the next page) as we are analyzing for $(y,z)$  for which $P_{Z|Y}(z|y)\geq P_{Z|Y}^{\min}$.   
\begin{claim}\label{claim:hoeffding:2}
If $N(x,y|\vec{x},\vec{y})\geq (1/4)~n\delta$ and $t_2>0$
\begin{IEEEeqnarray*}{rCl}
&\mathbb{P}\bigg(&  \frac{N(x,y,z|\vec{x},\vec{y},\vec{Z})}{N(x,y|\vec{x},\vec{y}) }- (P_{Z|Y}(z|y)-\delta'_0) \leq -t_2  \bigg)\\
&&\hspace{45mm}\leq e^{-\frac{n}{2}\delta t_2^2}.
\end{IEEEeqnarray*}
\end{claim}
\begin{IEEEproof}
The proof follows in a manner similar to that of Claim~\ref{claim:hoeffding}.
\end{IEEEproof}
Now summing over all possible $(x,y,z)$ in~\eqref{eq:1} and using Claims~\ref{claim:hoeffding} and~\ref{claim:hoeffding:2}, we have
\begin{IEEEeqnarray}{rCl}\label{eq:U:bound:unif}
\mathbb{P}\left( \vec{Z}\not \in \mathcal{T}^n_{\delta} (P_{X,Y,Z}|\vec{x},\vec{y})  \right)
 &\leq&  |\mathcal{X}| |\mathcal{Y}| |\mathcal{Z}|  \left(e^{-2n\delta t_1^2 }+e^{-\frac{n}{2}\delta t_2^2}\right) \nonumber\\
&\leq&  2~|\mathcal{X}| |\mathcal{Y}| |\mathcal{Z}|~e^{-\frac{n}{2}\delta t^2}, \label{eq:2}
\end{IEEEeqnarray}
where $t=\min(2t_1,t_2)$.
This shows that when $\bZ$ is chosen uniformly over $\mathcal{T}^n_{\delta_0}(P_{Y,Z}|\vec{y})$, the result holds. This
completes the first part of the proof.

 For the second part, we will now perturb the uniform distribution to
an arbitrary distribution $P_\bZ$ satisfying the conditions of the lemma. Under
$P_{\vec{Z}}(\vec{z})$, some non-zero probability (denoted by $\epsilon$) may
be assigned to the set of non-typical sequences, i.e., the complement of the
set $\mathcal{T}^n_{\delta_0}(P_{Y,Z}|\vec{y})$. 
Due to the perturbation, the probability of a typical
sequence in $\mathcal{T}^n_{\delta_0}(P_{Y,Z}|\vec{y})$ can also increase
by a factor of at most $2^{nh(\delta_0)}$, where $h(\delta_0)\rightarrow 0$ as
$\delta_0 \rightarrow 0$.  
Specifically, we know that
\begin{IEEEeqnarray*}{rCl}
\left|\mathcal{T}^n_{\delta_0}(P_{Y,Z}|\vec{y})\right|\leq
2^{n(H(Z|Y)+\tilde{g}(\delta_0))},
\end{IEEEeqnarray*}
where $\tilde{g}(\delta_0)=\left(\max_{y} H(Z|Y=y) \right)\cdot \delta_0$, and $\td{g}(\delta)\rightarrow 0$ as $\delta_0\rightarrow 0$. Note that $\td{g}(\delta_0)$ depends only on $P_{Z|Y}$. 
Hence, under the uniform distribution over
$\mathcal{T}^n_{\delta_0}(P_{Y,Z}|\vec{y})$,
\begin{IEEEeqnarray*}{rCl}
\mathbb{P}({\vec{Z}}=\vec{z})\geq 2^{-n(H(Z|Y)+\tilde{g}(\delta_0))}.
\end{IEEEeqnarray*}
By condition $(b)$ of the Lemma, the perturbation in the distribution can increase
the probability of any typical sequence by a factor of at most
\begin{IEEEeqnarray*}{rCl}
2^{-n(H(Z|Y)-g(\delta_0))}/2^{-n(H(Z|Y)+\tilde{g}(\delta_0))}=2^{n
h(\delta_0)}.
\end{IEEEeqnarray*}
Here $h(\delta_0)>0$ and $h(\delta_0)\rightarrow 0$ as $\delta_0\rightarrow 0$. 

Thus, the probability $\mathbb{P}\left( \vec{Z}\not \in \mathcal{T}^n_{\delta} (P_{X,Y,Z}|\vec{x},\vec{y})  \right)$ can now be bounded as 
follows. Given  $\vec{x}$ and $\vec{y}$, let us define the set  
\begin{equation*}
E=\left\{\vec{z}: (\vec{x},\vec{y},\vec{z})\not\in \mathcal{T}^n_{\delta}(P_{X,Y,Z})  \right\}=E_1\cup E_2,
\end{equation*}
where $E_1=E\cap \mathcal{T}^n_{\delta_0}(P_{Y,Z}|\vec{y})$ and $E_2=E\backslash E_1$.
Then, using the union bound and~\eqref{eq:U:bound:unif}, we have
\begin{IEEEeqnarray}{rCl}
\mathbb{P}(E)&\leq& \mathbb{P}(E_1)+\mathbb{P}(E_2)\nonumber\\
&\leq&\left(2~|\mathcal{X}| |\mathcal{Y}| |\mathcal{Z}| ~e^{-\frac{n}{2}\delta t^2}\right) 2^{nh(\delta_0)}+\epsilon\nonumber\\
&=& 2~|\mathcal{X}| |\mathcal{Y}| |\mathcal{Z}| ~e^{-n\left(\frac{1}{2}\delta t^2-h(\delta_0) \ln 2\right)}+\epsilon.\nonumber
\end{IEEEeqnarray}
We now choose a $\delta(\delta_0)$  large enough such that $K=\frac{1}{2}\delta t^2-h(\delta_0) \ln 2>0$ as well as all the other conditions on $\delta$ appearing in the proof are met.
This completes the proof of Lemma~\ref{lem:gen:markov:lemma}.
\section{Proofs of Lemmas~\ref{lem:binning:rate},~\ref{lem:enc},~\ref{lem:J:U} and~\ref{lem:Y:U:unit:expr}  }\label{app:lems:2}
\subsection{Proof of Lemma~\ref{lem:binning:rate}  }  
Let $M=m$ be the message and define the event (as a function of $\delta_1>0$)
\begin{equation}\label{eq:E:S}
E_0=\left\{\left| \|\vec{S}\|^2-n\sigma_S^2\right|>n\delta_0(\delta_1)\right\},
\end{equation}
where $0<\delta_0(\delta_1)< \delta_1$ (the exact choice of $\delta_0(\delta_1)$ will be discussed later in Claim~\ref{claim:bin}), and $\delta_0(\delta_1)\rightarrow 0$ as $\delta_1\rightarrow 0$. As $\bS$ is an i.i.d. Gaussian vector, where $S_i\sim\mathcal{N}(0,\sigma_S^2)$, $\forall i$, it follows that $\mathbb{P}(E_0)\rightarrow 0$ as $n\rightarrow \infty$ for given $\delta_0>0$. Next, let us define
\begin{equation}\label{eq:beta:1}
\beta:=\alpha \sqrt{\frac{\sigma_S^2}{P_U}}.
\end{equation}
Note that $\beta$ also depends on $\epsilon_1$ through the definition of $P_U$. We observe that 
\begin{IEEEeqnarray*}{rCl}
\td{R}&>&\frac{1}{2}\log\left(\frac{P_U}{P'}\right)\\
&\stackrel{}{=}& \frac{1}{2}\log\left(\frac{1}{1-\beta^2}\right),\yesnumber\label{eq:tdR:beta}
\end{IEEEeqnarray*}
as $P_U=P'+\alpha^2\sigma_S^2$ and by noting that $1-\beta^2=P'/P_U$ from~\eqref{eq:beta:1}. 
Our aim is to show that for any $\delta_1>0$,
\begin{IEEEeqnarray*}{rCl}
\mathbb{P}(\nexists k: |\left<\bU_{m,k}-\alpha\bS,\bS\right>|\leq n\delta_1  )\rightarrow 0,
\end{IEEEeqnarray*}
as $n\rightarrow \infty$. 
Note that
\begin{IEEEeqnarray}{rCl}\label{eq:bb:1}
&\mathbb{P}(&\nexists k: |\left<\bU_{m,k}-\alpha\bS,\bS\right>|\leq n\delta_1  )\nonumber\\
&\leq& \mathbb{P}(E_0)\nonumber\\
&&\>+\int\displaylimits_{\bs\in E_0^c}\hspace{-2mm}\mathbb{P}(\nexists k :|\left<\bU_{m,k}-\alpha\bs,\bs\right>|\leq n\delta_1  |\bS=\bs) dF_{\bS}(\bs),\IEEEeqnarraynumspace
\end{IEEEeqnarray}
where $F_{\bS}(\cdot)$ is the probability distribution function of $\bS$.
Recall from earlier that $\mathbb{P}(E_0)\rightarrow 0$ as $n\rightarrow \infty$. 
We now analyse the second term in the RHS of~\eqref{eq:bb:1}. Toward this, let us consider the following for any $\bs$ satisfying $|\|\bs\|^2-n\sigma_S^2|\leq n\delta_0(\delta_1)$ (i.e., $\bs\in E_0^c$). Then, 
\begin{IEEEeqnarray*}{rCl}
&\mathbb{P}(&\nexists k: |\left<\bU_{m,k}-\alpha\bs,\bs\right>|\leq n\delta_1 |\bS=\bs )\\
&=&\mathbb{P}( |\left<\bU_{m,k}-\alpha\bs,\bs\right>|> n\delta_1,\forall k )\\
&\stackrel{(a)}{=}&\prod_{k=1}^{2^{n\td{R}}} \mathbb{P}(|\left<\bU_{m,k}-\alpha\bs,\bs\right>|> n\delta_1 )\\
&\stackrel{}{=}&\left(\mathbb{P}(|\left<\bU_{m,1}-\alpha\bs,\bs\right>|> n\delta_1  )  \right)^{2^{n\td{R}}}\\
&=&  ( \mathbb{P}( \{\left<\bU_{m,1}-\alpha\bs,\bs\right>< -n\delta_1\} \\
&&\qquad \cup \{\left<\bU_{m,1}-\alpha\bs,\bs\right> > n\delta_1\} ))^{2^{n\td{R}}}    \\
&\stackrel{(b)}{\leq}&( \mathbb{P}( \left<\bU_{m,1}-\alpha\bs,\bs\right>< -n\delta_1 )\\
&&\>+\mathbb{P}(\left<\bU_{m,1}-\alpha\bs,\bs\right>\geq n\delta_1))^{2^{n\td{R}}}\\
&\stackrel{}{=}&( \mathbb{P}(\left<\bU_{m,1},\bs\right> <\alpha\|\bs\|^2 -n\delta_1 )\\
&&\>+\mathbb{P}(\left<\bU_{m,1},\bs\right> \geq\alpha\|\bs\|^2+ n\delta_1) )^{2^{n\td{R}}}\\
&\stackrel{}{=}&( \left(1-\mathbb{P}\left(\left<\bU_{m,1},\bs\right> \geq\alpha\|\bs\|^2 -n\delta_1  \right) \right)\\
&&\>+\mathbb{P}\left(\left<\bU_{m,1},\bs\right> \geq\alpha\|\bs\|^2+ n\delta_1\right) ) ^{2^{n\td{R}}}\yesnumber\label{eq:RHS2:ub:1}
\end{IEEEeqnarray*}
Here $(a)$ follows as $\bU_{m,k}$, $\forall k$, are independently chosen, while $(b)$ follows from the use of the union bound as well as relaxing the inequality in the second term.

To proceed, we require some additional results. We first state a lemma and then make a useful claim. 
\begin{lemma}\label{lem:shannon:lb}
Suppose $\vec{\hat{R}}$ is  chosen uniformly at random on the unit sphere surface. Then, for any unit vector $\vec{\hat{r}}$ and any $\gamma$ satisfying $0<\gamma< 1$, we have
\begin{equation*}\label{eq:shn}
\mathbb{P}\left( \left<\vec{\hat{r}},\vec{\hat{R}}\right> \geq \gamma\right)\geq  2^{-n \left( \frac{1}{2}\log{\frac{1}{1-\gamma^2}}+f(n) \right)},
\end{equation*}
where 
\begin{IEEEeqnarray*}{rCl}
f(n)= \frac{1}{2n} \log\left( 2\pi  n\gamma^2(1-\gamma^2)\left(\frac{ n \gamma^2}{n\gamma^2-(1-\gamma^2)}\right)^2  \right).
\end{IEEEeqnarray*}
There exists $n_0(\gamma)$ such that $f(n)\geq0$, $\forall n\geq n_0(\gamma)$,  and $\lim_{n\rightarrow \infty} f(n)=0$. 
\end{lemma}
\begin{IEEEproof}
The result directly follows from \cite[eqn.~(27)]{shannon-bstj1959}. To
see this,
let $\angle(\vec{\hat{r}},\vec{\hat{R}})$ denote the angle between the vectors $\vec{\hat{r}}$ and $\vec{\hat{R}}$. Then, from~\cite[eqn.~(27)]{shannon-bstj1959}, we know that
\begin{IEEEeqnarray*}{rCl}
\mathbb{P}\left(\angle(\vec{\hat{r}},\vec{\hat{R}})\leq \theta \right)\geq \left(1-\frac{1}{n}\tan^2{\theta}\right) \frac{1}{\sqrt{2\pi n}}\frac{\sin^{n-1}{\theta}}{\cos{\theta}}.
\end{IEEEeqnarray*}
Let us make the substitution $\gamma=\cos{\theta}$ in the above equation. Then, 
\begin{IEEEeqnarray*}{rCl}
&\mathbb{P}\Big(&\left<\vec{\hat{r}},\vec{\hat{R}}\right>\geq \gamma \Big)\\
&\geq& \left(1-\frac{1}{n}\frac{1-\gamma^2}{\gamma^2}\right) \frac{1}{\sqrt{2\pi n}}\sqrt{\frac{(1-\gamma^2)^{n-1}}{\gamma^2}}\\
&=& 2^{\log\left(\left(1-\frac{1}{n}\frac{1-\gamma^2}{\gamma^2}\right) \frac{1}{\sqrt{2\pi n}}\sqrt{\frac{(1-\gamma^2)^{n-1}}{\gamma^2}}\right)}\\
&=& 2^{-n \left( \frac{1}{2}\log{\frac{1}{1-\gamma^2}}+\frac{1}{2n} \log\left( 2\pi  n\gamma^2(1-\gamma^2)\left(\frac{ n \gamma^2}{n\gamma^2-(1-\gamma^2)}\right)^2  \right) \right)}.\label{eq:fn}
\end{IEEEeqnarray*}
Thus, we have shown that 
\begin{equation*}
\mathbb{P}\left(\left<\vec{\hat{r}},\vec{\hat{R}}\right>\geq \gamma\right) \geq 2^{-n \left( \frac{1}{2}\log{\frac{1}{1-\gamma^2}}+f(n) \right)},
\end{equation*} 
where $f(n)$ is as given in the lemma. It is easily verified from the expression for $f(n)$ that there exists $n_0(\gamma)$ such that $f(n)\geq 0$, $\forall n\geq n_0(\gamma)$. This completes the proof of the lemma. 
\end{IEEEproof}
We now make the following claim. The previous lemma is used in the proof of this claim. 
\begin{claim}\label{claim:bin}
There exists $\delta_0(\delta_1)$, where $\delta_0(\delta_1)\rightarrow 0$ as $\delta_1\rightarrow 0$ for $E_0$ as in~\eqref{eq:E:S}. Further, there exists $\td{\delta}_1(\delta_1)>0$, where $\td{\delta}_1(\delta_1)\rightarrow 0$ as $\delta_1\rightarrow 0$, such that for any $\bs\in E_0^c$, 
\begin{enumerate}[(i)]
\item there exists $n_0$,  such that $f_1(n)\geq 0$, $\forall n\geq n_0$  and $\lim_{n\rightarrow \infty} f_1(n)=0$, such that  
\begin{IEEEeqnarray*}{rCl}
&&\mathbb{P}(\left<\bU_{m,1},\bs\right>\geq \alpha \|\bs\|^2-n\delta_1 )\\
&&\geq 2^{-n \left( \frac{1}{2}\log{\frac{1}{1-(\beta-\td{\delta}_1)^2}}+f_1(n) \right)},
\end{IEEEeqnarray*}
\item we have 
\begin{IEEEeqnarray*}{rCl}
\mathbb{P}(\left<\bU_{m,1},\bs\right>\geq \alpha \|\bs\|^2+n\delta_1 ) \leq 2^{- \frac{(n-1)}{2}\left( \log{\frac{1}{1-(\beta+\td{\delta_1})^2}} \right)}.
\end{IEEEeqnarray*}
\end{enumerate}
\end{claim}
\begin{IEEEproof}
Consider any $\bs$ satisfying $|\|\bs\|^2-n\sigma_S^2|\leq n\delta_0(\delta_1)$ (where $\delta_0(\delta_1)$ is to be specified).
We begin with the proof of part (i).
\begin{IEEEeqnarray*}{rCl}
&\mathbb{P}(&\left<\bU_{m,1},\bs\right>\geq \alpha \|\bs\|^2-n\delta_1  )\\
&=&\mathbb{P}\left(\left<\frac{\bU_{m,1}}{\|\bU_{m,1}\|},\frac{\bs}{\|\bs\|}\right>\geq \frac{\alpha \|\bs\|}{\|\bU_{m,1}\|}-\frac{n\delta_1}{\|\bs\|\|\bU_{m,1}\|} \right)\\
&\stackrel{(a)}{=}&\mathbb{P}\left(\left<\hat{\bU}_{m,1},\hat{\bs}\right>\geq \frac{\alpha \|\bs\|}{\sqrt{n P_U}}-\frac{n\delta_1}{\|\bs\|\sqrt{n P_U}} \right)\\
&\stackrel{(b)}{\geq}&\mathbb{P}\Bigg(\left<\hat{\bU}_{m,1},\hat{\bs}\right>\geq \frac{\alpha \sqrt{n(\sigma_S^2+\delta_0)}}{\sqrt{n P_U}}\\
&&\hspace{25mm}-\frac{n\delta_1}{\sqrt{n(\sigma_S^2+\delta_0)}\sqrt{n P_U}} \Bigg)\\
&\stackrel{(c)}{\geq}&   \mathbb{P}\left(\left<\hat{\bU}_{m,1},\hat{\bs}\right>\geq \alpha\sqrt{\frac{\sigma_S^2}{P_U}}+\alpha\sqrt{\frac{\delta_0}{P_U}} -\frac{\delta_1}{\sqrt{P_U(\sigma_S^2+\delta_0)}} \right)    \\
&=&      \mathbb{P}\Bigg(\left<\hat{\bU}_{m,1},\hat{\bs}\right>\geq \alpha\sqrt{\frac{\sigma_S^2}{P_U}}\\
&&\hspace{25mm}-\left(\frac{\delta_1}{\sqrt{P_U(\sigma_S^2+\delta_0)}}-\alpha\sqrt{\frac{\delta_0}{P_U}}\right) \Bigg)        \\
&\stackrel{(d)}{=}&\mathbb{P}\left(\left<\hat{\bU}_{m,1},\hat{\bs}\right>\geq \alpha\sqrt{\frac{\sigma_S^2}{P_U}}-\td{\delta}_1 \right)\\
&\stackrel{(e)}{=}&\mathbb{P}\left(\left<\hat{\bU}_{m,1},\hat{\bs}\right>\geq \beta-\td{\delta}_1 \right)\\
&\stackrel{(f)}{\geq}&2^{-n \left( \frac{1}{2}\log{\frac{1}{1-(\beta-\td{\delta}_1)^2}}+f_1(n) \right)}.\label{eq:123:aa}
\end{IEEEeqnarray*}
We get $(a)$ since $\|\bU_{m,1}\|=\sqrt{n P_U}$, while $(b)$ follows as $|\|\bs\|^2-n\sigma_S^2|\leq n\delta_0$, $\forall \bs\in E_0^c$.  As $\sqrt{\sigma_S^2+\delta_0}<\sqrt{\sigma_S^2}+\sqrt{\delta_0}$ for $\delta_0>0$, we get $(c)$. Defining 
\begin{IEEEeqnarray*}{rCl}
\td{\delta}_1=\left(\frac{\delta_1}{\sqrt{P_U(\sigma_S^2+\delta_0)}}-\alpha\sqrt{\frac{\delta_0}{P_U}}\right)
\end{IEEEeqnarray*}
gives us $(d)$. Here we choose $\delta_0$ (as a function of $\delta_1$) small enough such that $\sqrt{\delta_0(\sigma_S^2+\delta_0)}<\delta_1$. As $\alpha\leq 1$, this implies that $\td{\delta}_1>0$.
We get $(e)$ from~\eqref{eq:beta:1}, while $(f)$ follows by using Lemma~\ref{lem:shannon:lb} with $\gamma=(\beta-\td{\delta}_1)$. Here it is easily verified using Lemma~\ref{lem:shannon:lb} that $f_1(n)$ is such that $\exists n_0$ such that $f_1(n)\geq 0$, $n\geq n_0$ and $\lim_{n\rightarrow \infty} f_1(n)=0$. This completes the proof of part (i). 

The proof of part (ii) proceeds along similar lines.
\begin{IEEEeqnarray*}{rCl}
&\mathbb{P}(&\left<\bU_{m,1},\bs\right>\geq \alpha \|\bs\|^2+n\delta_1 )\\
&\stackrel{(a)}{\leq}&\mathbb{P}\Bigg(\left<\hat{\bU}_{m,1},\hat{\bs}\right>\geq \frac{\alpha \sqrt{n(\sigma_S^2-\delta_0)}}{\sqrt{n P_U}}\\
&&\hspace{25mm}+\frac{n\delta_1}{\sqrt{n(\sigma_S^2+\delta_0)}\sqrt{n P_U}} \Bigg)\\
&\stackrel{(b)}{\leq}&     \mathbb{P}\Bigg(\left<\hat{\bU}_{m,1},\hat{\bs}\right>\geq \alpha\sqrt{\frac{\sigma_S^2}{P_U}}-\alpha\sqrt{\frac{\delta_0}{P_U}}\\
 &&\hspace{25mm}+\frac{\delta_1}{\sqrt{P_U(\sigma_S^2+\delta_0)}} \Bigg)      \\
&\stackrel{(c)}{=}&\mathbb{P}\left(\left<\hat{\bU}_{m,1},\hat{\bs}\right>\geq \beta+\td{\delta}_1 \right)\\
&\stackrel{(d)}{\leq}&2^{- (n-1)\left( \frac{1}{2}\log{\frac{1}{1-(\beta+\td{\delta_1})^2}} \right)}\label{eq:123:00}
\end{IEEEeqnarray*}
As $\|\bU_{m,1}\|=\sqrt{n P_U}$ and  $|\|\bs\|^2-n\sigma_S^2|\leq n\delta_0$, $\forall \bs\in E_0^c$, we get $(a)$. 
We get $(b)$ since $\sqrt{\sigma_S^2-\delta_0}>\sqrt{\sigma_S^2}-\sqrt{\delta_0}$ for $0<\delta_0<\sigma_S^2$, where the latter is trivially true. Recall that 
\begin{IEEEeqnarray*}{rCl}
\td{\delta}_1=\left(\frac{\delta_1}{\sqrt{P_U(\sigma_S^2+\delta_0)}}-\alpha\sqrt{\frac{\delta_0}{P_U}}\right),
\end{IEEEeqnarray*}
where $\td{\delta}_1>0$ given our choice of $\delta_0$. Using this and~\eqref{eq:beta:1}, we get $(c)$. Finally, Lemma~\ref{lem:csiszar:narayan} with $\gamma=\beta+\td{\delta}_1$ gives us $(d)$. 
This completes the proof of part (ii), and hence, establishes the claim. 
\end{IEEEproof}
Coming back to the proof, it follows from~\eqref{eq:RHS2:ub:1} and Claim~\ref{claim:bin} that for any $\bs$ such that $|\|\bs\|^2-n\sigma_S^2|\leq n\delta_0$,
\begin{IEEEeqnarray*}{rCl}
&\mathbb{P}(&\nexists k:|\left<\bU_{m,k}-\alpha\bs,\bs\right>|\leq n\delta_1 |\bS=\bs )\\
&\stackrel{}{\leq}&  \bigg( \left(1-2^{-n \left( \frac{1}{2}\log{\frac{1}{1-(\beta-\td{\delta}_1)^2}}+f_1(n) \right)} \right)\\
&&\>+2^{- (n-1)\left( \frac{1}{2}\log{\frac{1}{1-(\beta+\td{\delta}_1)^2}} \right)} \bigg) ^{2^{n\td{R}}}. \yesnumber\label{eq:cond:S}
\end{IEEEeqnarray*}
Note that the upper bound does not depend on $\bs$. We use this fact to now simplify the RHS of~\eqref{eq:bb:1} as follows.  
\begin{IEEEeqnarray*}{rcl}
&\int\displaylimits_{\bs\in E_0^c}&\mathbb{P}(\nexists k : |\left<\bU_{m,k}-\alpha\bs,\bs\right>|\leq n\delta_1  |\bS=\bs)~dF_{\bS}(\bs)\\
&\stackrel{(a)}{\leq}&  \bigg[ \left(1-2^{-n \left( \frac{1}{2}\log{\frac{1}{1-(\beta-\td{\delta}_1)^2}}+f_1(n) \right)} \right)\\
&&\hspace{15mm}+2^{- (n-1)\left( \frac{1}{2}\log{\frac{1}{1-(\beta+\td{\delta}_1)^2}} \right)} \bigg] ^{2^{n\td{R}}} \cdot \mathbb{P}(E_0^c)  \\
&\stackrel{}{\leq}&      \bigg[ 1-2^{-n \left( \frac{1}{2}\log{\frac{1}{1-(\beta-\td{\delta}_1)^2}}+f_1(n) \right)}\\
&&\hspace{15mm}+2^{- (n-1)\left( \frac{1}{2}\log{\frac{1}{1-(\beta+\td{\delta}_1)^2}} \right)}  \bigg] ^{2^{n\td{R}}}    \\
&\stackrel{}{=}&   \bigg[ 1-2^{-n \left( \frac{1}{2}\log{\frac{1}{1-(\beta-\td{\delta}_1)^2}}+f_1(n) \right)}\\
&&\cdot\bigg(1 -2^{- n\left( \frac{1}{2}\log{\frac{1-(\beta-\td{\delta}_1)^2}{1-(\beta+\td{\delta}_1)^2}}-f_1(n) -\frac{1}{2n}\log{\frac{1}{1-(\beta+\td{\delta}_1)^2}}\right)  }\bigg)  \bigg] ^{2^{n\td{R}}}    \\
&\stackrel{(b)}{=}&   \left[ 1-2^{-n \left( \frac{1}{2}\log{\frac{1}{1-(\beta-\td{\delta}_1)^2}}+f_1(n) \right)}\left(1 -2^{- nc(n)  }\right)  \right] ^{2^{n\td{R}}}    \\
&\stackrel{(c)}{\leq}&    \left[ 1-2^{-n \left( \frac{1}{2}\log{\frac{1}{1-(\beta-\td{\delta}_1)^2}}+\eta \right)}\left(1 -2^{- nc(n)  }\right)  \right] ^{2^{n\td{R}}}    \\
&\stackrel{(d)}{=}&\left[ 1-\mu(n)2^{-n \left( \frac{1}{2}\log{\frac{1}{1-(\beta-\td{\delta}_1)^2}}+\eta \right)} \right] ^{2^{n\td{R}}}\\
&\stackrel{(e)}{\leq}& e^{-2^{n\td{R}}\left[ \mu(n)2^{-n \left( \frac{1}{2}\log{\frac{1}{1-(\beta-\td{\delta}_1)^2}}+\eta \right)} \right]}\\
&\stackrel{}{=}& e^{-\mu(n) 2^{n\left(\td{R}-\left( \frac{1}{2}\log{\frac{1}{1-(\beta-\td{\delta}_1)^2}}+\eta \right)\right)}}.\IEEEeqnarraynumspace\yesnumber\label{eq:bb:12}
\end{IEEEeqnarray*}
Here~\eqref{eq:cond:S} gives $(a)$, and we get $(b)$ by defining 
\begin{IEEEeqnarray*}{rCl}
c(n)&:=& \frac{1}{2}\log{\frac{1-(\beta-\td{\delta}_1)^2}{1-(\beta+\td{\delta}_1)^2}}-f_1(n)\\
&&\hspace{15mm} -\frac{1}{2n}\log{\frac{1}{1-(\beta+\td{\delta}_1)^2}}.
\end{IEEEeqnarray*}
We get $(c)$ as follows. We choose $n$ large enough such that the exponent $c(n)>0$ as well as $0\leq f_1(n)\leq\eta$, for some $\eta>0$. The fact that such a choice of $n$ exists follows from part (i) of Claim~\ref{claim:bin} and  since 
\begin{IEEEeqnarray*}{rCl}
\log{\frac{1-(\beta-\td{\delta}_1)^2}{1-(\beta+\td{\delta}_1)^2}}>0.
\end{IEEEeqnarray*}
We discuss the choice of $\eta$ later, but note that we can choose any $\eta>0$.  This gives us $(c)$.  Next, we define $\mu(n):=1-2^{-nc(n)}$ to get $(d)$.   We know that for any $x\in [0,1]$ and any $k\geq0$, $(1-x)^{k}\leq e^{-kx}$. Now $(e)$ follows from noting that $0<\mu(n)<1$, $\forall n$, implies $0\leq\mu (n) \cdot 2^{-n l}\leq 1$, for any $l\geq 0$. 

Thus, given $\delta_1>0$ (and hence, $\td{\delta}_1(\delta_1)>0$) and from~\eqref{eq:tdR:beta}, it follows that we can choose an $\eta>0$ small enough 
such that 
\begin{IEEEeqnarray*}{rCl}\label{eq:bin:R:cond}
\td{R}>\left( \frac{1}{2}\log{\frac{1}{1-(\beta-\td{\delta}_1)^2}}+\eta \right).
\end{IEEEeqnarray*}
This guarantees that the RHS in~\eqref{eq:bb:12} goes to zero as $n\rightarrow\infty$. 
Using~\eqref{eq:bb:12} in~\eqref{eq:bb:1}, it then follows that
\begin{IEEEeqnarray*}{rCl}
\mathbb{P}(\nexists k: |\left<\bU_{m,k}-\alpha\bS,\bS\right>|\leq n\delta_1  )\rightarrow 0,
\end{IEEEeqnarray*}
as $n\rightarrow \infty$. This concludes the proof. 
\subsection{Proof of Lemma~\ref{lem:enc} }  
Given the message $M=m$, let us define the following events.
\begin{IEEEeqnarray*}{rCl}
E_0&=&\left\{\left| \|\vec{S}\|^2-n\sigma_S^2\right|>n\delta_0\right\}\\  \label{eq:E0}
E_1&=&\left\{\nexists\ k: \left|\left<\vec{U}_{m,k}-\alpha\vec{S},\vec{S}\right>\right|\leq n\delta_1 \right\}.\label{eq:E1}
\end{IEEEeqnarray*}
Here $\delta_0$, $\delta_1>0$ depend on $\delta_2$, and will be chosen such that they approach 0 as $\delta_2\rightarrow0$. Their choice will be specified later. Further,  recall the proof of Lemma~\ref{lem:binning:rate}, where $\delta_0$ is a function of $\delta_1$.
We use the same $\delta_0$ function here, and hence, only need to specify $\delta_1$.  As $\vec{S}$ is an i.i.d. Gaussian vector, where $S_i\sim\mathcal{N}(0,\sigma_S^2)$, $\forall i$, $\mathbb{P}(E_0|M=m)\rightarrow 0$ as $n\rightarrow \infty$ for $\delta_0>0$. From Lemma~\ref{lem:binning:rate}, it follows that for $\delta_1>0$, $\mathbb{P}(E_1|M=m)\rightarrow 0$ as $n\rightarrow \infty$. Let $\bU$ denote the codeword chosen. 

Let us define $E= E_0\cup E_1$. Conditioning on $E^c$ and noting that $\bU$ is chosen over the $n$-sphere with radius $\sqrt{n P_U}$, we have
\begin{IEEEeqnarray*}{rCl}
\|\vec{U}-\alpha \vec{S}\|^2 &=&\|\vec{U}\|^2+\alpha^2\|\vec{S}\|^2-2\alpha\left<\vec{U},\vec{S} \right>\\
&\stackrel{(a)}{\geq}&\|\bU\|^2+\alpha^2\|\bS\|^2-2\alpha (\alpha\|\bS\|^2+n\delta_1)\\
&=& \|\bU\|^2-\alpha^2 \|\bS\|^2-n(2\alpha\delta_1)\\
&\geq& n P_U-n\alpha^2(\sigma_S^2+\delta_0)-n(2\alpha\delta_1)\\
&=& n(P_U-\alpha^2\sigma_S^2) -n(\alpha^2\delta_0+2\alpha\delta_1)\\
&\stackrel{(b)}{=}& n P'-n\td{\delta},
\end{IEEEeqnarray*}
where $\td{\delta}=(\alpha^2\delta_0+2\alpha\delta_1)$, and $\td{\delta}\rightarrow 0$ as $\delta_0,\delta_1\rightarrow 0$. Here $(a)$ follows from Lemma~\ref{lem:binning:rate} as conditioned on $E_1^c$, we have 
\begin{IEEEeqnarray*}{rCl}
\alpha \|\bS\|^2-n\delta_1\leq\left<\vec{U},\vec{S}\right>\leq \alpha \|\bS\|^2+n\delta_1.
\end{IEEEeqnarray*}
We get $(b)$ from noting that $P'=P_{U}-\alpha^2\sigma_S^2$. 
Similarly, it can be shown that 
\begin{IEEEeqnarray*}{rCl}
\|\vec{U}-\alpha \vec{S}\|^2&\leq&  n(P'+\td{\delta}).
\end{IEEEeqnarray*}
We now ensure that $\delta_0$ and $\delta_1$ are chosen small enough such that
\begin{equation}\label{eq:delta:condition}
\delta_0+2\delta_1<\delta_2.
\end{equation}
As $\alpha\leq 1$, this implies that $\td{\delta}<\delta_2$. Hence, 
\begin{IEEEeqnarray*}{rCl}
\mathbb{P}\left(\left|\|\vec{U}-\alpha \vec{S}\|^2-n P' \right| > n\delta_2|M=m \right)&\leq& \mathbb{P}(E|M=m)\\
&\rightarrow& 0
\end{IEEEeqnarray*}
as $n\rightarrow \infty$. This completes the proof. 
\subsection{Proof of Lemma~\ref{lem:J:U}}
Let $M=m$ be the message and let $\bU$ denote the chosen codeword. We resolve the components of $\vec{J}$ and $\vec{U}$ along directions parallel and orthogonal to $\vec{S}$. We denote the latter components as $\vec{J^{\perp}}$ and $\vec{U^{\perp}}$ respectively. 
\begin{IEEEeqnarray*}{rCl}\label{eq:J:U}
\vec{J}&=&\left<\vec{J},\vec{\hat{S}} \right>\vec{\hat{S}}+\vec{J^{\perp}}\nonumber\\
\vec{U}&=&\left<\vec{U},\vec{\hat{S}} \right>\vec{\hat{S}}+\vec{U^{\perp}}.\nonumber
\end{IEEEeqnarray*}
Note that $\left<\vec{J^{\perp}},\vec{\hat{S}} \right>=0=\left<\vec{U^{\perp}},\vec{\hat{S}} \right>$, and thus, 
\begin{equation*}\label{eq:J:U:J:U:perp}
\left<\vec{J},\vec{U}\right>=\left<\vec{J},\vec{\hat{S}} \right>\left<\vec{\hat{S}},\vec{U} \right>+\left< \vec{J^{\perp}},\vec{U^{\perp}}\right>.
\end{equation*}
To prove this lemma, we need to show that for any $\delta_3>0$, 
\begin{IEEEeqnarray*}{rCl}
\mathbb{P}\left(\lvert \left<\vec{J^{\perp}},\vec{U^{\perp}} \right>\rvert > n\delta_3|M=m \right)\rightarrow 0
\end{IEEEeqnarray*}
as $n\rightarrow \infty$, i.e., $\vec{J^{\perp}}$ and $\vec{U^{\perp}}$ are nearly orthogonal for large enough $n$. 

To proceed, we introduce some notation. Let 
\begin{IEEEeqnarray*}{rCl}
\mathcal{S}^{n}\left(0,r\right)=\{\vec{w}\in \mathbb{R}^n:\|\vec{w}\|=r\}
\end{IEEEeqnarray*}
be the surface of an $n$-sphere centered at the origin and with radius $r$. For any $\vec{w}\in\mathbb{R}^n$, let $\mathcal{C}^{\perp}(\vec{w})$ denote the $(n-1)$ subspace orthogonal to $\vec{w}$. We now make the following claim.
\begin{claim}\label{claim:dp:avc:clm1}
Conditioned on $M=m$, $\vec{S}=\vec{s}$ and $\left<\vec{U},\vec{S}\right>= z$, the random vector $\vec{U}$ is uniformly distributed over  
\begin{equation}\label{eq:B}
\mathcal{B}_{z}(\vec{s})=\Big\{ z \frac{\vec{s}}{\|\vec{s}\|^2}+\vec{v}: \vec{v} \in \mathcal{S}^{n}\left(0,\rho_z(\vec{s})\right) \bigcap \mathcal{C}^{\perp}\left(\vec{s}\right)\Big\},
\end{equation}
where 
\begin{equation}\label{eq:rho:z:s}
\rho_z(\vec{s})=\sqrt{n P_U-\frac{z^2}{\|\vec{s}\|^2}}.
\end{equation}
\end{claim}
\begin{IEEEproof}
Given the symmetry of the codebook generation and the encoding, we know that the chosen codeword vector $\vec{U}$ is uniformly distributed over the set $\mathcal{S}^n(0,\sqrt{n P_U})$. Now conditioned on message $M=m$, state $\vec{S}=\vec{s}$ and $\left<\vec{U},\vec{S}\right>=z$, it follows that the codeword vector $\vec{U}$ is uniformly distributed over the set
\begin{equation}\label{eq:B:tilde}
\mathcal{\tilde{B}}_{z}(\vec{s})=\big\{ \vec{u}: \|\vec{u}\|=\sqrt{n P_U} \text{   and  } \left<\vec{u},\vec{s}\right>=z\big\}.
\end{equation}
To proceed further, we show that $\mathcal{B}_z(\vec{s})=\tilde{\mathcal{B}}_z(\vec{s})$. 
The claim then follows from observing that $\vec{U}$ is uniformly distributed over the set $\mathcal{\tilde{B}}_{z}(\vec{s})$. 
\begin{enumerate}[i)]
\item To show $\vec{u}\in \mathcal{\tilde{B}}_{z}(\vec{s})\Rightarrow \vec{u}\in \mathcal{B}_{z}(\vec{s})$.\\
Let $\vec{u}\in \mathcal{\tilde{B}}_{z}(\vec{s})$. Expressing $\vec{u}$ through its two components, one in the direction parallel to $\vec{s}$ and the other orthogonal to it, we get
\begin{equation*}
\vec{u}=\left<\vec{u},\vec{s}\right>\frac{\vec{s}}{\|\vec{s}\|^2}+\vec{u}^{\perp}.
\end{equation*} 
Note here that $\left<\vec{u}^{\perp},\vec{s}\right>=0$ and 
\begin{IEEEeqnarray*}{rCl}
\|\vec{u}^{\perp}\|=\sqrt{n P_U-\frac{z^2}{\|\vec{s}\|^2}}.
\end{IEEEeqnarray*}
Comparison with~\eqref{eq:B} completes the proof for the forward part.
\item To show $\vec{u}\in \mathcal{B}_{z}(\vec{s})\Rightarrow \vec{u}\in \mathcal{\tilde{B}}_{z}(\vec{s})$.\\
Consider some  vector $\vec{u}\in \mathcal{B}_z(\vec{s})$.  Using~\eqref{eq:B:tilde}, we can write 
\begin{IEEEeqnarray*}{rCl}
\vec{u}=z\frac{\vec{s}}{\|\vec{s}\|^2}+\vec{v}, 
\end{IEEEeqnarray*}
where 
\begin{IEEEeqnarray*}{rCl}
\vec{v}\in \mathcal{S}^{n}(0,\rho_z(\vec{s}))\bigcap \mathcal{C}^{\perp}\left(\vec{s}\right)
\end{IEEEeqnarray*}
and $\rho_z(\vec{s})$ is as given in~\eqref{eq:rho:z:s}. It can be easily verified that $\|\vec{u}\|=\sqrt{n P_U}$. Also, $\left<\vec{v,\vec{s}}\right>=0$, and hence, it can be immediately seen that $\left<\vec{u},\vec{s}\right>=z$. Thus, $\vec{u}\in\mathcal{\td{B}}_z(\vec{s})$. 
\end{enumerate}
This completes the proof of the claim.
\end{IEEEproof}
The following claim, which is equivalent to the lemma as discussed earlier, completes the proof. 
\begin{claim} \label{claim:dp:avc:clm2}
For any $\delta_3>0$,
\begin{equation*}
\mathbb{P}\left( \left| \left< \vec{J}^{\perp},\vec{U}^{\perp}\right> \right|> n\delta_3 \Big| M=m\right) \rightarrow 0,
\end{equation*}
as $n\rightarrow \infty$.
\end{claim}
\begin{IEEEproof}
We first prove the conditional version of this claim, where we condition on state $\vec{S}=\vec{s}$ and $\left<\vec{U},\vec{s}\right>=z$. From Claim~\ref{claim:dp:avc:clm1}, we know that  
\begin{IEEEeqnarray*}{rCl}
\vec{U}=z\frac{\bs}{\|\vec{s}\|^2}+\vec{V},
\end{IEEEeqnarray*}
where 
\begin{IEEEeqnarray*}{rCl}
\vec{V}\sim \text{Unif} \left( \mathcal{S}^{n}(0,\sqrt{\rho_z(\vec{s})}) \bigcap C^{\perp}\left(\vec{s}\right)\right)
\end{IEEEeqnarray*}
with $\rho_z(\vec{s})$ as given in~\eqref{eq:rho:z:s}. Now for $\delta_3>0$, we have
\begin{align}
&\mathbb{P}\bigg( \frac{\left| \left< \vec{J}^{\perp},\vec{U}^{\perp}\right> \right|}{n} >\delta_3 \bigg| M=m,\vec{S}=\vec{s}, \left<\vec{U},\vec{s}\right>=z  \bigg)\nonumber\\
&{=}\mathbb{P}\left( \frac{1}{n}\left| \left< \frac{\vec{J}^{\perp}}{\|\vec{J}^{\perp} \|},\frac{\vec{V}}{{\|\vec{V}\|}}\right> \right| >\frac{\delta_3}{{\|\vec{J}^{\perp} \| \|\vec{V}\|}} \middle| m,\vec{s}, z \right)\nonumber\\
&\stackrel{(a)}{\leq}\mathbb{P}\left( \frac{1}{n}\left| \left< \vec{\hat{J}}^{\perp},\vec{\hat{V}}\right> \right| >\frac{\delta_3}{ \sqrt{n\Lambda}\sqrt{n P_U} } \middle| m,\vec{s}, z \right)\nonumber\\ 
&\stackrel{}{=} \mathbb{P}\left( \left| \left< \vec{\hat{J}}^{\perp},\vec{\hat{V}}\right> \right| >\frac{\delta_3}{\sqrt{ \Lambda P_U}}  \middle| m,\vec{s},z\right).\nonumber\label{eq:J:V:gamma}
\end{align}
Here $(a)$ follows from noting that $\|\vec{J}^{\perp}\|\leq \|\vec{J}\|\leq \sqrt{n\Lambda}$  and $\|\vec{V}\|\leq \sqrt{n P_U}$. 

Since the shared randomness $\Theta$ is unavailable to the adversary, conditioned on $M=m$, $\bS=\vec{s}$ and $Z=z$, it follows that $\vec{J}^{\perp}$ and $\vec{V}$ are independent.  Also, both $\vec{J}^{\perp}$ and $\vec{V}$ lie in the $(n-1)$ hyperplane orthogonal to $\vec{s}$. Now using Lemma~\ref{lem:csiszar:narayan} with $\tilde{\delta}_3=\delta_3/\sqrt{\Lambda P_U}>0$, we have
\begin{IEEEeqnarray*}{rCl}
&\mathbb{P}\Big(& \left| \left< \vec{\hat{J}}^{\perp},\vec{\hat{V}}\right> \right| >\tilde{\delta}_3  \Big| m,\vec{s},z\Big)\\
&\leq& 2\left(2^{(n-1)\frac{1}{2} \log(1-\td{\delta}_3^2)    }\right)\,\,\,\forall \,m,\,\vec{s},\, z,\\
&=&2\left(2^{-(n-1)f(\td{\delta}_3)    }\right)\,\,\,\forall \,m,\,\vec{s},\, z,\yesnumber\label{eq:J:V:gamma'}
\end{IEEEeqnarray*}
where 
\begin{IEEEeqnarray*}{rCl}
f(\tilde{\delta}_3)&=&\frac{1}{2}\log\left(\frac{1}{1-\tilde{\delta}_3^2}\right)\\
&=&\frac{1}{2}\log\left(\frac{P_U\Lambda}{P_U\Lambda-\delta_3^2}\right)>0.
\end{IEEEeqnarray*}
Since the upper bound in~\eqref{eq:J:V:gamma'} tends to zero as $n\rightarrow \infty$, the conditional version of the claim follows. However, note that the bound in~\eqref{eq:J:V:gamma'} does not depend on $m$, $\vec{s}$ or $z$. Hence, the unconditioned version is also true, and the claim follows.
\end{IEEEproof}
\subsection{Proof of Lemma~\ref{lem:Y:U:unit:expr}}
Let $M=m$  be the message and let $\bU$ denote the chosen codeword. We know that
\begin{IEEEeqnarray}{rcl}\label{eq:Y:U:dot}
\left< \vec{Y},\vec{U}\right>&=&\left< \vec{U}+(1-\alpha)\vec{S}+\vec{J}+\vec{Z},\vec{U}\right>\nonumber\\
&=& \|\vec{U}\|^2+(1-\alpha)\left<\vec{S},\vec{U}\right>+\left<\vec{J},\vec{U} \right>+\left<\vec{Z},\vec{U} \right>
\end{IEEEeqnarray}
and
\begin{IEEEeqnarray}{rCl}
\|\vec{Y}\|^2	&=&\left< \vec{U}+(1-\alpha)\vec{S}+\vec{J}+\vec{Z},\vec{U}+(1-\alpha)\vec{S}+\vec{J}+\vec{Z}\right>\nonumber\\
&=& \|\vec{U}\|^2+(1-\alpha)^2\|\vec{S}\|^2+\|\vec{J}\|^2+\|\vec{Z}\|^2\nonumber\\
&&\>+2 ( \left<\vec{U},\vec{Z}\right>+\left<\vec{J},\vec{Z}\right>+\left<\vec{J},\vec{U}\right>)\nonumber\\
&&\>+2((1-\alpha)\left( \left<\vec{U},\vec{S}\right>+\left<\vec{J},\vec{S}\right>+\left<\vec{S},\vec{Z}\right>\right)).\label{eq:Y:Y}
\end{IEEEeqnarray}
Let us define the following events:
\begin{IEEEeqnarray*}{rCl}
E_0&=&\left\{\left|\|\vec{S}\|^2-n\sigma_S^2\right|>n\delta_0\right\},\\
E_1&=&\left\{\nexists\ k: \left|\left<\vec{U}_{m,k}-\alpha\vec{S},\vec{S}\right>\right|\leq n\delta_1 \right\},\\
E_2&=&\left\{\left|\|\bU-\alpha\bS\|^2-n P'\right|> n\delta_2 \right\},\\
E_3&=&\left\{\left| \left<\vec{J,\vec{U}}\right>-\left<\vec{J,\vec{\hat{S}}}\right>\left<\vec{\hat{S}},\vec{U}\right>\right|>n\delta_3\right\},\\
E_4&=&\left\{\left| \left<\vec{U,\vec{Z}}\right>\right|>n\delta_4\right\},\\
E_5&=&\left\{\left| \left<\vec{S,\vec{Z}}\right>\right|>n\delta_5\right\},\\
E_6&=&\left\{\left| \left<\vec{J,\vec{Z}}\right>\right|>n\delta_6\right\},\\
E_7&=&\left\{\left|\|\vec{Z}\|^2-n\sigma^2\right|>n\delta_7\right\}.
\end{IEEEeqnarray*}
Here $\delta_i>0$, $i=0,1,\dots,7$ depend on $\delta$, where $\delta_i$, $\forall i$, are such that they approach 0 as $\delta\rightarrow0$. The choice of $\delta_i$, $i=0,1,\dots,7$, will be specified later. Recall from the proof of Lemma~\ref{lem:binning:rate} that $\delta_0$ is a function of $\delta_1$. We choose the same $\delta_0$ function here, and hence, it is sufficient to specify $\delta_1$. Also, our choice of $\delta_0$, $\delta_1$ and $\delta_2$ will be such that $\delta_2<\epsilon_1$ as well as the condition~\eqref{eq:delta:condition} appearing in the proof of Lemma~\ref{lem:enc} is satisfied, thereby implying that $\mathbb{P}(E_2|M=m)\rightarrow 0$ as $n\rightarrow \infty$.  As $\bS$ is generated i.i.d., where $S_i\sim\mathcal{N}(0,\sigma_S^2)$, $\forall i$, we have $\mathbb{P}(E_0|M=m)\rightarrow 0$ as $n\rightarrow \infty$ for $\delta_0>0$.  From Lemma~\ref{lem:binning:rate}, it follows that $\mathbb{P}(E_1|M=m)\rightarrow 0$ as $n\rightarrow \infty$ for $\delta_1>0$. As discussed earlier, $\mathbb{P}(E_2|M=m)\rightarrow 0$ as $n\rightarrow \infty$ for $\delta_2>0$. 
Using Lemma~\ref{lem:J:U}, $\mathbb{P}(E_3|M=m)\rightarrow 0$ as $n\rightarrow\infty$ for  $\delta_3>0$. Since $\vec{Z}$ is independent of $\vec{U}$, $\vec{S}$ and $\vec{J}$, $\mathbb{P}(E_4|M=m)$, $\mathbb{P}(E_5|M=m)$ and $\mathbb{P}(E_6|M=m)\rightarrow 0$ as $n\rightarrow \infty$ for $\delta_4>0$, $\delta_5>0$ and $\delta_6>0$ respectively. $\vec{Z}$ is an i.i.d. Gaussian vector, where $Z_i\sim\mathcal{N}(0,\sigma^2)$, $\forall i$. Hence, for $\delta_7>0$, $\mathbb{P}(E_7|M=m)\rightarrow 0$ as $n\rightarrow \infty$. 
Let us define $E=\cup_{i=0}^7 E_i$ and let
\begin{IEEEeqnarray*}{rCl}
V&=&\left<\vec{\hat{J}},\vec{\hat{S}}\right>\\
W&=&\frac{1}{n}\|\vec{J}\|^2.\label{eq:W:def} 
\end{IEEEeqnarray*}
Since $\Big|\left<\vec{\hat{J}},\vec{\hat{S}}\right>\Big|\leq 1$, we have $V^2\leq 1$. It follows from $\|\vec{J}\|^2\leq n\Lambda$, that $0\leq W\leq \Lambda$. Note that $\mathbb{P}(E|M=m)$ approaches $0$ for large enough $n$ for $\delta_i$, $i=0,1,\dots,7$ as given above.
 
Recall that the codewords are chosen over the surface of an $n$-sphere of radius $\sqrt{n P_U}$. Thus, from~\eqref{eq:Y:U:dot} and~\eqref{eq:Y:Y} as well as   conditioned on the event $E^c$, 
\begin{IEEEeqnarray}{rCl}
\left<\vec{Y},\vec{U}\right>&\geq&n\left(P_U+(1-\alpha)\alpha\sigma_S^2+V\alpha\sqrt{W\sigma_S^2}-\delta_a\right),\IEEEeqnarraynumspace\label{eq:yu}
\end{IEEEeqnarray}
and 	
\begin{IEEEeqnarray}{rCl}
\left<\vec{Y},\vec{Y}\right>&\leq& n\Big(P_U+(1-\alpha)^2\sigma_S^2+W+\sigma^2+2(1-\alpha)\alpha\sigma_S^2\nonumber\\
&&\hspace{4mm} +2V\alpha\sqrt{W\sigma_S^2}+2(1-\alpha)V\sqrt{W\sigma_S^2}+\delta_b\Big).\IEEEeqnarraynumspace\label{eq:yy}
\end{IEEEeqnarray}
We know that
\begin{IEEEeqnarray}{rCl}
\big<\vec{\hat{Y}},\vec{\hat{U}}\big>=\frac{\left<\vec{Y},\vec{U}\right>}{\left<\vec{Y},\vec{Y}\right>}. \label{eq:yu:dot}   
\end{IEEEeqnarray}
\newcounter{storeeqcounter1}
\newcounter{tempeqcounter1}
Now substituting for $\left<\vec{Y},\vec{U}\right>$ from~\eqref{eq:yu} and $\left<\vec{Y},\vec{Y}\right>$ from~\eqref{eq:yy} in~\eqref{eq:yu:dot},  and noting that $P_U=P'+\alpha^2\sigma_S^2$ and $\alpha=P'/(P'+\Lambda+\sigma^2)$, we get~\eqref{eq:yu:hat} (given on top of the next page), 
%
%
\addtocounter{equation}{1}%
\setcounter{storeeqcounter1}%
{\value{equation}}%
%
\begin{figure*}[!t]
\normalsize
\setcounter{tempeqcounter1}{\value{equation}} 
\begin{IEEEeqnarray*}{rCl}
\setcounter{equation}{\value{storeeqcounter1}}
\big<\vec{\hat{Y}},\vec{\hat{U}}\big>&\geq& \frac{\left(P_U+(1-\alpha)\alpha\sigma_S^2+V\alpha\sqrt{W\sigma_S^2}-\delta_a\right)}{\sqrt{P_U\Big(P_U+(1-\alpha)^2\sigma_S^2+W+\sigma^2+2 (1-\alpha)\alpha\sigma_S^2+2V\sqrt{W\sigma_S^2} +\delta_b\Big)  }}\nonumber\\
&\stackrel{(a)}{=}& \frac{\sqrt{\alpha}\left(P'+\alpha \sigma_S^2+V \alpha \sqrt{W \sigma_S^2}-\delta_a\right)}{\sqrt{P_U\left(P'+\alpha\sigma_S^2+\alpha(W-\Lambda)+2V\alpha\sqrt{W\sigma_S^2}+\alpha\delta_b\right)  }},\label{eq:yu:hat}\yesnumber
\end{IEEEeqnarray*}%
\setcounter{equation}{\value{tempeqcounter1}} 
\hrulefill
\vspace*{4pt}
\end{figure*}
%
where $\delta_a$, $\delta_b>0$ and $\delta_a$, $\delta_b\rightarrow 0$ as $\delta_i\rightarrow 0$, $i=0,1,\dots,6$. 
Hence, conditioned on $E^c$, we have 
\begin{IEEEeqnarray*}{rcl}
&\Big<&\vec{\hat{Y}},\vec{\hat{U}}\Big>\\
&\geq& \frac{\sqrt{\alpha}\left(P'+\alpha \sigma_S^2+V \alpha \sqrt{W \sigma_S^2}\right)}{\sqrt{P_U\left(P'+\alpha\sigma_S^2+\alpha(W-\Lambda)+2V\alpha\sqrt{W\sigma_S^2}\right)  }}-\tilde{\delta},\IEEEeqnarraynumspace\yesnumber \label{eq:Y:U:a}
\end{IEEEeqnarray*}
where $\tilde{\delta}>0$ and $\tilde{\delta}\rightarrow0$ as $\delta_a$, $\delta_b\rightarrow 0$. 
It can be verified that there exists a choice of $\delta_i$, $i=0,1,\dots,7$, as functions of $\delta$, where $\forall i$, $\delta_i$ approaches 0 as $\delta\rightarrow 0$, such that, firstly, $\delta_0$, $\delta_1$ and $\delta_2$ are such that $\delta_2<\epsilon_1$ and they satisfy~\eqref{eq:delta:condition} as required in the proof of Lemma~\ref{lem:enc} earlier, and secondly,  $\td{\delta}$, which depends on $\delta_i$, $\forall i$, is such that $\td{\delta}<\delta$. 
\balance
Making this choice, conditioned on $E^c$, it follows from~\eqref{eq:Y:U:a} that
\begin{IEEEeqnarray*}{rcl}
&\Big<&\vec{\hat{Y}},\vec{\hat{U}}\Big>\\
&\geq& \frac{\sqrt{\alpha}\left(P'+\alpha \sigma_S^2+V \alpha \sqrt{W \sigma_S^2}\right)}{\sqrt{P_U\left(P'+\alpha\sigma_S^2+\alpha(W-\Lambda)+2V\alpha\sqrt{W\sigma_S^2}\right)  }}-\delta.\IEEEeqnarraynumspace\yesnumber\label{eq:Y:U:f}
\end{IEEEeqnarray*}
We now make the following claim. The proof of this claim is discussed later.
\begin{claim}\label{claim:f}
If 
\begin{IEEEeqnarray}{rCl}
f(v,w)=\frac{\sqrt{\alpha}\left(P'+\alpha \sigma_S^2+v \alpha \sqrt{w \sigma_S^2}\right)}{\sqrt{P_U\left(P'+\alpha\sigma_S^2+\alpha(v-\Lambda)+2v\alpha\sqrt{w\sigma_S^2}\right)}}\IEEEeqnarraynumspace\yesnumber\label{eq:f:def}
\end{IEEEeqnarray}
then for all $-1\leq v\leq 1$ and $0\leq w\leq \Lambda$,  
\begin{equation*}\label{eq:f:geq:f:0:Lambda}
f(v,w)\geq \theta,
\end{equation*}
where 
\begin{IEEEeqnarray*}{rCl}
\theta&=&f(0,\Lambda)\\
&=&\sqrt{\frac{\alpha(P'+\alpha\sigma_S^2)}{P_U}}.
\end{IEEEeqnarray*}
\end{claim}
Using the above claim in~\eqref{eq:Y:U:f}, conditioned on $E^c$, it follows that 
\begin{IEEEeqnarray*}{rCl}
\left<\vec{\hat{Y}},\vec{\hat{U}}\right>&\geq& \theta-\delta.\label{eq:Y:U:theta}
\end{IEEEeqnarray*}
Thus, we can conclude that 
\begin{IEEEeqnarray*}{rCl}
\mathbb{P}\left(\left<\vec{\hat{Y}},\vec{\hat{U}}\right>< \theta-\delta\big|M=m\right) &\leq& \mathbb{P}(E|M=m)\\
&\rightarrow& 0
\end{IEEEeqnarray*}
as $n\rightarrow \infty$. It only remains to prove Claim~\ref{claim:f} above. 
\begin{IEEEproof}[Proof of Claim~\ref{claim:f}]
We show that for $-1\leq v\leq 1$  and $0\leq w\leq \Lambda$,
\begin{equation}\label{eq:f:fmin}
f(v,w)\geq f(0,\Lambda).
\end{equation}
Let us first establish the simple fact that $f(v,w)\geq 0$. Consider the numerator term in~\eqref{eq:f:def}.
\begin{IEEEeqnarray*}{rCl}
&P'&+\alpha\sigma_S^2+v\alpha\sqrt{w\sigma_S^2}\nonumber\\
&= &P'+\alpha\left( \sigma_S^2+v\sqrt{w\sigma_S^2}\right) \nonumber\\
&\stackrel{(a)}{=} &P'+\frac{P'}{P'+\Lambda+\sigma^2}\left( \sigma_S^2+v\sqrt{w\sigma_S^2}\right) \nonumber\\
&= &\frac{P'}{P'+\Lambda+\sigma^2}\left(P'+\Lambda+\sigma^2+\sigma_S^2+v\sqrt{w\sigma_S^2}\right)\nonumber\\
&= &\frac{P'}{P'+\Lambda+\sigma^2}\\
&&\>\cdot~\left(P'+\left(\Lambda-w\right)+\sigma^2+\left(w+\sigma_S^2+v\sqrt{w\sigma_S^2}\right)\right)\nonumber\\
&\stackrel{(b)}{\geq} & \frac{P'}{P'+\Lambda+\sigma^2}\\
&&\>\cdot~\left(P'+\left(\Lambda-w\right)+\sigma^2+\left(w+\sigma_S^2-2\sqrt{w\sigma_S^2}\right)\right)\nonumber\\
&= &\frac{P'}{P'+\Lambda+\sigma^2}\left(P'+\left(\Lambda-w\right)+\sigma^2+\left(\sqrt{w}-\sigma_S\right)^2\right)\nonumber\\
&\stackrel{(c)}{\geq} &0.
\end{IEEEeqnarray*}
Here $(a)$ follows by substituting $\alpha=P'/(P'+\Lambda+\sigma^2)$. Then, $(b)$ follows since
$v\geq-1$, while $(c)$ follows from $w\leq \Lambda$.
Hence, we conclude that the numerator
of~\eqref{eq:f:def} is non-negative, and $f(v,w)\geq 0$.

As $f(v,\Lambda)\geq 0$ for $-1\leq v\leq 1$ and $0\leq w\leq\Lambda$, to show~\eqref{eq:f:fmin}, it is sufficient to prove 
\begin{equation}\label{eq:f:squared}
(f(v,w))^2\geq (f(0,\Lambda))^2,
\end{equation}
for $-1\leq v\leq1$ and $0\leq w\leq\Lambda$. Hence, using~\eqref{eq:f:def} in~\eqref{eq:f:squared}, we want to show that
\begin{IEEEeqnarray*}{rCl}
&&\left(\sqrt{\alpha}\frac{P'+\alpha \sigma_S^2+v \alpha \sqrt{w \sigma_S^2}}{\sqrt{P_U(P'+\alpha\sigma_S^2+\alpha(w-\Lambda)+2v\alpha\sqrt{w\sigma_S^2})  }}\right)^2\\
&&\hspace{3mm}{\geq} \left(\sqrt{\alpha}\frac{\sqrt{P'+\alpha \sigma_S^2}}{\sqrt{P_U  }}\right)^2\nonumber\\
&\Leftrightarrow&\left(\frac{P'+\alpha \sigma_S^2+v \alpha \sqrt{w \sigma_S^2}}{\sqrt{P'+\alpha\sigma_S^2+\alpha(w-\Lambda)+2v\alpha\sqrt{w\sigma_S^2}  }}\right)^2\\
&&\hspace{3mm} {\geq} \left(\sqrt{P'+\alpha \sigma_S^2}\right)^2\nonumber\\
&\Leftrightarrow&\frac{\left(P'+\alpha \sigma_S^2+v \alpha \sqrt{w \sigma_S^2}\right)^2}{P'+\alpha\sigma_S^2+\alpha(w-\Lambda)+2v\alpha\sqrt{w\sigma_S^2}  } \\
&&\hspace{3mm}{\geq} \ P'+\alpha \sigma_S^2\nonumber\\
&\Leftrightarrow&\left(P'+\alpha \sigma_S^2+v \alpha \sqrt{w \sigma_S^2}\right)^2\\
&&\hspace{3mm} {\geq} \left(P'+\alpha \sigma_S^2\right)\left(P'+\alpha\sigma_S^2+\alpha(w-\Lambda)+2v\alpha\sqrt{w\sigma_S^2} \right)\\
&\Leftrightarrow&\left(v \alpha \sqrt{w \sigma_S^2}\right)^2 {\geq} \left(P'+\alpha \sigma_S^2\right)\alpha(w-\Lambda).
\end{IEEEeqnarray*}
Since $w\leq\Lambda$, the RHS above is negative. However, $-1\leq v\leq 1$, and hence, $v^2\geq 0$. Thus,~\eqref{eq:f:squared} immediately follows and we conclude that $f(v,w)\geq f(0,\Lambda)$, for $-1\leq v\leq 1$ and $w\leq\Lambda$. This concludes the proof of the claim.
\end{IEEEproof}
This completes the proof of Lemma~\ref{lem:Y:U:unit:expr}. 

\section*{Acknowledgment}
The authors would like to thank the anonymous referees on an earlier version of the manuscript 
for their careful reading and many constructive suggestions. This has helped to improve
the quality of this manuscript. A. J. Budkuley thanks Anand D. Sarwate of Rutgers University for helpful early discussions and insightful suggestions on the problem.

A.~J.~Budkuley and B.~K.~Dey were supported in part by Bharti Centre for Communication, IIT Bombay, and in part by Information Technology Research Academy (ITRA), Government of India under ITRA-Mobile grant ITRA/15(64)/Mobile/USEAADWN/01. In addition, B.~K.~Dey was supported in part by the Department of Science \& Technology, Government of India under a grant SB/S3/EECE/057/2013. V.~M.~Prabhakaran was supported in part by the Department of Science \& Technology, Government of India through the Ramanujan Fellowship and in part by Information Technology Research Academy (ITRA), Government of India under ITRA-Mobile grant ITRA/15(64)/Mobile/USEAADWN/01.
\bibliographystyle{IEEEtran}
\bibliography{IEEEabrv,References}
\end{document}